\documentclass[a4paper,fleqn,usenatbib]{mnras}

\usepackage{newtxtext,newtxmath}
\usepackage[normalem]{ulem}

\usepackage[T1]{fontenc}
\usepackage{ae,aecompl}

\usepackage{graphicx}	
\usepackage{amsmath}	

\usepackage{amssymb}	
\usepackage{movie15}    
\usepackage{soul}       
\usepackage{orcidlink}

\newcommand{\delayedtau}{delayed-$\tau$}
\newcommand{\deltaGV}{$\Delta_{\mathrm{GV}}$}
\newcommand{\grizli}{{\textsc {Grizli}}}
\newcommand{\dynesty}{{\textsc {dynesty}}}
\newcommand{\bagpipes}{{\textsc {bagpipes}}}
\newcommand{\sextractor}{{\textsc {SExtractor}}}

\newcommand{\Oii}{[\ion{O}{ii}]$\lambda$3727}
\newcommand{\Oiii}{[\ion{O}{iii}]$\lambda$5007}
\newcommand{\Sii}{[\ion{S}{ii}]$\lambda$6725}
\newcommand{\BGR}{blue-cloud $\rightarrow$ green-valley $\rightarrow$ red-sequence}

\title[Across the Green Valley]{Across the Green Valley with {\it HST} grisms: colour evolution, crossing time-scales and the growth of the red sequence at $z=1.0-1.8$}
\author[Noirot et al.]{%
Ga\"el Noirot$^{1}$\thanks{e-mail: gnoirot@ap.smu.ca},
Marcin Sawicki$^{1}$\thanks{Canada Research Chair}\orcidlink{0000-0002-7712-7857},
Roberto Abraham$^{2}$\orcidlink{0000-0002-4542-921X},
Maru\v{s}a Brada\v{c}$^{3}$\orcidlink{0000-0001-5984-0395},
\newauthor
Kartheik Iyer$^{2}$\orcidlink{0000-0001-9298-3523},
Thibaud Moutard$^{4}$\orcidlink{0000-0002-3305-9901},
Camilla Pacifici$^{5}$\orcidlink{0000-0003-4196-0617},
Swara Ravindranath$^{5}$\orcidlink{0000-0002-5269-6527},
\newauthor
and Chris~J.~Willott$^{6}$\orcidlink{0000-0002-4201-7367}
\\
$^{1}$Institute for Computational Astrophysics and Department of Astronomy \& Physics, Saint Mary's University, 923 Robie Street, Halifax, NS B3H 3C3, Canada\\
$^{2}$Dunlap Institute for Astronomy and Astrophysics, University of Toronto, 50 St George St, Toronto, ON M5S 3H4, Canada\\
$^{3}$University of California Davis, Department of Physics and Astronomy, One Shields Ave., Davis, CA 95616, USA\\ 
$^{4}$Aix Marseille Universit\'e, CNRS, CNES, LAM - Laboratoire d’Astrophysique de Marseille, 38 rue F. Joliot-Curie, F-13388, Marseille, France\\
$^{5}$Space Telescope Science Institute, 3700 San Martin Drive, Baltimore, MD 21218, USA\\
$^{6}$NRC Herzberg, 5071 West Saanich Rd, Victoria, BC V9E 2E7, Canada\\
}

\date{Accepted XXX. Received YYY; in original form ZZZ}

\pubyear{2022}

\begin{document}
\label{firstpage}
\pagerange{\pageref{firstpage}--\pageref{lastpage}}
\maketitle

\begin{abstract}
We measure the colour evolution and quenching time-scales of $z=1.0-1.8$ galaxies across the green valley. We derive rest-frame $NUVrK$ colours and select blue-cloud, green-valley and red-sequence galaxies from the spectral energy distribution modelling of CANDELS GOODS-South and UDS multi-band photometry. Separately, we constrain the star-formation history (SFH) parameters (ages, $\tau$) of these galaxies by fitting their deep archival {\it HST} grism spectroscopy. 
We derive the galaxy colour-age relation and show that only rapidly evolving galaxies with characteristic delayed-$\tau$ SFH time-scales of $<0.5$~Gyr reach the red sequence at these redshifts, after a period of accelerated colour evolution across the green valley.
These results indicate that the stellar mass build-up of these galaxies stays minimal after leaving the blue cloud and entering the green valley (i.e., it may represent $\lesssim 5\%$ of the galaxies' final, quiescent masses).
Visual inspection of age-sensitive features in the stacked spectra also supports the view that these galaxies follow a quenching sequence along the blue-cloud $\rightarrow$ green-valley $\rightarrow$ red-sequence track.
For this rapidly evolving population, we measure a green-valley crossing time-scale of $0.99^{+0.42}_{-0.25}$~Gyr and a crossing rate at the bottom of the green valley of $0.82^{+0.27}_{-0.25}$~mag/Gyr.
Based on these time-scales, we estimate that the number density of massive ($M_\star>10^{10} M_\odot$) red-sequence galaxies doubles every Gyr
at these redshifts, in remarkable agreement with the evolution of the quiescent galaxy stellar mass function.
These results offer a new approach to measuring galaxy quenching over time and represent a pathfinder study for future {\it JWST}, {\it Euclid}, and {\it Roman Space Telescope} programs.
\end{abstract}

\begin{keywords}
galaxies: general -- galaxies: evolution -- galaxies: high-redshift
\end{keywords}



\section{Introduction}\label{sec:introduction}

The bimodal colour distribution of galaxies has been fairly well studied and characterized up to relatively high redshifts (at least up to z$\sim$3; e.g., \citealp{Whitaker2011, Fang2018}). Blue galaxies have young stellar populations and are experiencing active episodes of star formation while red galaxies are older and predominantly quenched.
The quiescent (red and dead) galaxy population is thought to build via the cessation of star formation in star-forming galaxies through in-situ and/or external quenching mechanisms \citep[e.g.,][]{Peng2010, Peng2012, Darvish2016, Huertas-Company2016, Bower2017, Nantais2017}.
Proposed in-situ mechanisms include internal processes preventing the accretion of cold gas from the intergalactic medium (IGM), or preventing the cooling or collapse of available gas in the interstellar medium (ISM) (through, e.g., stellar feedback, AGN feedback, starvation, or bulge growth; \citealp{Martig2009, Peng2015, Zolotov2015, Pontzen2017, Trussler2020}). On the other hand, proposed environmentally-driven quenching mechanisms generally include processes that  remove, heat, or abruptly consume available gas (through, e.g., mergers, harassment, strangulation, ram-pressure or tidal stripping; \citealp{Gunn1972, Moore1996, Abadi1999, Bosch2008, Zolotov2015, Poggianti2017, Pontzen2017}). Some {studies} have suggested that for $M_{\star}\gtrsim10^{9.5} M_{\odot}$ galaxies, environmental quenching starts playing a major role at $z\lesssim1$, whereas in-situ quenching might be dominant at $z\gtrsim1$ and correlates with stellar mass (\citealp{Darvish2016}; see also \citealp{Peng2010}). While the contribution of external mechanisms might not be as strong at $z\gtrsim1$ compared to $z\lesssim1$, at all cosmic epochs various combinations of the different quenching processes can be interlinked and may in reality be at play in the quenching of galaxies (e.g., \citealp{Zolotov2015, Bower2017}; including interplay between external and in-situ mechanisms, such as, for instance, AGN feedback triggered by merger events; \citealp{Hopkins2008, Yuan2010, Pontzen2017, Rodriguez2019}). Given this complexity and the multiplicity of quenching pathways, there is still to date no clear consensus on the relative contributions of the different quenching mechanisms over time.

Cosmological hydrodynamical simulations have however shown that galaxy quenching time-scales vary greatly \citep{Sales2015, Nelson2018, Rodriguez2019, Wright2019} and may be generally categorized into two classes, fast or slow, depending on galaxy properties (typically stellar or halo mass), cosmic epoch, and the quenching mechanisms involved \citep[e.g.,][]{Rodriguez2019, Wright2019}. Constraining quenching time-scales from observations at different cosmic epochs and using different samples of galaxies (i.e., selected in stellar or halo masses, local density, morphology, etc) may therefore help us better disentangle the relative contributions of the physical processes responsible for the cessation of star-formation over cosmic time.
Over the last decade, a number of observational studies have started to constrain the time-scales associated with galaxy quenching in the low, intermediate and high-redshift Universe. 
By constraining galaxy star-formation histories (SFHs) through the spectral energy distribution (SED) modelling of multi-wavelength broadband photometry, \citet{Carnall2020} have shown for instance that massive quiescent galaxies already in place at $2<z<5$ must have quenched rapidly and at early times. Other {studies}, at slightly lower redshifts ($z\sim1-2$), have shown that there exist a diverse range of possible quenching time-scales for $M_{\star}\gtrsim10^{10} M_{\odot}$ galaxies at these redshifts, from $\sim 10^{2}$~Myr to $\lesssim2$~Gyr (\citealp{Belli2019, Estrada2020, Wild2020}; where rapid quenching is often associated with galaxies experiencing a post-starburst phase). On the other hand, observational studies probing the local Universe typically find longer quenching time-scales of the order of several Gyr \citep[e.g.,][]{Schawinski2014, Peng2015, Trussler2020}, although rapid quenching scenarios remain possible \citep[e.g.,][]{Schawinski2014}.
At low and intermediate redshifts (i.e., $z<1$), these rapid and slow quenching time-scales have been shown to be associated with different quenching pathways through the green valley \citep[e.g.,][]{Schawinski2014, Moutard2016b}. 

The green valley is thought of as the transitional region between the star-forming and quiescent populations and is often defined using rest-frame colour diagnostics used to identify star-forming, dusty star-forming and quiescent galaxies.
Indeed, these populations (star-forming, dusty star-forming, and quiescent galaxies) have for more than a decade been identified based on their rest-frame colours using diagnostics such as the rest-frame $UVJ$ \citep[e.g.,][]{Labbe2005, Wuyts2007, Williams2009, Fang2018}, or, more recently, the rest-frame $NUVrK$ \citep[e.g.,][]{Arnouts2013, Moutard2016b, Vergani2018} {colour-colour diagrams.} To distinguish between the three populations, these methods rely on the presence or absence of age-sensitive and dust-sensitive features in the SEDs of galaxies. Specifically, filters bracketing the rest-frame $4000$~\AA\ region (e.g., $U-V$, $NUV-r$) are sensitive to galaxy stellar ages via ({\it i}) the strong UV and blue continuum excess of young, massive O and B stars ($T>10^4~K$, $t<10^8$~yrs) which produce blue $U-V$ or $NUV-r$ colours, and ({\it ii}) the strong Balmer and $4000$\AA\ breaks caused by hydrogen, calcium, and metallic absorption lines in the spectra of older stellar populations below rest-frame $4000$~\AA, which produce red $U-V$ or $NUV-r$ colours \cite[e.g.,][]{Bruzual1983, Hamilton1985, Jaschek1995, Bruzual2003}. However, strong dust {attenuation} at short wavelengths in dusty star-forming galaxies can also mimic the typical red $U-V$ or $NUV-r$ colours of quiescent populations, resulting in an old-vs.-dusty degeneracy \citep[e.g.,][]{Brammer2009}. To counter this potential degeneracy, filters bracketing redder wavelengths (e.g., $V-J$, $r-K$) are used in these colour-colour diagrams to discriminate between dust-absorbed and quiescent populations as only the dusty star-forming galaxies will show red $V-J$ or $r-K$ colours produced by their dust attenuation curve attenuating the underlying blue stellar continuua \citep[e.g.,][]{Williams2009, Patel2011, Arnouts2013, Fang2018}.

The main, but crucial, difference between the $NUVrK$ and the shorter wavelength-baseline $UVJ$ diagnostics is that {$NUV-r$ colours} {offer a wider dynamic range in colour between young and old stellar populations compared to $U-V$ colours, in part due to the stronger blue continuum excess of young stellar populations in the rest-frame $NUV$. $NUV-r$ colours are therefore} more sensitive to stellar population ageing (and especially the ageing of young stellar populations) than are $U-V$ colours (e.g., \citealp{Bruzual1993, Bruzual2003, Salim2005, Arnouts2013}). For this reason, the $NUVrK$ is better than the $UVJ$ diagram at resolving the separation between the star-forming and quiescent populations, and allows the identification of the so-called green-valley galaxies which accumulated evidence shows are a transitional population between the star-forming blue cloud and the quiescent red sequence (e.g., green-valley galaxies are intermediate between the blue and red populations in terms of sSFR, \citealp[e.g.,][]{Siudek2018, Moutard2020b}, or other properties, e.g., see \citealp{Salim2014} for a review).

In this paper, we aim at measuring the quenching (transition) time-scale of galaxies through the green valley (i.e., the green-valley crossing time-scale) in the cosmic noon Universe, and testing the hypothesis that the quiescent galaxy population builds-up via the quenching of star-forming galaxies. We make use of the rest-frame $NUVrK$ diagram to photometrically classify blue-cloud, green-valley, and red-sequence galaxies, and we compare our measured quenching (i.e., crossing) time-scales to the growth of the quiescent stellar mass function reported in the literature to test our hypothesis.

The paper is structured as follows. Seciton~\ref{sec:parentdata} describes our parent photometric and spectroscopic datasets. In Section~\ref{sec:SampleSelection}, we describe our photometric sample selection as well as our broadband SED-fitting procedure, including rest-frame colours and our green-valley definition. In Section~\ref{sec:grism_sample}, we describe our {\it HST} grism spectroscopic sample selection as well as the {procedure we use to fit the} grism data, while Section~\ref{sec:delayedtaumodels} provides details of the SFH models we use in this work. Section~\ref{sec:results} {reports and discusses our results}, including the determination of spectroscopic SFH populations (Section~\ref{sec:tauDistribution}), the analysis of correlations between photometric colours and spectroscopically-derived galaxy ages and SFH time-scales (Section~\ref{sec:agetau}), visual inspection of the stacked spectra of our different galaxy populations (Section~\ref{sec:stackspectra}), the derivation of the galaxy colour-age relation (Section~\ref{sec:colourage}), measurements of the green-valley crossing time-scales (Section~\ref{sec:gvcrossingtime}), and predictions of the growth of the red sequence (Section~\ref{sec:SMFgrowth}). We summarize our results in Section~\ref{sec:conclusion}. Throughout {the paper} we use AB magnitudes and a flat $\Lambda$CDM cosmology with $\Omega_{\rm M} = 0.3$, $\Omega_{\Lambda} = 0.7$, and $H_0 = 70 $\,km\,s$^{-1}$\,Mpc$^{-1}$.

\section{Parent Datasets}\label{sec:parentdata}
\subsection{Photometry}

Our parent photometric dataset consists of the CANDELS GOODS-South \citep{Guo2013} and UDS \citep{Galametz2013} multi-wavelength catalogues. 
While the CANDELS observations alone consist of {\it HST} imaging in $5$ to $9$ filters depending on field and depth, the GOODS-S and UDS multi-wavelength catalogues also include ancillary ground- and space-based observations in a total of $17$ and $19$ bands, respectively, from the near-UV to the mid-infrared. The photometry reaches $5\sigma$ limiting depths of $\gtrsim26$~mag (AB) in all these bands except the two reddest {\it Spitzer}/IRAC  channels at $5.8\micron$ and $8.0\micron$ ($\sim23.7$ and $\sim22.3$ mag for GOODS-S and UDS, respectively). The catalogues contain  a total of $70862$ sources, including $34930$ and $35932$ in GOODS-S and UDS, respectively.

For both fields, source detection was performed by the CANDELS team with \sextractor\ \citep{Bertin1996} on the WFC3 F160W images using the so-called cold+hot modes \citep{Galametz2013}.  Photometry in the additional {\it HST} bands was then obtained using \sextractor\ in dual-image mode after PSF-matching the images to the F160W data. For the lower-resolution ground-based and {\it Spitzer} imaging, photometry was derived using the TFIT software \citep{Laidler2007} that uses a template fitting method with prior information from the high resolution data to accurately recover fluxes in the low resolution images.

The \citet{Guo2013} and \citet{Galametz2013} catalogues 
{include galaxy physical properties and a number of rest-frame colours derived from SED fitting}, but do not contain $NUV$ rest-frame colours required to trace galaxy evolution in the $NUVrK$ diagram. Additionally, the new generation of SED-fitting codes now available include more robust treatments of the underlying physics as well as better exploration of the parameter spaces and posterior estimation through full Bayesian approaches. For these reasons, we therefore carry out our own SED fitting, using the CANDELS PSF-matched photometry as inputs but deriving new photometric redshifts, $NUVrK$ rest-frame colours, and galaxy stellar masses following the procedure described in Sec.~\ref{sec:broadebandSED}.

\subsection{Spectroscopy}\label{sec:spec_data}

Our parent spectroscopic dataset consists of publicly available {\it HST} WFC3/IR G102~\&~G141 grism spectroscopy over the GOODS-S and UDS footprints. The G102 grism covers the wavelength range $\lambda = 0.80-1.15$~\micron~with a throughput $>10\%$ at a low spectral resolution of $R=210$, while the G141 covers the range  $\lambda = 1.08-1.70$~\micron~with a throughput $>10\%$ at a low spectral resolution of $R=130$. Dispersions of the G102~\&~G141 grisms are $24.5$~\AA\,pix$^{-1}$ and $46.5$~\AA\,pix$^{-1}$, respectively. These grisms were chosen as they cover strong age- and metallicity-sensitive spectral features at our redshifts of interest. These features are the $4000$\AA\ break, and metallic and hydrogen lines (i.e., Ca H \&\ K, H$\delta$, G-band, H$\gamma$, H$\beta$, Mg, Na) which cover the rest-frame wavelength range $\lambda = 4000-6000$~\AA. This corresponds to visibilities within $\lambda = 8000-12000$~\AA~at $z=1.0$ and $\lambda = 11200-16800$~\AA~at $z=1.8$, fully covered by the G102~\&~G141 grisms. At slightly lower and higher redshifts, not all features {are} visible within the G102~\&~G141 wavelength windows, but a number of them are still visible depending on redshift.

We extract our spectroscopic dataset from the database of the Grism Redshift and Line Analysis software \citep[\grizli,][]{Brammer2019} hosted on G.~Brammer’s Amazon Web Services repository\footnote{Access available on reasonable request by contacting G.~Brammer.}. This database consists of catalogues (including redshift fits) and enhanced data products (spectra, emission-line maps, etc) of reprocessed archival {\it HST} grism spectroscopic observations in a number of extragalactic fields including CANDELS \citep{Grogin2011, Koekemoer2011}, 3D-HST \citep{Brammer2012, Momcheva2016}, WISPS \citep[WFC3 Infrared Spectroscopic Parallel Survey;][]{Atek2010}, HFF \citep[Hubble Frontier Fields;][]{Lotz2017}, CARLA \citep[Clusters Around Radio-Loud AGN;][]{Noirot2016, Noirot2018}, and others. Reprocessing of these data (data query, reduction, spectral extraction, and additional products) had been performed with \grizli\ version {\tt 1.0-6-gb542b34} at the time of extraction.

We refer the reader to the official \grizli\ webpages\footnote{\url{https://github.com/gbrammer/grizli}} for a detailed overview of the processing steps \citep[see also, e.g.,][]{Abramson2020}. In short, after initial query of the archive, the different processing steps include: exposure file association, bad pixel and cosmic-ray rejection, persistence masking, relative exposure-level astrometric alignment, flat-fielding, direct imaging and grism sky-background subtraction, fine astrometric alignment to the Gaia Data Release 2 \citep[][]{Gaia2018} or other available reference catalogue, direct image mosaicking, and \sextractor\ source detection and matched-aperture photometry on the different filters available. The source catalogue is then used to associate sources to their spectral traces, and perform morphology-dependant spectral modelling and contamination removal at the exposure level. Next, 2D spectra are extracted at the individual exposure level as well as stacked in the resampled drizzled space. Redshift fits, following a template-fitting approach similar to that of \textsc{EAZY} \citep{Brammer2008}, are then performed on the 2D spectra and emission line maps are created.

From this \grizli\ database, we first extract the \grizli\ catalogues for all sources in the GOODS-S and UDS footprints ($15501$ and $34638$ sources, respectively), and only extract spectral products for those in our final spectroscopic sample that we describe in Sec.~\ref{sec:grism_sel}. They are the spectra we use with our grism data fitting procedure that we describe in Sec.~\ref{sec:grism_fit} and from which we derive galaxy physical parameters and SFHs independently from our broadband SED-fitting.

\section{Photometric Sample}
\label{sec:SampleSelection}

\subsection{Sample selection}
From the full CANDELS GOODS-S and UDS catalogues, we first select sources with good photometry. These are sources that are neither contaminated by bright nearby objects, saturated, truncated, nor suffer other photometric issues (i.e., have $FLAGS = 0$, and $PhotFlag = 0$ in the CANDELS catalogues). We additionally reject sources with stellarity indexes $> 0.25$ as measured with \sextractor\ on the F160W detection band, and exclude objects identified as AGN in \citet{Xue2011} and \citet{Ueda2008} for GOODS-S and UDS, respectively ($AGNFlag = 0$). From this first selection of non-AGN, extragalactic objects with good photometry, we then further reduce our sample to objects brighter than 26~AB mag in the $H$-band, measured by \sextractor\ $MAG\_AUTO$ on the F160W data, and use an additional cut of $\rm SNR > 3$ in the F160W band to ensure robust detections. As we are interested in investigating the colour evolution of galaxies in the late cosmic noon epoch for which near-infrared {\it HST} grism data cover spectral features of interest ($4000$\AA\ break, hydrogen and metallic absorption lines, hydrogen and oxygen emission lines, etc), we finally use a redshift cut of $0.9 < z_{best} < 1.8$. Here, $z_{best}$ is the best redshift estimate available in the CANDELS catalogue, i.e., either a spectroscopic redshift from the literature, if available, or the photometric redshift as estimated by the CANDELS team from a hierarchical Bayesian approach that takes into account and combines the redshift probability distributions derived by six different CANDELS team members (\citealp{Santini2015}; see also \citealp{Dahlen2013} for details on the individual methods).  After applying these steps, we have our primary photometric sample which consists of a total of $13279$ sources, including $5569$ in GOODS-S and $7710$ in UDS, respectively.

\subsection{Broadband SED-fitting}
\label{sec:broadebandSED}

We use the CANDELS PSF-matched broadband photometry of our GOODS-S and UDS sample to homogeneously derive rest-frame colours and physical parameters of our sources. 

\subsubsection{Galaxy physical properties}

The broadband SED-fitting technique has for many years now provided a way to estimate physical parameters of high-$z$ galaxies from their broadband fluxes \citep[e.g.,][]{Sawicki1998, Papovich2001, Sawicki2001, Labbe2005, Shapley2005, Erb2006, Arnouts2007, Gawiser2007, Yabe2009, Sawicki2012a, Sawicki2012b}.  In recent years, many improvements have been made in SED-fitting codes to alleviate parameter-space gridding shortcomings as well as to optimize likelihood convergence and better characterize parameter posterior probabilities. This new generation of codes (e.g., \textsc{galmc}: \citealp{Acquaviva2011}; \textsc{beagle}: \citealp{Chevallard2016}; \textsc{prospector}: \citealp{Leja2017, Johnson2021}; \textsc{dense basis}: \citealp{Iyer2017, Iyer2019}; \bagpipes: \citealp{Carnall2018, Carnall2019b}) is based on optimized Bayesian inference algorithms (e.g., \textsc{multinest}: \citealp{Feroz2009,Feroz2019}; \textsc{emcee}: \citealp{Foreman2013}; \textsc{george}: \citealp{Ambikasaran2015}; \dynesty: \citealp{Speagle2020}) and also allows for the exploration of flexible star-formation histories. In addition, {these new codes} also include treatment of a number of physical components such as, e.g., nebular emission lines, multi-component dust emission models, etc, that were often not considered in older codes.

To take advantage of these recent improvements, in this work we use the Bayesian inference SED-fitting code \bagpipes\ \citep{Carnall2018} to derive new photometric redshifts, rest-frame colours, and stellar masses of the sources in our photometric dataset. The code uses the (\citealt{Bruzual2003}; BC03) stellar population synthesis models, \citet{Kroupa2001} initial mass function (IMF), \citet{DraineLi2007} dust emission models, and Cloudy \citep{Ferland2017} photoionization prescriptions, and performs parameter-range exploration and posterior estimation using \textsc{pymultinest} \citep{Buchner2014}, the \textsc{python} wrapper of \textsc{multinest}, a nested-sampling \citep{Skilling2004} algorithm. \textsc{Multinest}, similarly to other nested-sampling algorithms (e.g., \dynesty; \textsc{ultranest}: \citealp{Buchner2021}), continuously explores and reduces the parameter space on the fly (see \citealp{Feroz2009,Feroz2019,Buchner2014} for details).  \bagpipes\ also includes several parametrizations of star formation histories which determine the overall shape of when and how many stars formed over time (see, e.g., \citealp{Pacifici2015, Iyer2017, Lee2018, Carnall2019a, Leja2019, Lower2020} for the pros and cons of choosing different SFH models). In this paper, we adopt the delayed-$\tau$ SFH model of the form 
\begin{equation}
\label{eq:delayedtau}
SFR(t) \propto t \exp(-t/\tau),
\end{equation}
where $t$ is time since the onset of star formation ({i.e.,} ``age'') and $\tau$ is the characteristic time at which the SFR reaches its maximum. Section~\ref{sec:delayedtaumodels} discusses the \delayedtau\ model and its characteristic time-scales in more detail. We defer the exploration of other SFHs, including more flexible SFHs, to a future paper. Flexible (i.e., ``non-parametric'') SFHs are indeed better suited to recover rejuvenation or multiple burst episodes, which cannot be investigated with our current models. While this is a limitation of our current study, it is not expected to significantly affect our results and conclusions (see, e.g., \citealp{Chauke2019}). 

We fit each object in our sample with a set of six free parameters with uniform priors. The parameters we fit are: 
(a) redshift, $z$, free to vary in the range $0.75$ to $1.95$, 
(b) age, from $0.1$~Gyr to the age of the Universe at redshift $z$, 
(c) $\tau$, from $0.05$ to $3$~Gyr, 
(d) the total mass of stars formed throughout the object's star-formation history, $M_{T}$, from $6$ to $13$ in~log$_{10}(M_{\odot})$, 
(e) metallicity, $Z$, from $0.1$ to $2.5$~Z$_{\odot}$ {(with Z$_{\odot} = 0.02$)}, and
(f) dust attenuation in the V band, $A_V$, from $0$ to $2$~mag, parametrized using the default \citet{Calzetti2000} dust attenuation law. While all these six parameters are fit, we are mostly interested here in the stellar masses\footnote{Note that for a given object, its total mass $M_{T}$ is the integral of its SFH and therefore includes the mass of material recycled back to the interstellar medium (ISM), while its stellar mass $M_\star$ is the current (living) stellar mass at the redshift of observation.} ($M_\star$), photometric redshifts, and derived galaxy rest-frame colours. Note that we derive new photometric redshifts for all galaxies in our sample, including galaxies with available spectroscopic redshifts in the CANDELS catalogues. Also, while the Universe is about $5.8$~Gyr old at $z=1.0$, we do not attempt to fit $\tau$ parameters beyond values of $\tau = 3.0$~Gyr. As discussed in Sections~\ref{sec:tauDistribution}~\&~\ref{sec:agetau}, we cannot well discriminate between the SFHs of galaxies with large $\tau$ values of about $>1.5$~Gyr, and only treat this group as a single ‘large-$\tau$’ population. Moreover, we are mostly interested in our analysis in galaxy populations with smaller $\tau$ values for which the posteriors are well constrained. Using a larger $\tau$ prior would therefore not affect our conclusions.

\subsubsection{Rest-frame colours}\label{sec:restframecolours}

Rest-frame colours can be directly derived from the posterior SED models by integrating the rest-frame posterior SEDs over the filters of interest. In the past, this often led to colour issues arising from the non-optimized step-wise sampling of the parameter space or other modelling assumptions (no nebular emission, etc). To mitigate such SED-fitting code limitations and properly derive rest-frame colours, previous works used correction terms based on the observed-frame colours closest in (rest) wavelength to the rest-frame colours of interest\footnote{Note that this has the advantage to minimize the $k$-correction term.} (e.g., \citealp{Rudnick2003, Ilbert2005, Taylor2009, Williams2009, Patel2011, Ilbert2013, Moutard2016b}). Although \bagpipes\ and other recent codes do not suffer from this so-called gridding of parameter space and better model the underlying physics compared to the previous generation of codes, rest-frame colour estimation is still often unrealistic and can lead to unrealistically small scatter, underestimated uncertainties, and boxy patterns in colour-colour space that are likely due to improper propagation of parameter uncertainties (e.g., redshift), limitations from the explored parameter ranges, or modelling assumptions {(see details in Appendix~\ref{app:colours})}. To alleviate any such undesirable effects and derive realistic rest-frame colours, we therefore follow a procedure similar to that of the standard corrections applied to derive rest-frame colours in the literature (e.g., \citealp{Hogg2002, Dahlen2005, Dahlen2013, Ilbert2005, Ilbert2010, Ilbert2013, Moutard2016b, Beifiori2017, Moutard2020a}), which we here apply to each iteration of the fitting procedure to obtain the full posterior of our corrected rest-frame colours with proper propagation of all parameter uncertainties.

\begin{figure}
    \centering
    \includegraphics[width=8.3cm]{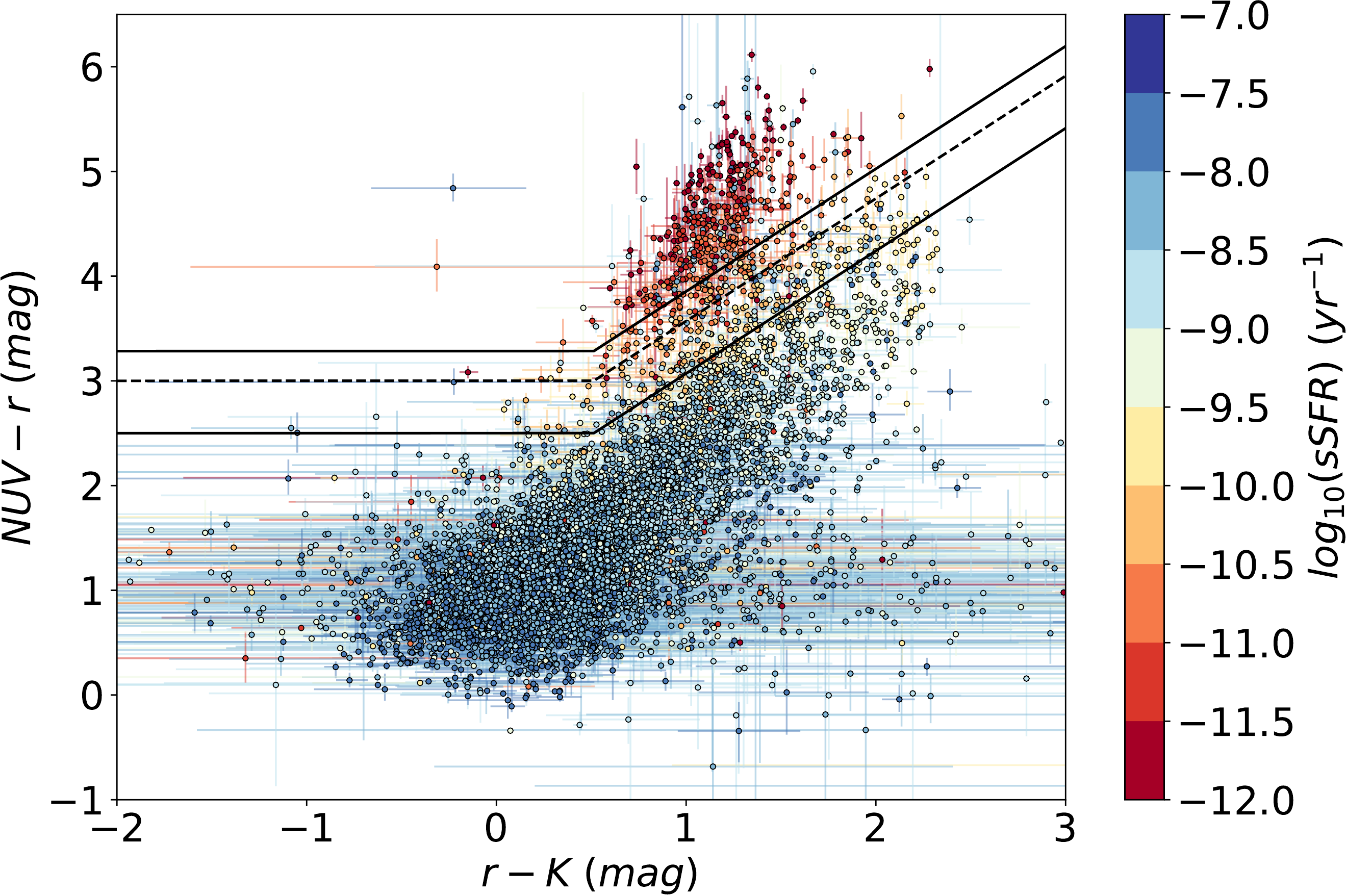}
    \includegraphics[width=8.3cm]{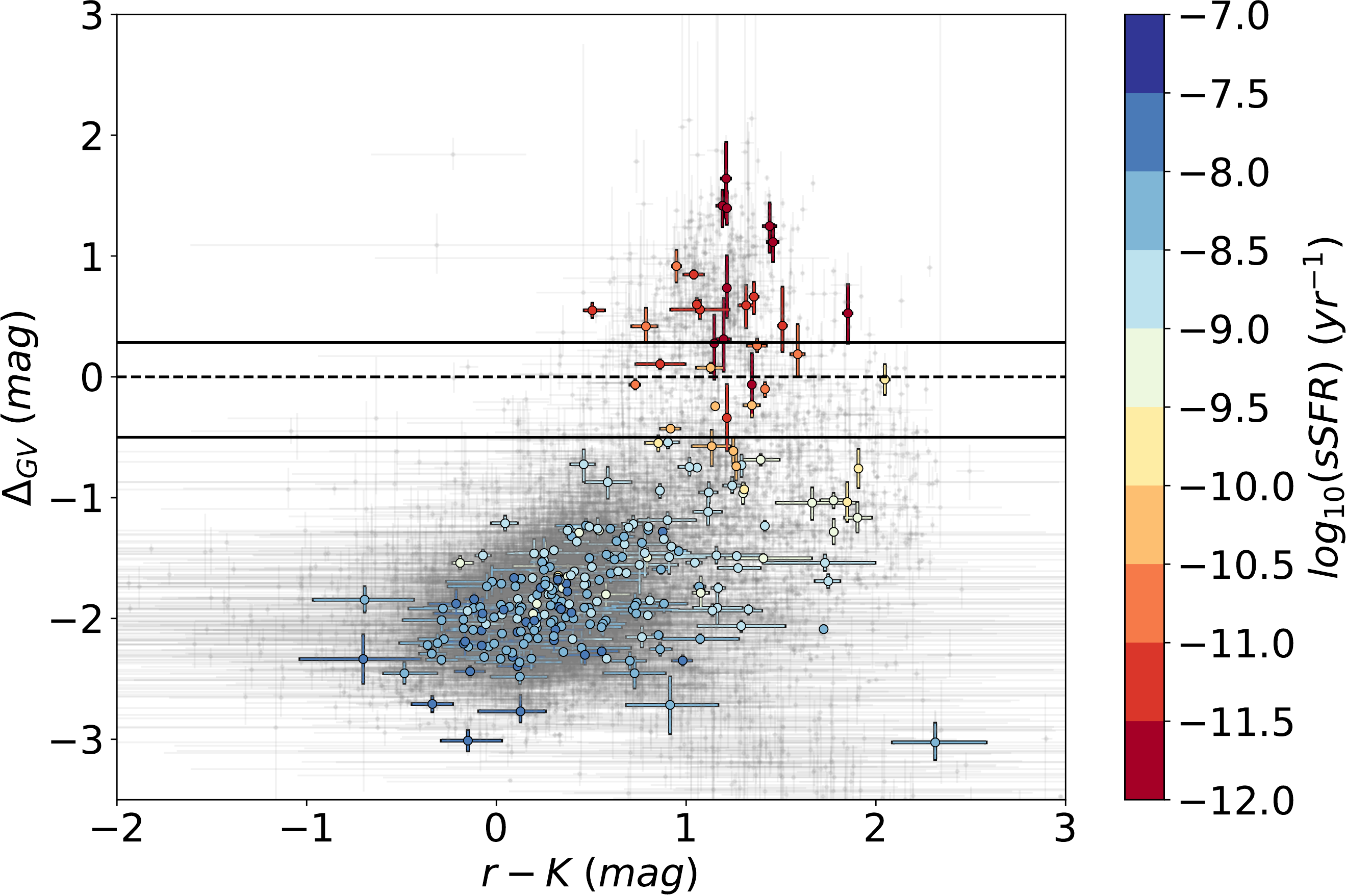}
    \caption{Rest-frame $NUVrK$ colours.  The top panel shows the distribution of rest-frame photometric colours of galaxies in the $NUVrK$ diagram. colours of points encode galaxy specific star formation rates measured from broadband photometry alone. The dashed black line shows the bottom of the green valley, while the solid black lines show the green-valley boundaries (defined in Sec.~\ref{sec:greenvalley}).   
     Bottom panel: as in the top panel, but now showing the distance from the bottom of the green valley, $\Delta_{GV}$, as a function of $r-K$ colour.  The top panel shows all galaxies with photometric information  ($10787$ sources), while in the bottom panel only galaxies with grism data ($265$ sources) are shown as coloured points, while the full photometric sample is shown in gray.}
    \label{fig:nuvrk}
\end{figure}

Specifically, for each iteration of the fitting procedure, we select the three observed bands that best correspond to the rest-frame $NUV$, $r$, and $K$ at the redshift under consideration, and {apply a $k$-correction term based on the model SED to derive corrected} rest-frame colours following the approach described in \citet{Hogg2002} {and in Appendix A in \citet{Ilbert2005}}. 
{Specifically, the $k$-correction term is evaluated from the magnitudes of the redshifted SED in the observed bands best probing the rest-frame bands of interest and the magnitudes of the rest-frame SED in the rest-frame bands of interest. As noted above, by selecting the observed bands that best correspond to the rest-frame bands of interest, these approaches have the advantage to minimize the $k$-correction term as well as the contribution from the model SED. To derive realistic rest-frame quantities, we also take the uncertainties of the observed photometry into account when applying the $k$-correction term. To do that, we generate observed-magnitude normal distributions for each source using their reported (i.e., catalogue) values and associated errors, and randomly sample from these distributions when applying the correction.}
After applying this correction to each iteration of the fitting procedure, this gives us the full posterior distribution of corrected $NUV$, $r$, and $K$ rest-frame colours for all the sources in our sample. Note that we  use the same $NUVrK$ rest-frame filter set as in the COSMOS2015 catalogue \citep{Laigle2016}; namely: GALEX $NUV$, Subaru SuprimeCam $r$, and VISTA $Ks$.

We then only keep in our final sample sources with sufficient SNR, selected in the following way. We select the three observed bands that best correspond to the rest-frame $NUV$, $r$, and $K$ at the redshift of each source, and keep all the sources for which these observed bands each have a $\rm SNR > 3$. For cases where a band has a $\rm SNR < 3$, we also consider SNR constraints over longer wavelength baselines. For each rest-frame band ($NUV$, $r$, and $K$) we select the three observed bands that best bracket the rest-frame at the redshift of each source. In our final sample we keep the sources that satisfy: (a) at least two of the three best-bracketing bands have $\rm SNR>1.5$, or (b) the three bracketing observed-bands have a combined $\rm SNR>10$ from the posterior SED. Following \citet{Moutard2016b}, we only use our colour correction method if the observed-bands best probing the rest-frame have errors $< 0.3$~mag. This ensures that our colours are not too noisy and relatively well constrained even if the observed SNR is relatively low. When errors are $> 0.3$~mag, we use the posterior SED colours without correction.
In total, our final photometric catalogue consists of $10787$ sources ($4071$ in GOODS-S, and $6716$ in UDS), including $69\%$ (GOODS-S) and $59\%$ (UDS) with $\rm SNR > 3$ and errors $<0.3$~mag in all the bands of interest.

The top panel of Fig.~\ref{fig:nuvrk} shows the rest-frame $NUVrK$ colour-colour diagram of our final photometric sample, colour-coded by specific star-formation rate (sSFR=SFR/$M_\star$) derived from our photometric fitting. The diagram shows the galaxy bimodal colour distribution as well as the correspondence between galaxy colour and sSFR. As mentioned in Sec.~\ref{sec:introduction}, the advantage of the $NUVrK$ diagram over shorter wavelength-baseline diagnostics such as the $UVJ$ diagram is that it better resolves the separation between the blue and red populations, and allows for a clear identification of the so-called green-valley galaxies. We describe our procedure to identify these populations in the next section.

\subsection{Green-valley definition and $\Delta_{GV}$ distance}\label{sec:greenvalley}

\subsubsection{Green-valley definition}\label{sec:greenvalleydefinition}

In this work, and similarly {to previous studies that use the $UVJ$ or $NUVrK$ diagrams (e.g., \citealp{Williams2009, Arnouts2013, Ilbert2015, Mortlock2015, Moutard2016b, Vergani2018, Moutard2020b})}, we define {the separation between star-forming and quiescent galaxies} as the lowest density region between the two clusters formed by the blue and red populations in {colour-colour space (i.e., the $NUVrK$ diagram shown in the top panel of Fig.~\ref{fig:nuvrk} in our work)}. {Using the $NUVrK$ diagram,} \citet{Moutard2016b} {determined the separation (i.e., the lowest density region) between the blue and red populations in six redshift intervals} in the range $z=0.2-1.0$, {and show that} the $NUV-r$ normalization {of the red/blue separation} linearly varies with time at a rate of $2.9\times10^{-2}$~mag/Gyr {over the redshift range considered}. Extrapolated to higher redshifts, this corresponds to a $NUV-r$ reddening of {the lowest density region between the blue and red populations of} $6.4\times10^{-2}$~mag from $z=1.8$ to $z=1.0$, which represents less than $8\%$ of the width of the green-valley region that we derive here (see below). We therefore do not sub-divide our galaxy sample into multiple redshift intervals to construct our $NUVrK$ diagram.

{These previous works determine the normalization of the red/blue separation based on finding the minimum galaxy number density between the two populations either visually or using Gaussian fits to the bimodal distribution, and determine the slope of the separation visually.}
To best determine the lowest density region between the blue and red populations {(i.e., determine the slope {\it and} normalization of the separation without arbitrary choices)}, we here first divide the colour-colour plane in four equal-width bins of $r-K$ colour over which the red and blue populations overlap (i.e., $r-K = 0.9 - 1.3$ in this work). In each bin, we then fit the $NUV-r$ colour distribution with a double Gaussian model and determine the lowest density point as the minimum between the two Gaussians. This gives us four density minima between the blue and red populations which we then fit with a first order polynomial in the colour-colour plane. For $r-K < 0.9$ where the blue and red populations do not overlap, we use the same slope and normalization as in \citet{Moutard2016b} at similar redshift. By construction, this fit, shown as the dotted line in the top panel of Fig.~\ref{fig:nuvrk}, represents the bottom of the green valley and does not depend on arbitrary choices of slope or normalization to separate the blue and red populations in the region where both overlap.
The bottom of the green valley is thus defined by
\begin{equation}\label{eq:greenvalley}
    (NUV - r) = 
    \begin{cases}
    1.171 \times (r-K) + 2.399 & \text{for $(r-K)>0.51$}\\
    3.00 & \text{otherwise.}
    \end{cases}
\end{equation}
However, owing to the fact that the colour distribution is bimodal, defining boundaries to the green-valley region around this density minimum based on colour alone remains imperfect, as is also the case for other methodologies in the literature that are based on bimodal distributions between the red and blue populations (e.g., sSFR, SFR--mag, or SFR--$M_{\star}$ plane definitions). Here, we define the green-valley boundaries as the density values around the four density minima that reach $35\%$ of the peak of each Gaussian on the red and blue sides independently, and constrain the linear fits of these density boundaries in the colour-colour plane to have the same slopes as the bottom of the green valley. We show the green-valley boundaries based on this definition as solid lines in Fig.~\ref{fig:nuvrk}, which overlap well with the low density gap of the colour distribution. Our boundaries are parametrized by an $NUV-r$ offset of $+0.3$~mag on the red side and $-0.5$~mag on the blue side of the bottom of the green valley defined in Eq.~\ref{eq:greenvalley}.
This boundary definition, although somewhat arbitrary, is consistent with other robust colour-based definitions in the literature at these redshifts (e.g., \citealp{Moutard2016b}) as well as with definitions based on, e.g., sSFR. 

Among the limited number of studies that have defined green-valley samples at the redshifts probed here, sSFR thresholds {approximately} range from $\log_{10}(sSFR) \simeq -10.5$ to $\log_{10}(sSFR) \simeq -9.5~yr^{-1}$ (e.g., \citealp{Pandya2017, Jian2020}), which is similar to the typical sSFR values of the green-valley galaxies in our sample based on our colour definition, as can be seen in Fig.~\ref{fig:nuvrk}. {Note that these green-valley definitions are however also based on somewhat arbitrary choices to define the width of the green-valley region. For instance, \citet{Pandya2017} adopt a somewhat arbitrary threshold between $0.4$ and $1.6$~dex below the fit to the star-forming main sequence in the SFR--$M_{\star}$ plane, while \citet{Jian2020} adopt a somewhat arbitrary width of $\pm0.2$~dex around their separation between the star-forming and quiescent populations in the SFR--$M_{\star}$ plane. We therefore cannot adequately compare these methods to our own green-valley definition, but note the tight correlation between $NUVrK$ colours and sSFR in our sample as well as the reasonably good correspondence between the sSFR values of our green-valley galaxies and those defined in \citet{Pandya2017} and \citet{Jian2020} (see top panel in Fig.~\ref{fig:nuvrk}). As already mentioned and shown in several studies, the $UVJ$ diagram does not well distinguishes between star-forming, transitioning, and quiescent galaxies near the transition region between the different populations (see, e.g., Figs.~2 in \citealp{Siudek2018} and \citealp{Moutard2020b}). This is in contrast to the $NUVrK$ diagram where the populations are well separated and have little overlap in e.g., sSFR. We refer the reader to \citet{Siudek2018}, \citet{Moutard2020b} and references therein for more details on these comparisons.}

\subsubsection{Distance from the bottom of the green valley,  $\Delta_{GV}$}\label{sec:greenvalleydistance}

In an attempt to overcome potential issues from these imperfect boundary definitions, we define a new parameter, $\Delta_{GV}$, which is the $NUV-r$ colour difference between a given point in the $NUVrK$ plane and the bottom of the green valley at that $r-K$ colour. In other words, $\Delta_{GV}$ is the vertical distance between a galaxy and the bottom of the green valley and encodes how likely a given object is a transitioning (i.e., green-valley) galaxy, independently from any green-valley boundary definition. We show $\Delta_{GV}$ as a function of $r-K$ colour in the bottom panel of Fig.~\ref{fig:nuvrk} for our final grism spectroscopic sample that we describe in Sec.~\ref{sec:grism_sample}. This {diagram} can be thought of as a transformation of the $NUVrK$ colour-colour diagram where the bottom of the green valley is set at $\Delta_{GV}=0$. It is expected that when galaxies quench, they move predominantly vertically (upward) in this diagram, with \deltaGV\ encoding their progress along their quenching track and $\Delta_{GV}=0$ marking the bottom of the green valley. In this diagram, our green-valley boundaries are located at $\Delta_{GV} = -0.5$~mag and $\Delta_{GV} = 0.3$~mag on the blue and red sides of the plane, respectively.

The additional benefit of the $\Delta_{GV}$ parameter is its robustness against dust reddening (i.e., dust {attenuation}, $A_V$). As their dust content increases, and assuming a \citet{Calzetti2000} dust law, (dusty) star-forming galaxies move predominantly diagonally in the $NUVrK$ (or $UVJ$) colour-colour diagram, following the slanted locus of the blue-cloud galaxies (i.e., parallel to the separation between the blue cloud and the red sequence; \citealp{Wuyts2007, Williams2009, Patel2012}). By transforming the $NUV-r$ axis with respect to the bottom of the green valley, $\Delta_{GV}$ colours are therefore not expected to evolve with $A_V$. In the bottom panel of Figure~\ref{fig:nuvrk}, dust content is only expected to displace galaxies horizontally, towards redder $r-K$ colours (supporting that quenching alone is expected to move galaxies vertically in this diagram, and that $\Delta_{GV}$ encodes their progress along their quenching track). In that respect, $\Delta_{GV}$ represents a dust-corrected colour, allowing us to derive the ($\Delta_{GV}$) colour evolution of galaxies as a function of their age using monotonic parametrizations of the colour-age relation (see Sec.~\ref{sec:colourage}).

\begin{figure*}
  \centering
    \includegraphics[width=8.0cm]{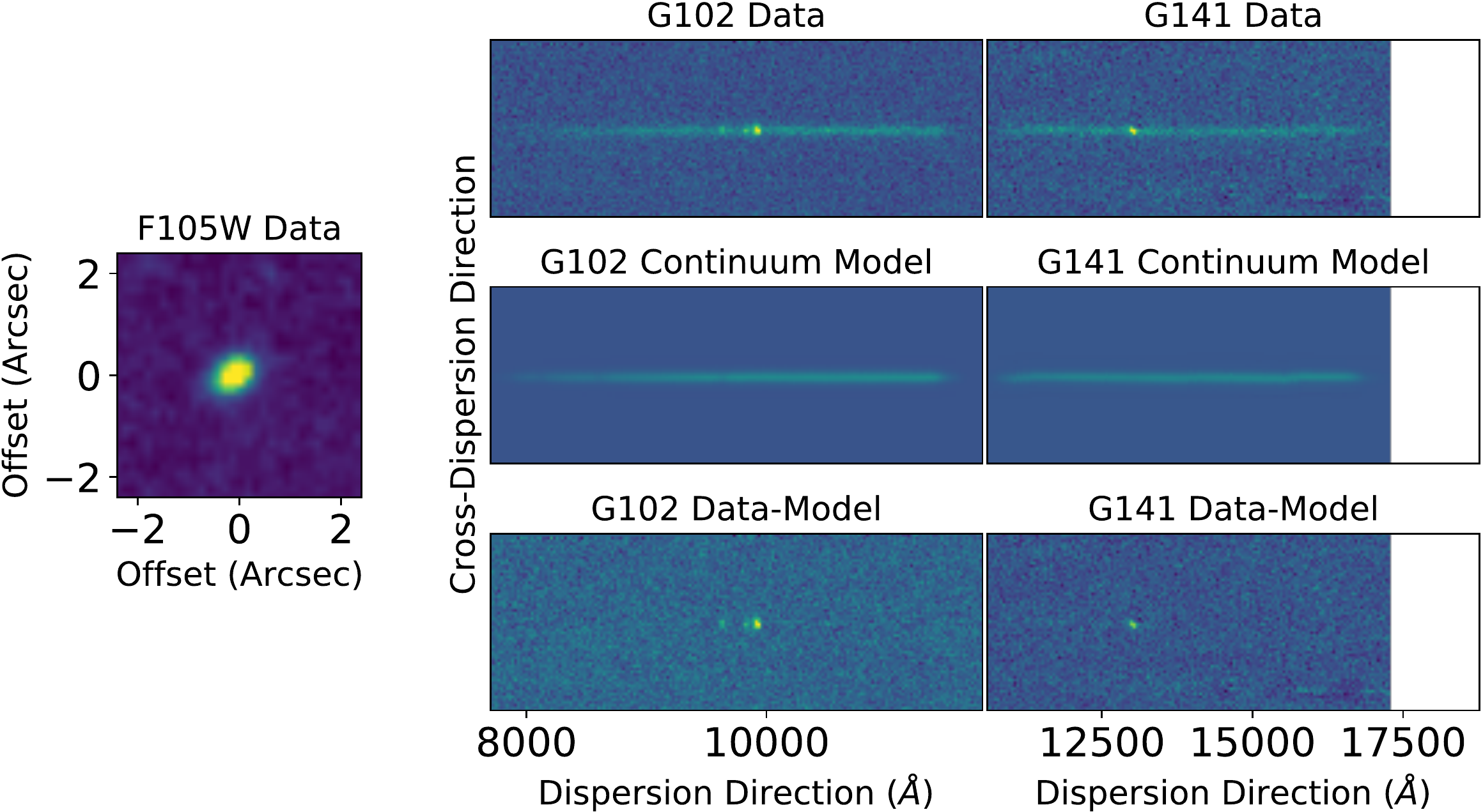}
    \hspace{0.6cm}
    \includegraphics[width=8.0cm]{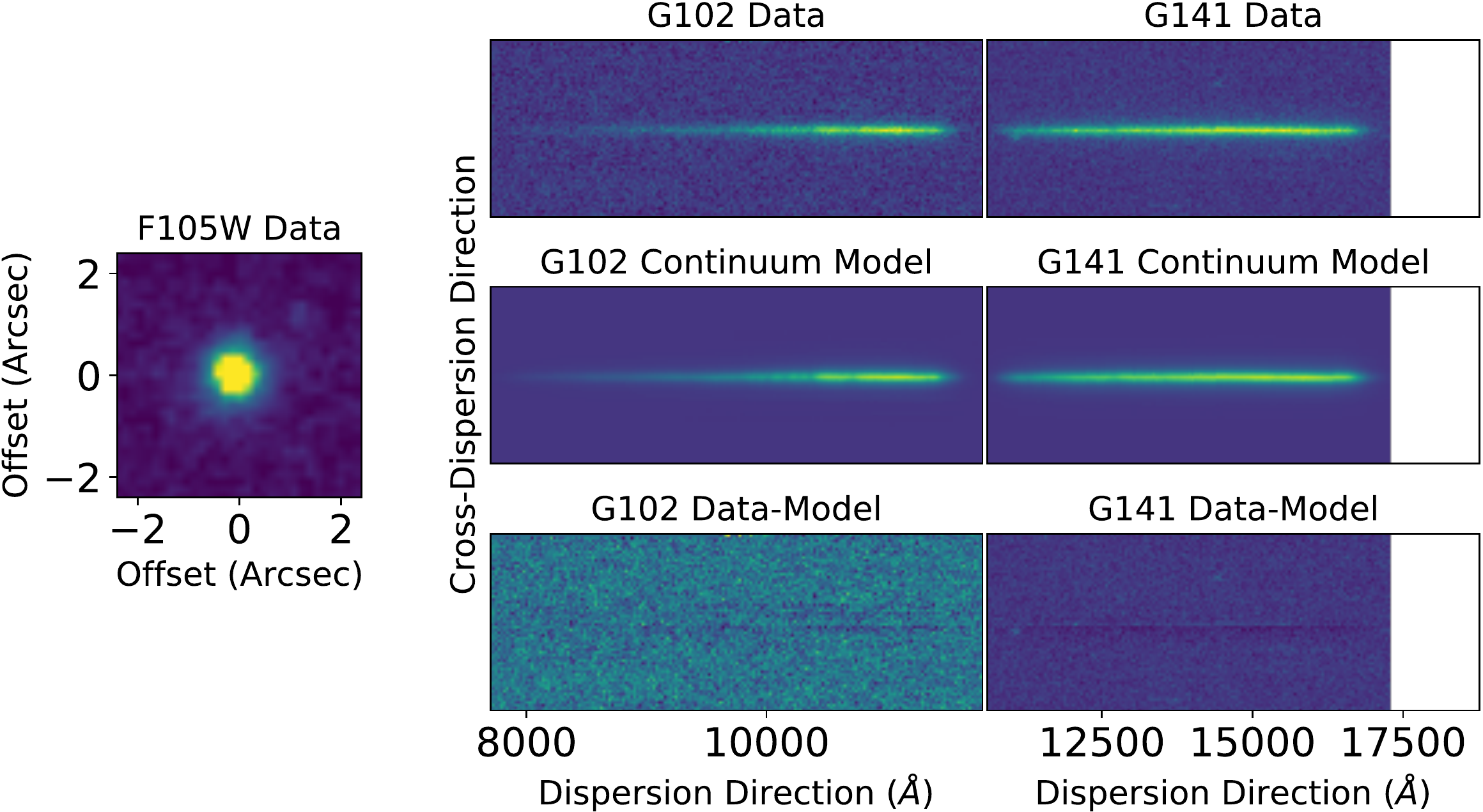}
    \caption{Example 2D grism spectra of our final spectroscopic sample. The left-hand side panels show the direct image (F105W) and corresponding 2D grism spectrum (G102 \& G141) of an example star-forming galaxy in our final spectroscopic sample. The panels on the right-hand side show the same but for an example quiescent galaxy. In each grism panel, the top rows show the stacked, drizzled grism data, the middle rows show the \grizli\ continuum models for each grism, and the bottom rows the data minus the continuum models. In the bottom row of the star-forming source (left panels) only the emission lines remain visible after subtraction of the model, and only the noise remains visible for the quiescent source (right panels). This illustrates the excellent \grizli\ modelling of the continuum for these sources.
    }
    \label{fig:grism2D}
\end{figure*}

\section{Spectroscopic Sample}\label{sec:grism_sample}

\subsection{Sample selection}\label{sec:grism_sel}

We first match our grism spectroscopic data in GOODS-S and UDS (see Sec.~\ref{sec:spec_data}) to our final photometric sample using $1.0\arcsec$ sky-apertures and verify that no additional astrometric correction is needed. From this matched sample, we then only select sources that have photometric and \grizli\ grism redshifts estimates {that are consistent} within $15\%$ of $(1+z)$. {We} further reduce the sample to sources which have exposure times between $20,000$ and $30,000$ seconds (i.e., $\sim10$ to 15 {\it HST} orbits) in the G102 grism to ensure sufficient SNR for the age- and metallicity-sensitive spectral features that are visible in the G102 grism in our redshift range.  Note that almost all of these sources were also observed with the G141 grism, {over} 2 to 15 orbits, which extends their {spectral} coverage to longer wavelengths; these G141 data are also included in our analysis when available. Finally, we only keep in our sample galaxies which have reliable \grizli\ grism redshifts as defined in the following way: ({\it i}) the \grizli\ redshift fitting has converged ($\rm status = 6$), {({\it ii}) the reduced chi-square of the redshift fitting is $\chi^2_{\nu}<1.5$}, and ({\it iii}) the \grizli\ redshift quality is $q_z > -1.5$, where $q_z$ typically varies between $-5$ and $0$, with values closer to $0$ indicating higher redshift quality. 
To determine this \grizli\ $q_z > -1.5$ threshold, we calibrated $q_z$ using sources from our full grism dataset that have available spectroscopic redshifts from the literature provided in the \grizli\ catalogues. Comparing the grism and spectroscopic redshifts of these sources, we find that a quality of $q_z > -1.5$ corresponds to $\gtrsim 60\%$ of the sources having grism and spectroscopic redshifts within $2\%$ of $(1+z_{spec})$, and $\gtrsim 75\%$ within $15\%$ of $(1+z_{spec})$. Higher $q_z$ values naturally increase redshift purity but at the cost of sample size. We therefore choose a threshold of $q_z > -1.5$ as a trade-off between the two.

\begin{figure}
    \centering
    \includegraphics[width=8.3cm]{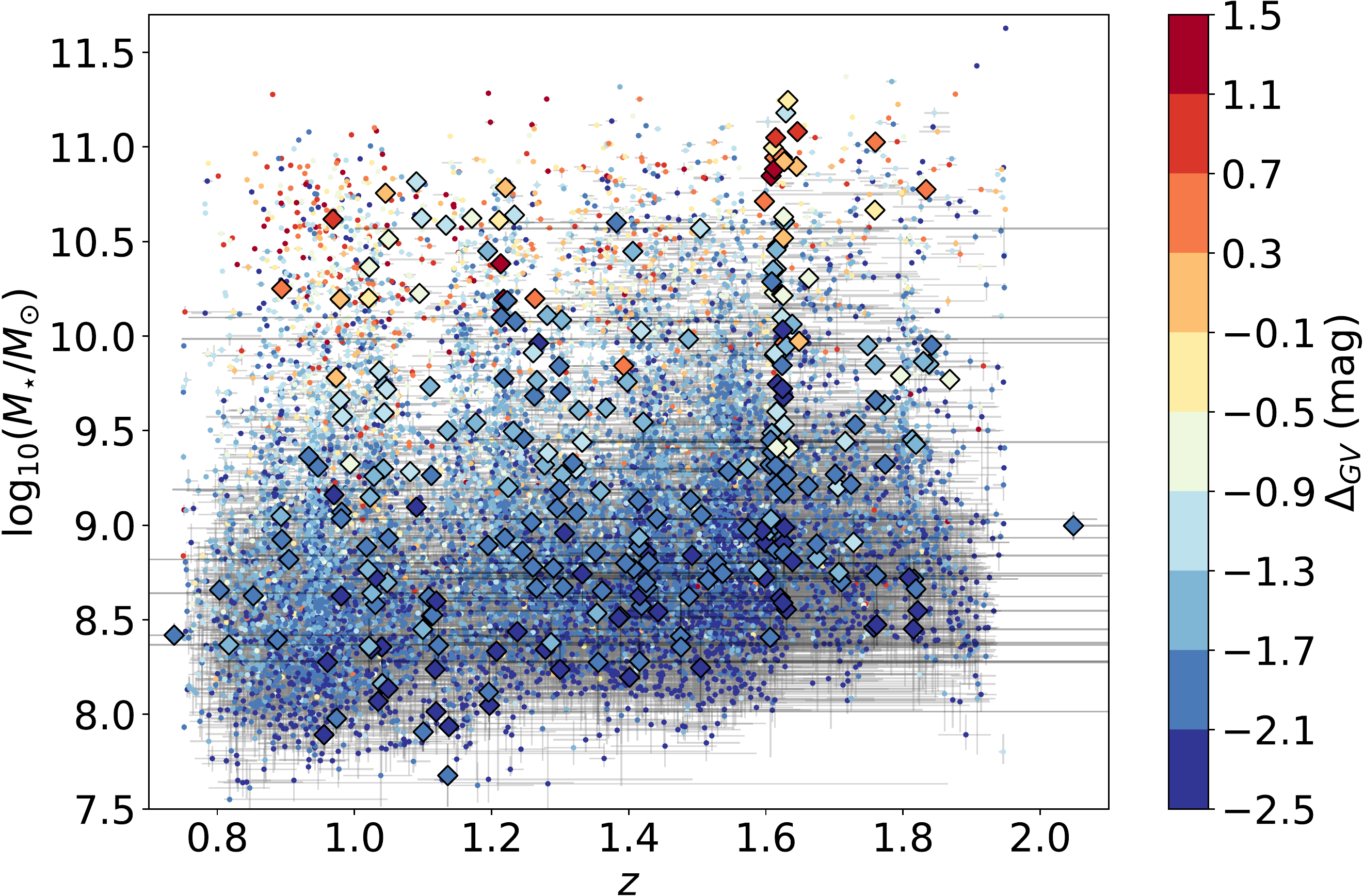}
    \caption{The stellar mass vs.~redshift distribution of our final photometric and spectroscopic samples, colour coded by $\Delta_{GV}$. For our final photometric sample (small points), the redshift is the photometric redshift derived from our SED-fitting procedure using \bagpipes, while we show the \grizli\ grism redshift for sources in our final spectroscopic sample (shown as diamonds).}
    \label{fig:finalsamples_massz}
\end{figure}

After applying the aforementioned quality cuts, our sample consists of a total of $265$ sources ($156$ in GOODS-S and $109$ in UDS) and constitutes the final grism sample that we use in Sec.~\ref{sec:grism_fit} to derive the SFHs and physical parameters of our sources independently from their photometric physical parameters and rest-frame colours {that we} derived in Sec.~\ref{sec:broadebandSED}. The bottom panel of Figure~\ref{fig:nuvrk} shows the $\Delta_{GV}$-colour diagram of our final grism sample (coloured circles, colour-coded by sSFR) overlaid on our final photometric sample (grey points). In Figure~\ref{fig:grism2D} we show example 2D grism spectra from a star-forming (left panels) and a quiescent source (right panels) in our final spectroscopic sample. We show both the G102 and G141 data (top rows in each grism panel), as well as the \grizli\ continuum models and the residuals, in the middle and bottom rows, respectively. The Figure shows the excellent modelling of the continuum for these two sources. As seen in the bottom panel, only the emission lines (namely, H$\beta$, \Oiii, and H$\alpha$) remain visible for the star-forming source, while only the noise is seen in the residuals of the quiescent source.

{In Figure~\ref{fig:finalsamples_massz}}, we show the stellar mass vs.~redshift distribution of our final photometric (small points) and spectroscopic (diamonds) samples, colour-coded by $\Delta_{GV}$. We see the expected {evolving limiting stellar mass as function of redshift, and the expected} correlation between mass and colour, whereby more massive galaxies have on average redder colours \citep[e.g.,][]{Schawinski2014, Powell2017}. {We also} note the presence of two known $z\simeq1.6$ overdensities in our sample, namely Cl~J0218.3-0510, located in the UDS footprint at $z=1.62$ \citep{Papovich2010, Santos2014, Hatch2016, Krishnan2017}, and Cl~J0332-2742, located in the GOODS-S footprint at $z=1.61$ \citep{Castellano2007, Kurk2009, Salimbeni2009}. We do not remove these two overdensities from our sample as we here aim at capturing universal quantities from the full population of quenching pathways at those redshifts.
{As seen in Figure~\ref{fig:finalsamples_massz}, both photometric and spectroscopic samples cover the stellar mass range $M_{\star} \sim 10^{7.5} - 10^{11.5} M_{\odot}$. We use the full mass range when appropriate and also adopt a threshold of $M_{\star} = 10^{10} M_{\odot}$ to define a mass complete sample (see, e.g., Sec.~\ref{subsec:redseqgrowth}). \citet{Grazian2015} derive the strict stellar mass completeness limit as a function of redshift up to $z=8$ for a CANDELS GOODS-S and UDS $H$-band $< 26$~mag (AB) selected sample. They derive the completeness limit of the sample using maximally old galaxies from their template library, with formation redshift $z_{form} = 20$, dust extinction $E(B-V)=0.1$, metallicity of $Z = 0.2$~$Z_{\odot}$, and an exponentially declining SFH with declining time-scale of $0.1$~Gyr. As seen in their Fig.~1, this completeness limit stays well below $M_{\star} = 10^{10} M_{\odot}$ at all redshifts up to $z=2$. We therefore use this threshold of $M_{\star} = 10^{10} M_{\odot}$ to ensure a conservative mass completeness over our redshift range for all galaxy populations in our photometric sample.}
{To assess the impact of our spectroscopic selection function on the stellar mass completeness, we also perform a KS-test statistic between the final photometric and spectroscopic samples. We use a sliding stellar mass threshold in the range $M_{\star} = 10^8 - 10^{11.25} M_{\odot}$ in steps of $0.01 \log(M_{\star}/M_{\odot})$ and evaluate the KS-test statistic between the photometric and spectroscopic stellar mass distributions for galaxies with masses above that threshold at each step. For thresholds up to $M_{\star} \sim 10^{9.2} M_{\odot}$, we find p-values lower than $0.01$ and can therefore confidently reject the null hypothesis that the distributions are identical. For thresholds above $M_{\star} \sim 10^{9.2} M_{\odot}$, we however cannot confidently reject the null hypothesis (i.e., we find p-values $>0.01$). This indicates that our photometric mass completeness threshold of $M_{\star} = 10^{10} M_{\odot}$ also ensures the mass completeness of our spectroscopic sample.}

\subsection{Fitting of the grism data}\label{sec:grism_fit}

{Several} broadband SED-fitting codes allow the user to fit 1D spectroscopic data (e.g., \bagpipes, \textsc{beagle}, \textsc{dense basis}, \textsc{prospector}). However, these {codes} are generally not well adapted for the fitting of slitless grism spectroscopy which requires a source-dependent modelling of the model spectra to account for the spectral smearing of adjacent pixels in the direct imaging. To date, \grizli\ is one of only {a handful of}  publicly-available codes that perform forward-modelling of input model spectra and their fitting to the 2D grism data in the native slitless grism frames. It is the code that we use to fit our grism data. We use \grizli\ version {\tt 1.0.dev1365} at the time of this analysis.

To generate model spectra for \grizli, we use the \textsc{Python} version of the Flexible Stellar Population Synthesis (FSPS) code \citep{Conroy2009, Conroy2010}. FSPS includes the Padova \citep{Marigo2008}, MIST \citep{Paxton2011, Paxton2013, Paxton2015, Choi2016, Dotter2016}, PARSEC \citep{Bressan2012}, BaSTI \citep{Pietrinferni2004}, GENEVA \citep{Ekstrom2012}, and BPASS \citep{Eldridge2009, Stanway2018} stellar tracks and isochrones, as well as the MILES \citep{Sanchez2006} and BaSeL \citep{Westera2002} spectral templates. Here, we use the default combination of MIST and MILES tracks and templates, {which assume solar metallicities of Z$_{\odot} = 0.0142$ \citep{Asplund2009} and Z$_{\odot} = 0.019$, respectively}\footnote{{Note that while the assumed solar metallicities are different, FSPS interpolates the spectral library to the isochrone grid at fixed Z/Z$_{\odot}$}.}. FSPS also includes Cloudy photoionization for nebular emission \citep{Ferland2017}, the \citet{DraineLi2007} dust emission models, and the \citet{Villaume2015} AGB circumstellar dust models. For consistency with our broadband SED-fitting, we assume the \citet{Kroupa2001} IMF and \citet{Calzetti2000} dust {attenuation} model, as well as delayed-$\tau$ SFHs. In total, we use a set of four parameters in the fitting: age, $\tau$, metallicity, and dust attenuation in the $V$ band, $A_V$, that we define in the same ranges and with the same uniform priors as for the photmetric SED-fitting in Sec.~\ref{sec:broadebandSED}. We do not fit again for redshift -- which we fix to the \grizli\ redshift -- and we use the same photoionization parameter as we did with \bagpipes\ (i.e., $\log_{10}(u)=-2$). Similarly to \bagpipes, we also use a nested-sampling algorithm \citep{Skilling2004}, here \dynesty\ \citep{Speagle2020}, to perform and optimize the exploration of the parameter space, the fitting convergence, and posterior estimation. Within \dynesty, we use uniform parameter sampling to explore the parameter space, and the multiple bounding ellipsoids technique \citep{Feroz2009} to iteratively reduce the prior volume. In the following, all the physical parameters we report are those derived from our spectroscopic (grism) fitting, except galaxy colours and stellar masses that we independently derive from our broadband SED fits (see  Sec.~\ref{sec:broadebandSED}). This {independent fitting} ensures that correlations between our spectroscopic and photometric measurements, if any, are primarily driven by the data and do not suffer from improper weighting between the datasets or from model-driven parameter correlations {that could have occurred} had we attempted to fit our photometric and (independent) spectroscopic datasets together.

\section{SFH models}\label{sec:delayedtaumodels}

\begin{figure}
    \centering
    \includegraphics[width=7cm]{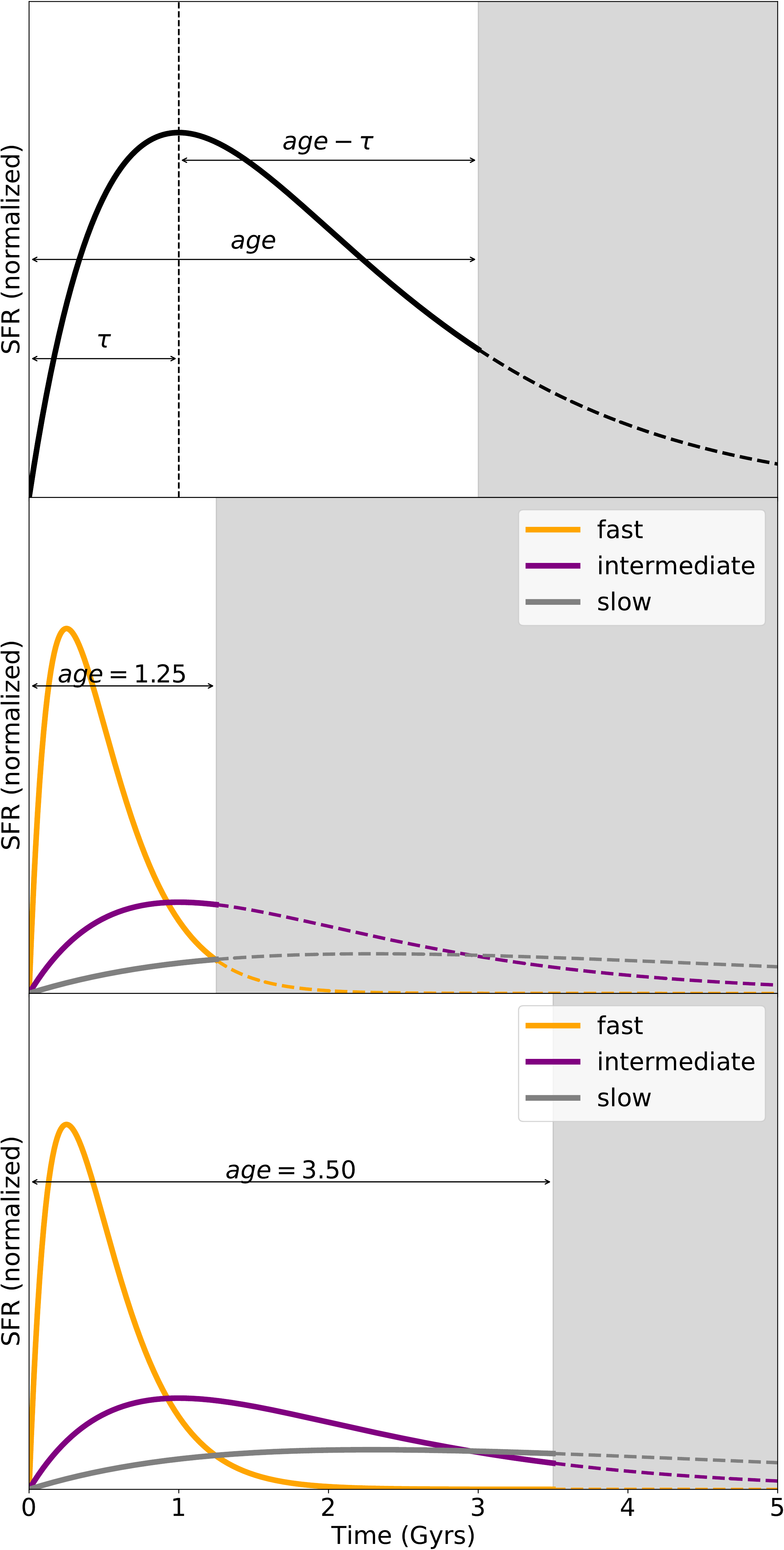}
    \caption{Delayed-$\tau$ models.  The top panel illustrates the meaning  of $\tau$, age, and age-$\tau$ for a generic \delayedtau\ star formation history. In particular, age means time since the onset of star formation, $\tau$ is the time between the onset of star formation and the peak of star formation, and age-$\tau$ is the time since the peak of star formation.  The middle panel shows models with three different $\tau$ values (fast, intermediate, and slow) observed at age=1.25~Gyr, slightly after the time of peak star formation in the intermediate-$\tau$ model. The lower panel shows the same three models but observed at an older age, 3.50~Gyr, about 1~Gyr after the time of peak star formation in the slow-$\tau$ model. In the middle and lower panels, the fast models ($\tau=0.25$~Gyr) are shown in orange, the intermediate models ($\tau=1.0$~Gyr) in purple, and the slow models ($\tau=2.3$~Gyr) in grey.
    The models are normalized by their total area under the curve (i.e., they will have formed the same total amount of stellar mass by $t \to +\infty$). In these two panels, the three models are example SFHs corresponding to the three $\tau$-populations identified in Fig.~\ref{fig:grism_tausel}.
    }
    \label{fig:sfh_cartoon}
\end{figure}

As mentioned in the previous sections, we use delayed-$\tau$ models (Eq.~\ref{eq:delayedtau}) to parametrize the star-formation histories in our spectroscopic (and photometric) SED-fitting procedures. To better understand the results of our grism SED-fitting, we here clarify the physical meaning of the parameters of these models, namely $\tau$ and age. In the top panel of Fig.~\ref{fig:sfh_cartoon} we show a representative \delayedtau\ model. Age represents the time between the onset of star formation and the epoch of observation (i.e., the age of the oldest stellar population), while $\tau$ represents the time between the onset of star formation and the time at which the SFR peaks (as can easily be shown by differentiating $\rm SFR(\it t)$ with respect to $t$ in Eq.~\ref{eq:delayedtau}). 
Additionally, age$-\tau$ represents the time since the peak of star formation for galaxies that have already passed that peak (i.e., galaxies on the declining phase of their SFH), or the remaining time to the peak for those that haven't done so yet (i.e., galaxies still on their rising phase). 

{Next}, for a galaxy that has passed its peak of star formation (i.e., ${\rm age} - \tau > 0$), $\tau/ \rm age$ gives the fraction of its life to date that the galaxy has spent increasing its star formation rate (i.e., spent in its rising phase), and $({\rm age}-\tau) / \rm age$ gives the fraction of the galaxy's current age that the galaxy has spent with its SFR declining (i.e., spent in its declining phase). Similarly $({\rm age}-\tau) / \tau$ is the ratio of times in the declining and increasing modes of star formation. 

In the middle and bottom panels of Fig.~\ref{fig:sfh_cartoon} we additionally show three \delayedtau\ models, normalized by their total integrated SFRs, with characteristic (i.e., rising-phase) time-scales of $\tau=0.25, 1.0$ and $2.3$~Gyr shown in orange, purple, and grey, respectively.  The middle and bottom panels show these three galaxies seen at $t=1.25$~Gyr and 3.5~Gyr, respectively. This shows that at similar ages, and/or SFRs, galaxies can be at various stages of their evolution (rising, near the peak, declining, or nearly or completely quenched) depending on their respective characteristic SFH time-scale, $\tau$.  Galaxies with similar values of $\tau$ may form an evolutionary sequence that connects galaxies of the same SFH (i.e., the same $\tau$) observed at different times in their histories\footnote{See also \citet{Abramson2016} for a detailed discussion of how log-normal SFHs may reproduce a number of observables in the Universe while only connecting galaxies through their $\tau_{\rm log-normal}$ SFH parameter.}. Although ideally, e.g., given a statistically large enough sample, formation redshifts ($z_f = z_{obs} - \rm age$) and SFH normalizations (e.g., total masses at a given $t$) should also be considered when connecting galaxies on such SFH-based evolutionary sequences.

\begin{figure}
    \centering
    \includegraphics[width=8.3cm]{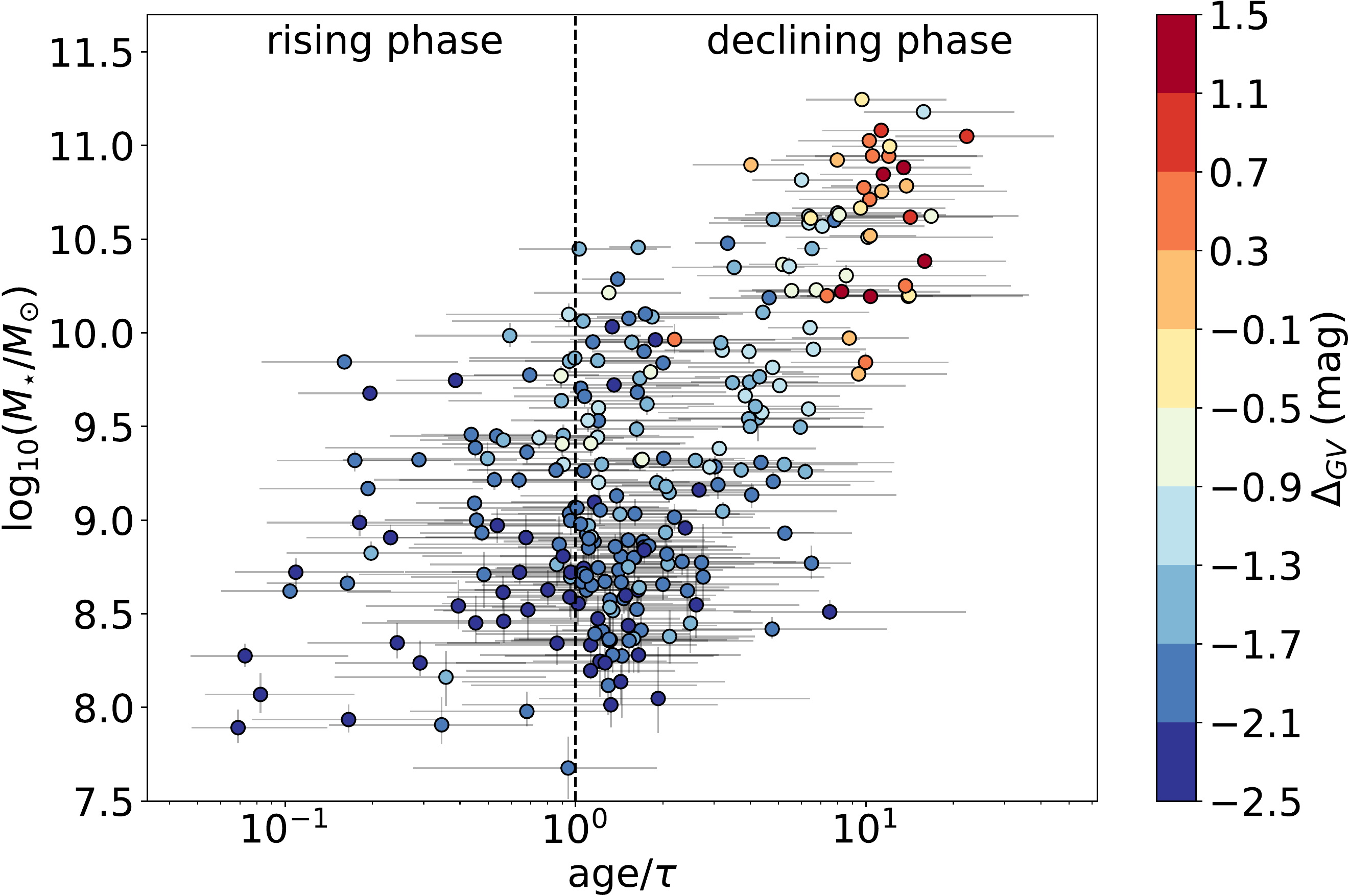}
    \caption{The stellar mass vs.~age$/\tau$ distribution of our final spectroscopic sample, colour-coded by $\Delta_{GV}$. Galaxies with age$/\tau < 1$ are on the rising phase of their SFHs, while galaxies with age$/\tau > 1$ are on the declining phase of their SFHs. While the distribution is consistent with the build-up of stellar mass as galaxies age and become red, it also reveals the potential effect of our selection function as well as galaxy downsizing (see text in Sec.~\ref{sec:delayedtaumodels} for details).}
    \label{fig:finalsamples_massagetau}
\end{figure}

Figure~\ref{fig:finalsamples_massagetau} shows the stellar mass vs.~age/$\tau$ distribution of our final spectroscopic sample, colour-coded by $\Delta_{GV}$. For each galaxy, we derive its age/$\tau$ posterior distribution by randomly drawing 500 times from the age and $\tau$ spectroscopic-fitting posterior distributions and evaluating age/$\tau$ at each iteration. In the Figure, the data points and uncertainties represent the median and $68\%$ interval of the posterior distributions, respectively. We use the same approach to derive the other age -- $\tau$ quantities in the rest of the paper (i.e., $\rm age-\tau$, $\rm (age-\tau$)/age ). In the Figure, galaxies with age$/\tau < 1$ are on the rising phase of their SFHs, while galaxies with age$/\tau > 1$ are on the declining phase of their SFHs. We see that galaxies with photometric blue colours span a large range of stellar masses from about $10^8$ to $10^{10.5} M_{\odot}$ ($\sim2.5$~dex) and age/$\tau$ from about $10^{-1}$ to $10$ ($\sim2$~dex) which includes both rising and declining phase galaxies, while galaxies with photometric red colours have masses spanning $\sim1.5$~dex from about $10^{9.8}$ to $10^{11.3} M_{\odot}$ and age/$\tau$ spanning $\sim 0.5$~dex around $\rm age/\tau=10$ (i.e., these only include galaxies in the declining phase of their SFHs, with little remaining star-formation activity). While this is consistent with the gradual build-up of stellar mass as galaxies age and become red and quiescent, and also showcases the consistency between our spectroscopic and photometric measurements, we also see the effect of our selection function as we may be missing faint, low-mass ($\lesssim 10^{10} M_{\odot}$) red galaxies in our final sample. Additionally, the apparent lack of high-mass ($\gtrsim 10^{10.5} M_{\odot}$) blue galaxies is consistent with the so-called downsizing of galaxies \citep[e.g.,][]{Cowie1996, Treu2005, Cimatti2006, Neistein2006, Fontanot2009} and may indicate that we do not capture the star-forming progenitors of the most massive quiescent galaxies in our sample. However, as mentioned in the previous paragraph, galaxies with similar values of $\tau$ (galaxies with similar SFHs) and seen at different stages of their evolution may be connected into consistent evolutionary sequences, a connection that  we attempt to implement in Section~\ref{sec:colourage} in order to mitigate these potential bias in our sample.

\section{Results}
\label{sec:results}

\begin{figure}
  \centering
    \includegraphics[width=8.3cm]{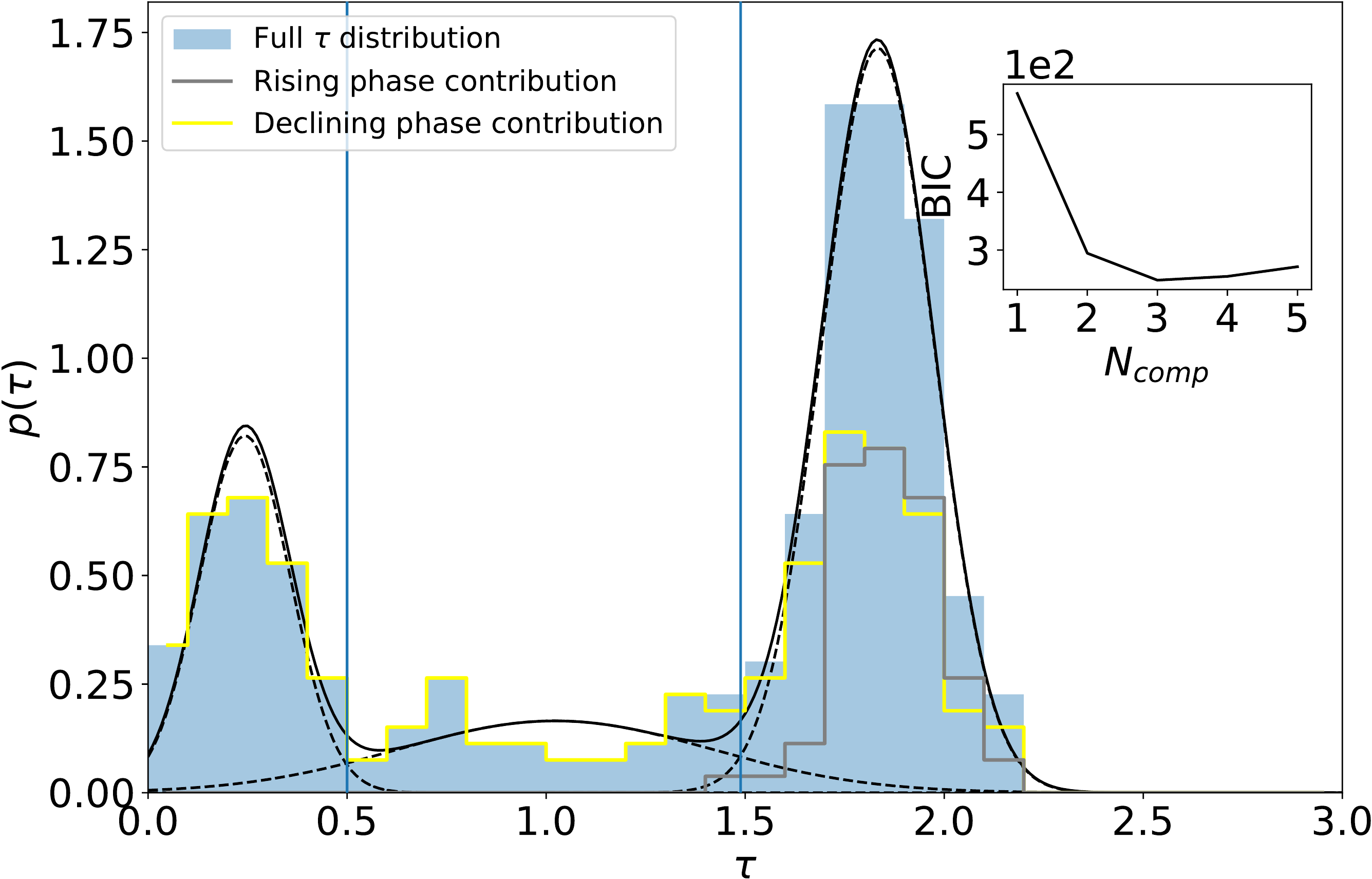}
    \caption{Determination of the $\tau$ populations. The blue histogram represents the $\tau$ probability density distribution of our final sample, derived using the median of the $\tau$ posterior for each galaxy. The black solid line is the best-fitting Gaussian mixture model to the data, with the individual components shown as dashed lines. The best-fitting model is a three-component model, as determined by the Bayesian Information Criterion (BIC) shown in the inset panel.  As the inset shows, the three-component model better describes the data than other models with one to five components.  The vertical lines show the intersection between the components. We define three populations based on these separations: ‘fast’, which are objects with $\tau<0.5$~Gyr, ‘intermediate’ which are those with $0.5<\tau<1.5$~Gyr, and ‘slow’ which are objects with $\tau>1.5$~Gyr. See the middle and bottom panels of Fig.~\ref{fig:sfh_cartoon} for example SFHs of these three populations.
    The yellow histogram, overlaid on the blue, represents the contribution of galaxies in the declining phase of their SFHs (i.e., $\rm age - \tau > 0$) with respect to the total distribution. The grey histogram represents the contribution of rising-phase galaxies (i.e., $\rm age - \tau < 0$).
    }
    \label{fig:grism_tausel}
\end{figure}

\subsection{Distribution of spectroscopic $\tau$ values}\label{sec:tauDistribution}

Figure~\ref{fig:grism_tausel} shows the probability density distribution of the spectroscopically-derived $\tau$ values of galaxies in our final sample, derived using the median of the $\tau$ posterior for each galaxy. The distribution of (median) $\tau$ values is clearly multi-modal. To identify different populations within the distribution, we fit the distribution with a Gaussian Mixture Model \citep{Dempster1977} with up to 5 components and use the Bayesian Information Criterion (BIC; \citealp{Schwarz1978, Liddle2007}) to determine the best fitting model, which here is the three-component model, $N_{\rm {comp}} = 3$. The three populations identified by this procedure are $\tau < 0.5$~Gyr, $0.5< \tau < 1.5$~Gyr, and $\tau > 1.5$~Gyr, which we refer to as the ‘fast’, ‘intermediate’ and ‘slow’ populations, respectively. Most of our galaxies are in the fast ($25\%$) and slow ($60\%$) populations, with only a small number in the intermediate population ($15\%$). 
In the Figure, we also show the contribution of galaxies in the declining phase of their SFHs (i.e., $\rm age - \tau > 0$; yellow histogram in the Figure) with respect to the total $\tau$ probability density distribution shown in blue. We observe that declining-phase galaxies almost entirely account for the total distribution at $\tau < 1.5$~Gyr. This can be expected, as by definition, a larger fraction of rapidly evolving galaxies at any redshift is expected to be observed after the peak of their SFHs compared to slowly evolving galaxies simply due to their shorter rising time-scales. On the other hand, the distribution for galaxies on the rising phase of their SFHs (shown in grey in the Figure), is uni-modal and is comprised almost exclusively of slowly evolving galaxies (i.e., galaxies with $\tau > 1.5$~Gyr). This can again be expected, as rapidly evolving galaxies may reach the peak of their SFHs within relatively short time-scales, and only remain visible in the declining phase of their SFHs thereafter. Inversely, only slowly evolving galaxies may remain visible on the rising phase of their SFHs for relatively long time-scales. 

However, as we also discuss in Section~\ref{sec:agetau}, we have to caution the reader of the poor constrains imposed on the $\tau$ values of rising-phase galaxies (comprised almost exclusively of $\tau>1.5$~Gyr galaxies) as well as declining-phase galaxies with large $\tau$ values, showing systematically large $\tau$ uncertainties. {In contrast,} the $\tau$ values of declining-phase galaxies with intermediate and small $\tau$ values are better constrained by the data. {These effects are seen in} the full $\tau$ posterior distributions of the galaxies in our sample, where galaxies with small and intermediate (median) $\tau$ values (comprised exclusively of galaxies observed in the declining phase of their SFHs) have on average well constrained, {Gaussian-shaped posteriors that span a few Myr,} while galaxies with large (median) $\tau$ values have on average flat posterior distributions over the range of $\tau\sim 1.5-3.0$~Gyr with a rapid cut-off at smaller $\tau$s. This indicates that while our grism fitting procedure is able to discriminate between short and long SFHs, it unfortunately cannot well discriminate between two SFHs with long, albeit different, $\tau$ time-scales. While this might come as a potential bias, we note that we do not attempt in this work to further discriminate our sample beyond the three $\tau$ populations derived here, and that most of our analysis in the next Sections focuses on the fast $\tau$-population for which SFH parameters are relatively well constrained.
The three models in the middle and bottom panels of Fig.~\ref{fig:sfh_cartoon} are schematic representations of these three $\tau$-populations (i.e., fast, intermediate, and slow), normalized by their total integrated SFHs in the Figure (i.e., to $t \to +\infty$).

\subsection{Age-$\tau$ diagrams}\label{sec:agetau}

\begin{figure}
  \centering
    \includegraphics[width=8.3cm]{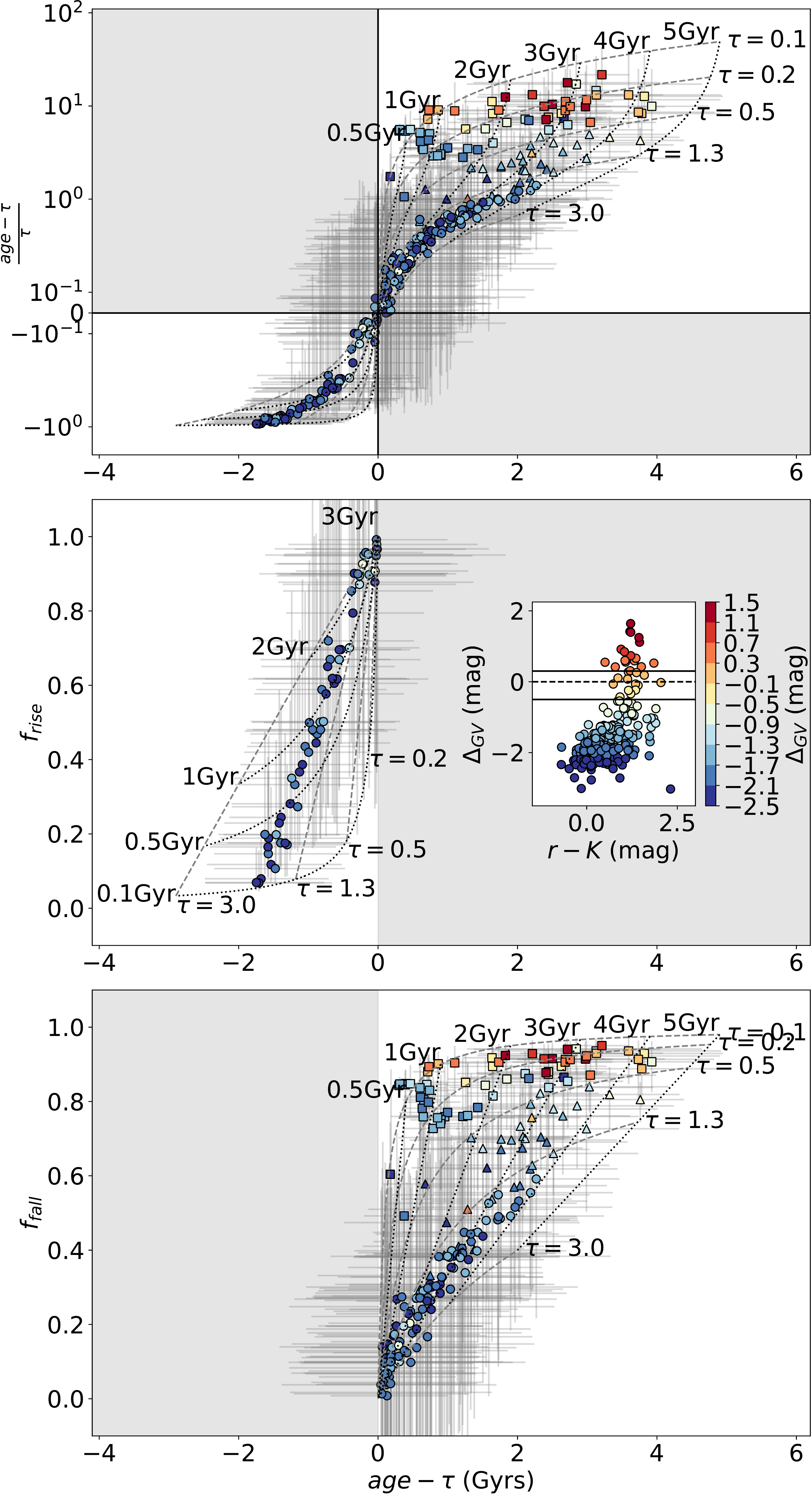}
    \caption{
    Age -- $\tau$ diagram.
    {\it Top}: ($\rm age - \tau) / \tau$ vs.~$\rm age - \tau$. For positive $age - \tau$, this panel shows how the declining time-scale ($\rm age - \tau$) compares to the rising time-scale ($\tau$) as a function of time since the peak of star-formation. For negative $\rm age - \tau$, it shows the remaining fractional time (in negative values) for a galaxy to reach its peak of star-formation.
    {\it Middle}: This panel, related to the top panel's lower left quadrant, shows the completed fractional time to the peak of star-formation ($f_{rise} = \rm age/\tau$) for galaxies in their rising phase ($\rm age - \tau < 0$).
    {\it Bottom}: Related to the top panel's upper right quadrant, this panel shows the fractional time a galaxy has spent on its declining phase with respect to its age ($f_{fall} = (\rm age - \tau) / age$). In this panel, we only show galaxies that are past their peak of star-formation.
    In all panels, data points are colour coded by their distance to the green valley, $\Delta_{\rm GV}$, which is shown in the inset as a function of $r-K$ colour. Additionally, squares indicate the fast population as shown in Fig.~\ref{fig:grism_tausel} ($\tau<0.5$~Gyr), triangles the intermediate population ($0.5<\tau<1.5$~Gyr), and circles the slow population ($\tau>1.5$~Gyr).
    In all panels, we also show tracks of constant $\tau$ as dashed lines ($\tau = 0.1, 0.2, 0.5, 1.3, 3.0$~Gyr) and age isochrones as dotted lines (age $= 0.1, 0.5, 1, 2, 3, 4, 5$~Gyr).
    }
    \label{fig:agetau}
\end{figure}

In Figure~\ref{fig:agetau} we show the relation between age and $\tau$ (both measured from spectroscopy) and galaxy $NUVrK$ colours (measured from photometry). All panels are shown as a function of $\rm{age}-\tau$ (in Gyr), which measures the time before (if negative) or after (if positive) the peak of a galaxy's SFH. Furthermore, in all panels the points are colour-coded by their distance to the green valley, $\Delta_{GV}$, which is shown as a function of $r-K$ colour in the inset of the middle panel. In all panels, we also represent the three $\tau$-populations identified in Sec.~\ref{sec:tauDistribution} (i.e., fast, intermediate, and slow) with squares, triangles, and circles, respectively.
We also show model tracks of constant $\tau$ as dashed lines and age isochrones as dotted lines.  By definition, stellar populations with \delayedtau\ star formation histories move along the constant-$\tau$ tracks from left to right. Consequently, in this picture, galaxies located on lines of constant $\tau$ may be thought to form an evolutionary sequence.  

In the top panel of Fig.~\ref{fig:agetau} we first investigate how the elapsed (or remaining) time since (until) the peak of star-formation compares to the rising time-scale (i.e., $({\rm age} - \tau) / \tau$).  By definition, galaxies in the lower left quadrant have rising SFRs, and those in the upper right quadrant have declining SFRs. In this context, we see a clear correspondence between the $\tau$ and age tracks, spectroscopic $\tau$ and age measurements, and the photometric colours. First, the lower left quadrant (galaxies with rising SFRs) contains only galaxies that are photometrically classified as being in the blue cloud (i.e., star-forming). In other words, our spectroscopic measurements confirm that galaxies with rising SFRs are all photometrically classified as star-forming. 
In contrast, the upper right quadrant, which shows galaxies with declining SFRs, is populated by galaxies of all photometric types: blue cloud, green valley, and red sequence.  In other words, galaxies with declining star formation histories can be found in all regions of the $NUVrK$ diagram, and may still be strongly star-forming, already quiescent, or anywhere in between. However, galaxies populate different regions of the age-$\tau$ diagram depending on their photometric colours.
Specifically, photometrically-classified blue-cloud galaxies (blue symbols) cover the full range of ages for the slow $\tau$ tracks, and only cover young ages on the faster tracks. Inversely, photometrically-classified red-sequence galaxies (red symbols) mainly populate the faster $\tau$ tracks from intermediate to the oldest ages, while green-valley galaxies (yellow and orange points) only cover intermediate ages for the faster $\tau$ tracks and older ages for slower tracks. 

Overall, all three photometric galaxy populations (blue, green, red) are present on the diagram and tend to follow trends indicative of a clear evolutionary sequence along fixed $\tau$ tracks. In particular, for the faster SFHs (small $\tau$ values) we see galaxies on a photometric colour sequence (correlated to spectroscopic age) from the blue cloud to the green valley and then to the red sequence. For the slower tracks (large $\tau$ values), the absence of red galaxies indicate that we {are seeing} galaxies whose SFRs are already declining but which, given their long decline times, have not yet had time to leave the blue cloud within the relatively short Hubble time at these redshifts (the Universe is 3.6~Gyr old at $z=1.8$, 4.2~Gyr old at $z=1.5$, and 5.8~Gyr old at $z=1.0$). Finally, we note that at least for the faster tracks ($\tau < 0.5$~Gyr),  galaxies seem to be going through their transitional phase (i.e., leaving the blue cloud) when ($\rm age-\tau$) is roughly almost an order of magnitude larger than $\tau$; i.e., galaxies move away from the {(photometric)} blue cloud when they have spent almost $\sim 10\times$ more time in the declining phase than in the rising phase of their SFHs.

\begin{figure*}
  \centering
    \includegraphics[width=14cm]{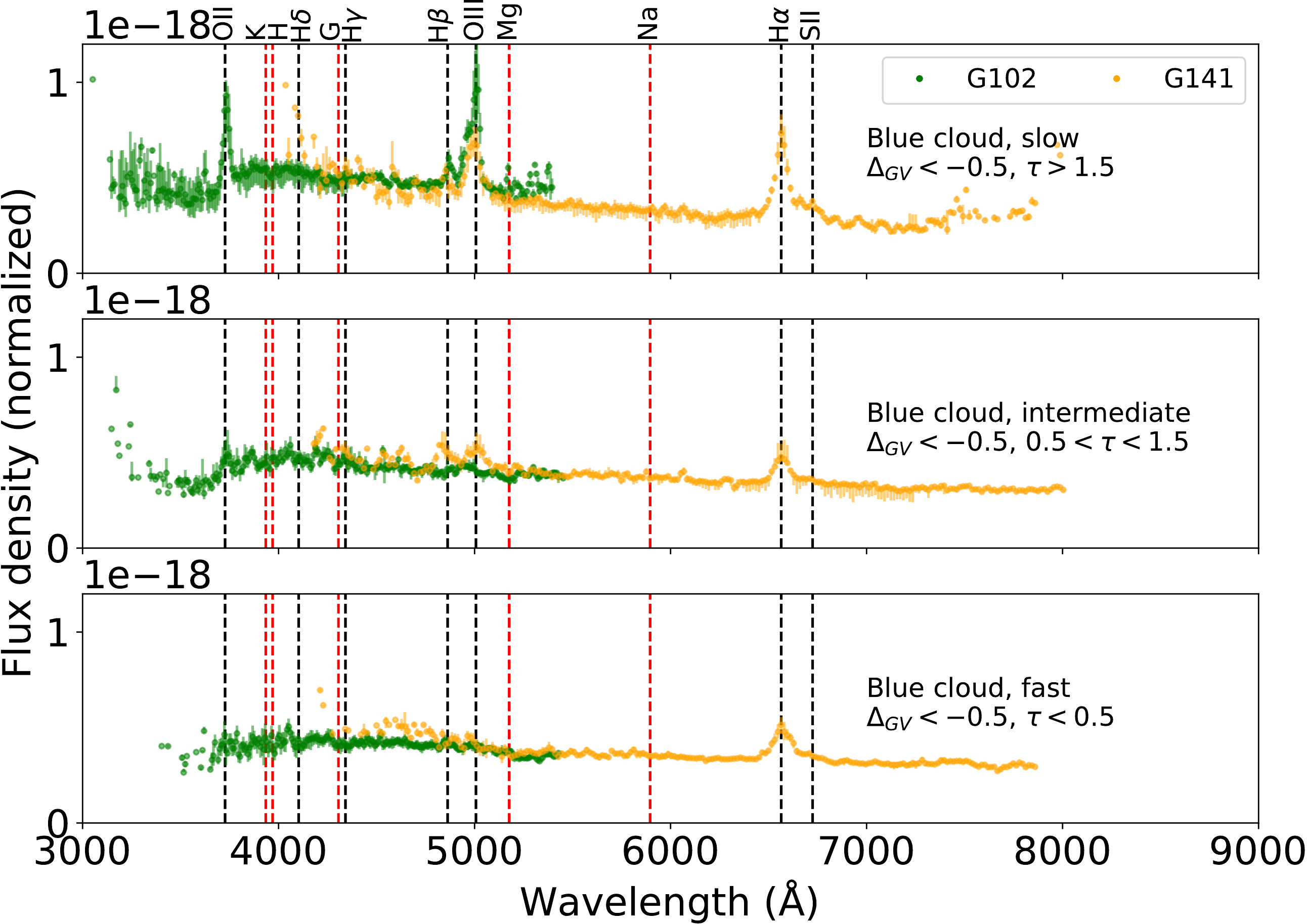}
    \caption{Rest-frame inverse-variance weighted stacked spectra of our blue-cloud ($\Delta_{GV}<-0.5$~mag) galaxies, grouped by their evolutionary time-scales, $\tau$. We show the stacked spectra of the fast ($\tau<0.5$~Gyr), intermediate ($0.5<\tau<1.5$~Gyr), and slow ($\tau>1.5$~Gyr) populations in the bottom, middle, and top panels, respectively. Green points represent the G102 data while we show the G141 data in orange. Vertical dashed lines indicate common emission (black) and absorption (red) spectral features. The spectra show emission lines characteristic of star-formation activity, and lack spectral features associated with old stellar populations, consistent with these galaxies being predominantly young (i.e., dominated by light from young stars) and star-forming (see Sec.~\ref{sec:stackspectra} for details).}
    \label{fig:stacksSF}
\end{figure*}

In the middle and bottom panels of Figure~\ref{fig:agetau} we investigate the fractional times that galaxies spend in either the rising or declining phases of their SFHs. In the middle panel we focus on galaxies in their rising phase, and show the completed fractional time to the peak of star formation, which we define as $f_{rise} = \rm age/\tau$. Galaxies with $f_{rise} \sim 0$ just started to form stars, while galaxies with $f_{rise} \sim 1$ are just reaching the peak of their star formation. 
Galaxy positions cannot be well constrained for these objects, as is indicated by the large error bars and the alignment of points in the diagram. 
Following our discussion in Sec.~\ref{sec:tauDistribution}, we conclude that, at least with the present spectroscopic data, we cannot precisely determine the star formation history (i.e., $\tau$ parameter) of a galaxy in its rising phase of star formation.

In the bottom panel of Figure~\ref{fig:agetau} we focus on the star formation histories of declining-phase galaxies. In contrast to rising-phase galaxies, we show the fractional time a galaxy has spent in its declining phase with respect to its total age. We define this as $f_{fall} = \rm (age-\tau)/age$. 
An object with $f_{fall} = 0.5$ has spent as much time in its rising phase as in its declining phase. As can be seen in the Figure, such $f_{fall}\sim0.5$ objects remain in the (photometric) star-forming blue cloud. In other words, galaxies that have spent a significant amount of their lifetime with declining star-formation rates may still be photometrically classified as star-forming.   
Roughly, galaxies seem to move away from the photometric blue cloud only when $f_{fall} \gtrsim 0.8$, which corresponds to mass fractions of $\gtrsim 0.95$ of their total final masses as evaluated by integrating over their total SFHs (i.e., to $\rm age \to +\infty$).
This seems to indicate that stellar mass build-up {is minimal} once a galaxy leaves the blue cloud and enters the green valley; this phase contributes just a few percent of the galaxy's total, final expected mass. 
This suggests that lower bound thresholds for quenching time-scale definitions based on the fraction of mass a galaxy has formed (often measured from their SFHs) should be carefully chosen and may need to be as high as $95\%$ to correspond to transition time-scales derived from colour definitions such as in this work.

\begin{figure*}
  \centering
    \includegraphics[width=14cm]{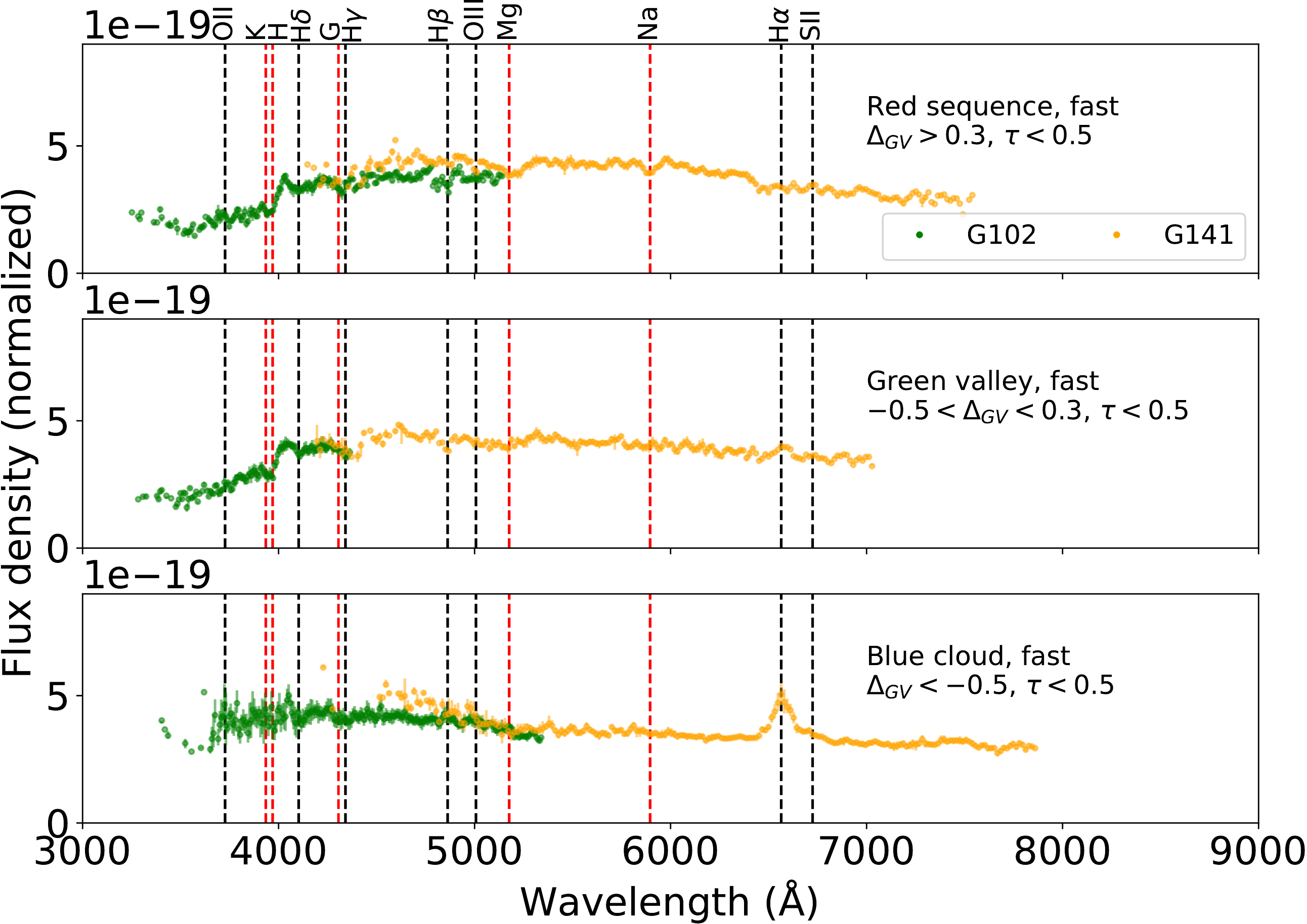}
    \caption{Rest-frame inverse-variance weighted stacked spectra of our rapidly-evolving ($\tau < 0.5$ Gyr) galaxies in the red sequence (top panel), green valley (middle panel) and the blue cloud (bottom panel). As in Fig.~\ref{fig:stacksSF}, green points represent the G102 data while we show the G141 data in orange, and the vertical dashed lines indicate common emission (black) and absorption (red) spectral features. The stacked spectra present a clear quenching time-sequence from bottom to top, with spectral features indicative of the stellar population ageing and star-formation decline along this sequence (see Sec.~\ref{sec:stackspectra} for details).
     }
    \label{fig:stacksBGR}
\end{figure*}

\subsection{Spectroscopic properties of our different populations}\label{sec:stackspectra}

Figures~\ref{fig:stacksSF}~and~\ref{fig:stacksBGR} present the stacked rest-frame spectra of our different galaxy populations, grouped with respect to (spectroscopic) $\tau$ and (photometric) $\Delta_{GV}$ colour. To create the stacks, we first interpolate all G102 and G141 spectra to a common rest-frame wavelength grid of $\Delta\lambda_{\rm rest} = 8$~\AA\ and $\Delta\lambda_{\rm rest} = 16$~\AA, respectively, and normalize them to the source observed flux densities in the F105W-band. We additionally mask all spectral regions that either have a $\rm SNR<3$ per resolution element or have contamination from neighbouring objects {that is} above $25\%$ of the source flux. For each rest-frame wavelength covered by at least five sources, we then average-stack the spectra by inverse-variance weighting, and bootstrap resample 500 times to estimate the uncertainties in our stacking. We repeat this procedure for each population of interest.

In Figure~\ref{fig:stacksSF} we examine the stacked spectra of the (photometric) blue-cloud galaxies (i.e., $\Delta_{GV} < -0.5$~mag) with respect to our three (spectroscopic) $\tau$-populations as defined in Sec.~\ref{sec:tauDistribution} (i.e., fast, $\tau<0.5$~Gyr; intermediate, $0.5<\tau<1.5$~Gyr; and slow, $\tau>1.5$~Gyr). As expected of star-forming galaxies, the blue-cloud galaxies of the fast, intermediate, and slow $\tau$-populations (shown in the bottom, middle and top panels of the Figure, respectively) reveal spectral emission lines (H$\alpha$/\Sii, \Oiii/H$\beta$, and \Oii) indicative of on-going star-formation activity \citep[e.g.,][]{Kennicutt1998, Kewley2004, Moustakas2006}. Moreover, the spectra do not show significant $4000$\AA\ breaks or strong metallic absorption lines indicative of more evolved stellar populations \citep[e.g.,][]{Bruzual1983, Hamilton1985, Jaschek1995, Bruzual2003}, which confirms that blue $NUVrK$-selected galaxies are indeed star-forming, and predominantly young (i.e., dominated by light from young stars). The slow $\tau$-population, however, shows much stronger emission lines than the intermediate and fast blue-cloud galaxies, with the intermediate population only displaying moderate H$\alpha$ and moderate to weak \Oiii\ \& \Oii, and the fast population only revealing moderate H$\alpha$ emission (and perhaps some weak absorption lines). Unless {these spectral features are} due to AGN contamination, this trend {is} consistent with galaxies on slow tracks being visible over longer times near the peak of their star formation compared to rapidly evolving galaxies, and therefore showing on average higher star-formation rates. As seen in Figure~\ref{fig:agetau}~ (as well as in Fig.~\ref{fig:finalmass} in the next Section), this seems to be supported by our data where all of our blue-cloud galaxies on fast tracks are seen on the declining phase of their SFHs while blue-cloud galaxies on slower tracks are also seen around or before the peak of their SFHs. Consequently, a higher fraction of our fast blue-cloud galaxies are near {the transition} to the green-valley region compared to the slow population, which might further reduce the average emission line strength of our fast blue-cloud population. 
It is beyond the scope of this paper to further explore these potential effects. 

Next, in Figure~\ref{fig:stacksBGR}, we examine the stacked spectra of our (spectroscopic) fast-$\tau$ galaxies in the (photometric) blue-cloud, green-valley, and  red-sequence regions\footnote{Note that we cannot investigate the stacked spectra of our intermediate and fast $\tau$ galaxies in the green-valley and red-sequence regions as our sample {contains} too few or no such galaxies. This can be seen in Fig.~\ref{fig:agetau} and is further observed in Fig.~\ref{fig:grism_dGV_vs_age} as well.}. 
As also seen in Fig.~\ref{fig:stacksSF}, the (fast-$\tau$) blue-cloud galaxies (bottom panel in the Figure) show moderate H$\alpha$ emission and no other significant spectral features, consistent with ongoing star formation activity. On the other hand, the green-valley galaxies (middle panel) already present a significant $4000$\AA\ break and several moderate to weak absorption lines (namely, Ca H \&\ K, H$\delta$, G-band and/or H$\gamma$, and H$\beta$), and only reveal weak to very weak H$\alpha$ emission.
The appearance of the $4000$\AA\ break in the green-valley spectrum is expected, since spectral synthesis models show that the $4000$\AA\ break appears rather quickly ($\sim100$~Myr  after the cessation or decline of star-formation activity) and subsequently grows only slowly with time \citep[see, e.g., Fig.~4 in][]{Bruzual1993}. The absorption lines seen in the green-valley spectrum have also been shown to be associated with the presence of a significant number of cool, late-type stars in galaxy spectra \citep[i.e., typically A to G types,][]{Jaschek1995}, while the level of H$\alpha$ emission seen in the stacked spectra indicates only low-level, residual star-formation activity, unless significantly dust-absorbed. The presence of these spectral features is consistent with the ageing of the stellar populations and the ongoing quenching of previously blue, star-forming galaxies now moved into the green valley. In the top panel of Figure~\ref{fig:stacksBGR}, which shows the stacked spectra of the red-sequence population, we again see the presence of the $4000$\AA\ break and the previously identified absorption lines characteristic of relatively old stellar populations. In contrast to the other panels, we do not see any residual emission lines commonly associated with star-formation activity, and we identify the presence of moderate to strong Mg and Na absorption lines, not seen in the other stacks. While Mg and Na are also commonly associated with old stellar populations, they typically show strong absorption in cooler, later star-types than the previously mentioned lines \citep[roughly F to K-types, and K to M-types, respectively, although they weakly start to appear in the stellar photospheres of slightly earlier stellar types;][]{Jaschek1995}. Together, this is consistent with the further ageing of the stellar populations for the red-sequence population compared to the blue cloud and the green valley, as well as the further (and likely near complete) cessation of star formation.

Overall, these behaviours are consistent with the gradual stellar population ageing and SFR decline of our fast $\tau$-population from the blue cloud, through the green valley, and on to the red sequence. Consequently, by securely resolving the green valley using the $NUVrK$ diagram and spectroscopically examining the galaxies therein, we support the picture whereby green-valley galaxies are intermediate in evolutionary state between the star-forming and quiescent populations and, indeed, lie on a \BGR\ quenching time-sequence.

\subsection{The galaxy colour-age relation}\label{sec:colourage}

\begin{figure}
    \centering
    \includegraphics[width=8.3cm]{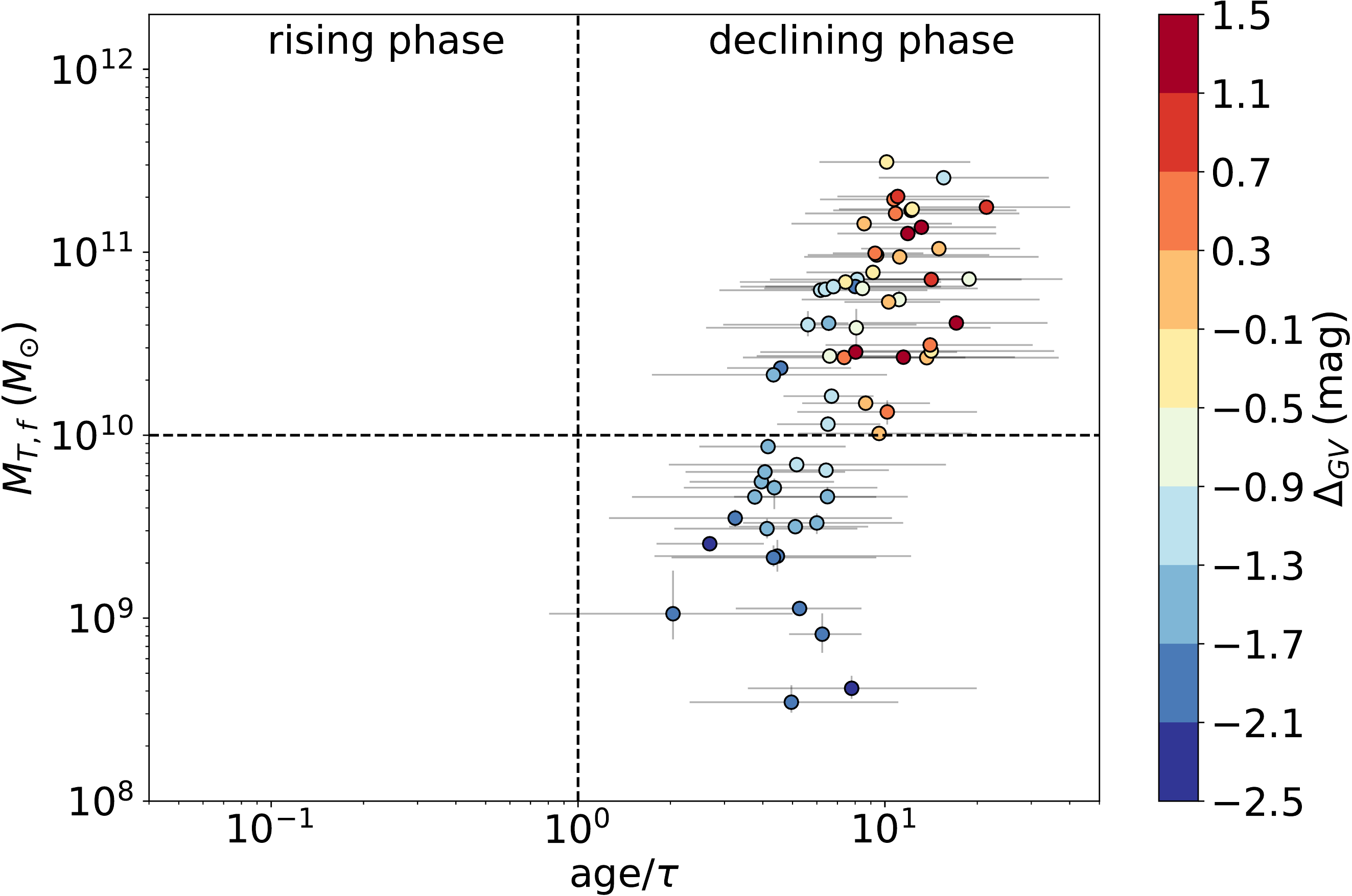}
    \caption{Evolutionary sequence selection in terms of final total mass, $M_{T,f}$, vs.~$\rm age/\tau$ for the fast-$\tau$ population, colour-coded by $\Delta_{GV}$. In this diagram, galaxies move horizontally from left to right as they age and become red. Galaxies that have similar final total masses, $M_{T,f}$, form a consistent evolutionary sequence. Because of limited statistics, we create two samples only. Sample~1 (S1) which imposes no selection in final total mass and includes all galaxies with $M_{T,f}>10^{8} M_{\odot}$, and Sample~2 (S2) which only includes galaxies with final total masses $M_{T,f}>10^{10} M_{\odot}$. We apply the same selection criteria to create S1 and S2 samples for the intermediate and slow $\tau$-populations.}
    \label{fig:finalmass}
\end{figure}

\begin{figure}
  \centering
    \includegraphics[width=8.3cm]{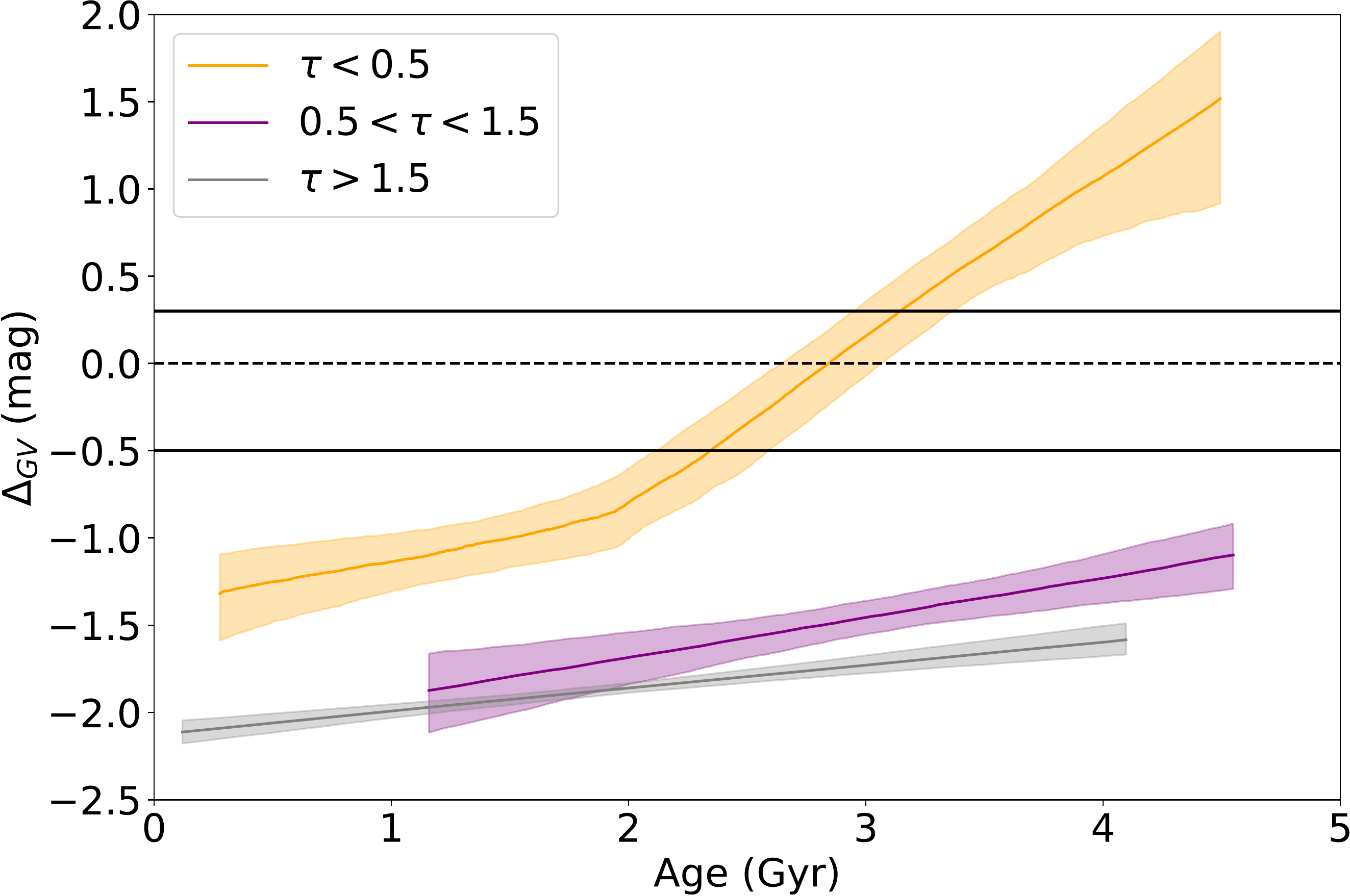}
    \caption{Distance to the bottom of the green valley, \deltaGV, versus age for each $\tau$ population (fast, intermediate and slow, shown in orange, purple, and grey, respectively). For each population, the solid line and shaded area represent the median and $68\%$ interval of the derived colour-age relation. The relation for the fast-$\tau$ population (orange) is derived using a three-slope model, while we use single-slope models to derive the relations of the intermediate (purple) and slow (grey) populations (see Sec.~\ref{sec:colourage} for details). In this model, the fast-$\tau$ galaxies take $\sim$0.8~Gyr to cross the green valley, starting at age~$\sim$~2.4 Gyr. Slower-$\tau$ galaxies show shallow colour evolution, and stay within the blue cloud at the redshifts probed here.}
    \label{fig:grism_dGV_vs_age}
\end{figure}

As seen in Figure~\ref{fig:agetau}, there is a clear relationship between galaxy age and colour at fixed $\tau$, which is also supported by our visual inspection of the stacked spectra in the previous Section (i.e., Fig.~\ref{fig:stacksSF}~\&~\ref{fig:stacksBGR}). 
For each of the three $\tau$-populations defined in Sec.~\ref{sec:tauDistribution}, we quantitatively derive the colour evolution of {its} galaxies, quantified by \deltaGV, as a function of their age. 
In an attempt to better connect galaxies on consistent evolutionary sequences, we further divide each $\tau$-population into bins of galaxy final total masses, $M_{T,f}$, as defined by integrating over {the individual galaxies'} total SFHs (i.e., to $\rm age \to +\infty$)\footnote{While $M_{T}$, defined in Sec.~\ref{sec:broadebandSED}, represents the total mass formed up to the redshift of observation (i.e., $M_{T} = \int_{0}^{\rm age}$SFR($t$)$dt$), $M_{T,f}$ represents here the final total mass formed after a given galaxy will have ceased forming new stars according to its SFH; i.e., $M_{T,f} = \int_{0}^{+\infty}$SFR($t$)$dt$.}. 

Figure~\ref{fig:finalmass} shows our selection for the fast $\tau$-population, in terms of final total masses, $M_{T,f}$, as a function of $\rm age/\tau$. In the Figure, galaxies move horizontally from left to right as they age and become red. We create two samples, a first statistical sample that includes all galaxies with $M_{T,f} > 10^{8} M_{\odot}$ (hereafter S1), and a refined sample that only includes galaxies with $M_{T,f} > 10^{10} M_{\odot}$ (hereafter S2).  
We use the same thresholds to create similar samples for the intermediate and slow $\tau$-populations, but in the following (i.e., Fig.~\ref{fig:grism_dGV_vs_age}) we only show results for the $M_{T,f} > 10^{8} M_{\odot}$ samples as they do not differ from the higher mass samples for these two $\tau$-populations.  
While our S1 samples are statistically more significant than our S2 samples, they contain galaxies with a large range of final masses which could present a potential bias (e.g., as seen in Figure~\ref{fig:finalmass}, our sample does not capture the red descendants of our fast-$\tau$, $M_{T,f} < 10^{10} M_{\odot}$ blue galaxies). On the other hand, our refined, S2 samples ensure that galaxies not only follow similar $\tau$-tracks (i.e., have similar SFH shapes) but also follow more consistent stellar mass build-up (i.e., have similar SFH normalizations), albeit at the cost of sample size. 
Note that for the fast-$\tau$ population, the S1 sample is comprised of $65$ sources, including $37$, $12$, and $16$ in the blue cloud, green valley, and red sequence, respectively. On the other hand, the S2 sample is lacking the $M_{T,f} < 10^{10} M_{\odot}$ sources of the S1 sample, which are all blue-cloud galaxies, and is comprised of $45$ sources, including $17$, $12$, and $16$ in the blue cloud, green valley, and red sequence, respectively. For both S1 \&~S2 fast-$\tau$ samples, the stellar mass ($M_{\star}$) and final total mass ($M_{T,f}$) distributions are almost identical, only offset by an approximately constant $\sim0.2$~dex due to {\it (i)} the mass difference between living stellar mass ($M_{\star}$) and total mass formed ($M_{T}$), and {\it (ii)} the minimal remaining stellar mass build-up for galaxies that have already spent a significant amount of time on the declining phase of their SFHs, which includes all fast-$\tau$ galaxies in our sample (see Sec.~\ref{sec:agetau} and Fig.~\ref{fig:finalmass}, respectively).

To derive each colour-age relation, we fit linear models to the data with the Markov Chain Monte Carlo (MCMC) {\sc {python}} sampler {\sc {emcee}} \citep{Foreman2013}. For the fast $\tau$-population, which covers the full range of photometric colours (blue-cloud, green-valley, and red-sequence galaxies), we first fit the S1 galaxies with a model (hereafter Model 1) consisting of three contiguous segments defined by a total of six free parameters: the slopes of the three segments, the zero-point of the first segment, and the two break-points between the three segments. This choice of modelling allows {us} to fit the colour-age relation irrespective of our boundaries for what constitutes the blue cloud, the green valley and the red sequence. We only use single slope models for the intermediate and slow $\tau$-populations which only cover the range of blue photometric colours. In all fits, we also include a nuisance parameter and we use 32 walkers for a total of 20,000 steps with a burn-in of 4000 and a thinning of 10. In Figure~\ref{fig:grism_dGV_vs_age} we show the medians (solid lines) and 68\% intervals (shaded areas) of the models for the fast, intermediate and slow $\tau$-populations in orange, purple and grey, respectively. As seen in the Figure, the fast population shows the most colour evolution with time (i.e., with galaxy age). First, this population leaves the blue cloud at age $\sim$2.4~Gyr, shows {an acceleration} in colour change slightly before its galaxies enter the green valley, and finally reaches the red sequence at age $\sim$3.2~Gyr with some flattening of the colour evolution (albeit marginal in this model).
As expected, the other two $\tau$-populations show flatter trends over the full range of galaxy ages, with changes of less than \deltaGV$\sim$1 mag within $\sim4$~Gyr of evolution. In the time available at these redshifts, these slower populations never reach the green valley or the red sequence, as was also seen in Figure~\ref{fig:agetau}. 

\begin{figure}
  \centering
    \includegraphics[width=8.3cm]{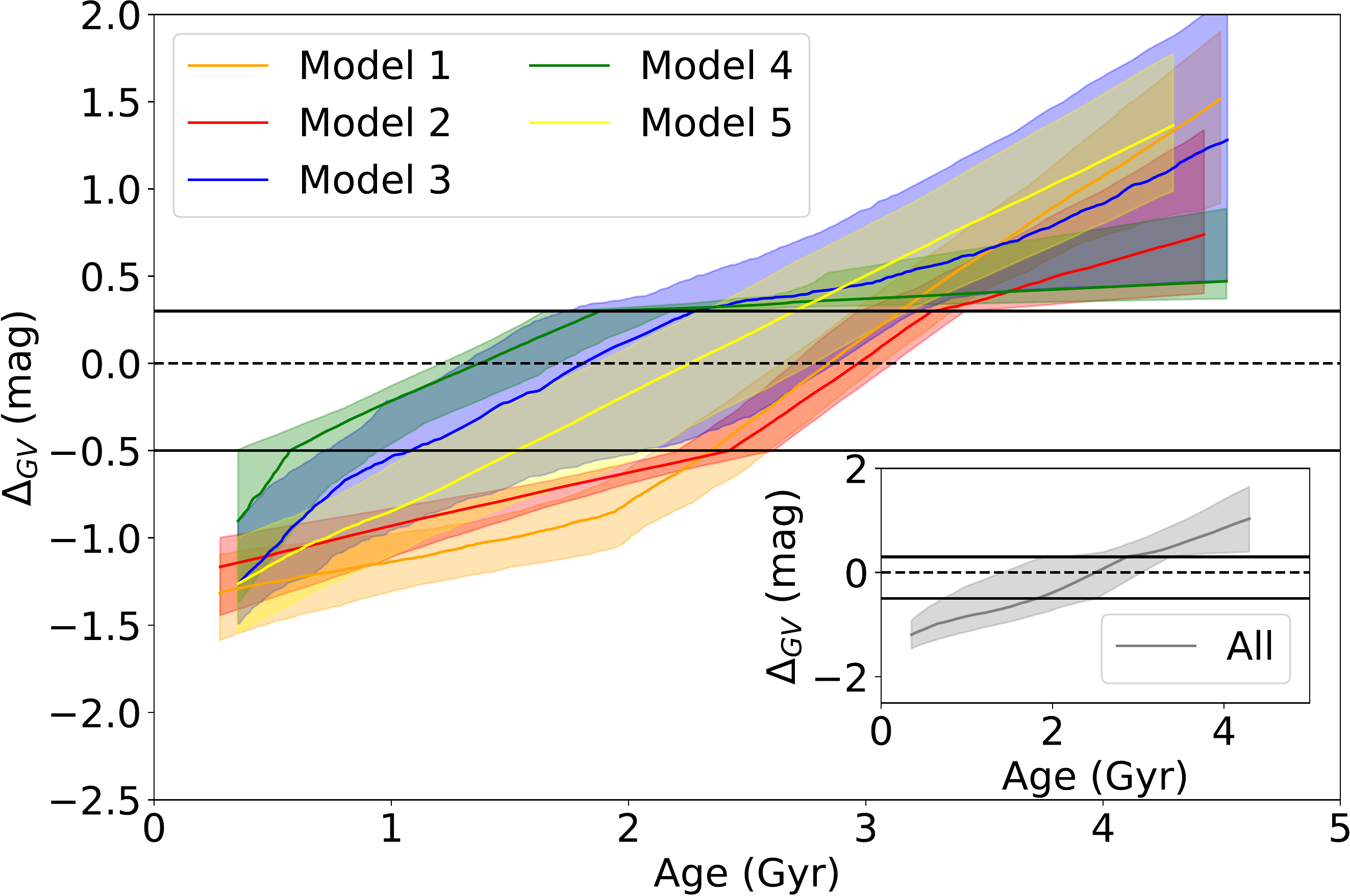}
    \caption{Distance to the bottom of the green valley, \deltaGV, versus age (same as Fig.~\ref{fig:grism_dGV_vs_age}), but for the five fast-$\tau$ models explored in Sec.~\ref{sec:colourage}. As in Fig.~\ref{fig:grism_dGV_vs_age}, the lines and shaded areas represent the median and $68\%$ interval of the derived colour-age relation for each model, through MCMC fitting of the data. Models 1 \&\ 2 are based on the fitting of the fast-$\tau$ S1 sample, while Models 3, 4, \&\ 5 are based on the fitting of the S2 sample (see Fig.~\ref{fig:finalmass} and Sec.~\ref{sec:colourage} for details on the Samples and Models). The inset shows the derived colour-age relation from all five models combined. Again, the line and grey shaded area represent the median and $68\%$ interval of the relation. The relation is steady in the blue cloud, accelerates in the green valley, and flattens in the red sequence.}
    \label{fig:grism_dGV_vs_age_models}
\end{figure}

To explore systematic effects in our derived colour-age relation of the fast-$\tau$ population, we also fit the data in the following ways: (a) we first adjust our Model 1 parameters by constraining the model break-points to our green-valley boundaries as defined in Sec.~\ref{sec:greenvalley} (i.e., $\Delta_{GV} = -0.5$ and $0.3$~mag); this imposes that each photometric region defined here (blue-cloud, green-valley, and red-sequence) is fitted with a single slope, we call this Model 2; (b) we also fit S2 galaxies (i.e., $M_{T,f} > 10^{10} M_{\odot}$) following the approaches of Model 1 and Model 2, hereafter referring to them as Model 3 and Model 4, respectively; (c) last, we also fit S2 galaxies with a single slope model, hereafter Model 5.
Figure~\ref{fig:grism_dGV_vs_age_models} shows the derived colour-age relations from the five models.
Model 2, shown in red, is consistent with Model 1 in the blue-cloud and green-valley regions and shows a flatter slope in the red sequence. Despite the smaller number of galaxies in the S2 sample, especially at young ages (i.e., $\lesssim2$~Gyr) in the blue cloud ($7$ sources compared to $18$ in the S1 sample), Models 3, 4, and 5 have similar slopes to those of Models 1 and 2 in the green valley, albeit slightly shallower and {entering the green valley at} younger ages. Additionally, Models 3 and 4, similarly to Model 2, show a flattening of the colour-age relation at $\Delta_{GV} \gtrsim 0.3$~mag, consistent with galaxies experiencing no further significant colour evolution and resting in the red sequence after quenching. Albeit not informative with respect to potential differences between the colour regions, the simpler single-slope Model 5, shown in yellow in  Figure~\ref{fig:grism_dGV_vs_age_models}, is globally consistent with the range of colours and ages of the other models. The inset in the Figure shows the median and $68\%$ interval of all five models combined, derived by randomly drawing 500 samples from the posterior distribution of each colour-age relation. It shows and confirms that, on average, the galaxy colour-age relation is steady in the blue cloud, accelerates in the green valley, and subsequently flattens in the red sequence.

From the combined colour-age relation (inset panel in Fig.~\ref{fig:grism_dGV_vs_age_models}), we can measure the average time required for a fast-$\tau$ galaxy to reach the lower boundary of the green-valley region (i.e., $\Delta_{GV} = -0.5$~mag). We measure {this time to be} $1.8^{+0.7}_{-1.1}$~Gyr. For our fast-$\tau$ population, we also measure a median $\tau$ value of $\sim 0.25$~Gyr. Taken at face value, this indicates that,  on average, galaxies reaching the lower boundary of the green valley have already spent a significant amount of time on the declining phase of their SFHs (i.e., with declining SFRs), about $\sim 1.5$~Gyr at the redshifts probed here. This represents the time between the peak of a galaxy's SFR and when it enters the green-valley region as defined in our work, and roughly represents $1.5/1.8 \sim 80\%$ of the galaxies' lifetime (i.e., age) {at the time that} they enter the green valley. This is consistent with our previous results from Sec.~\ref{sec:agetau} (bottom panel in Fig.~\ref{fig:agetau}), where we find that galaxies on fast $\tau$-track only leave the photometric blue cloud after they have spent up to $\sim 80\%$ of their lifetime (i.e., age) on the declining phase of their SFHs.

\subsection{Green-Valley crossing time}\label{sec:gvcrossingtime}

\subsubsection{Green-Valley crossing time and crossing rate}\label{sec:crossingTimeMeasurement}

\begin{figure}
  \centering
    \includegraphics[width=8.3cm]{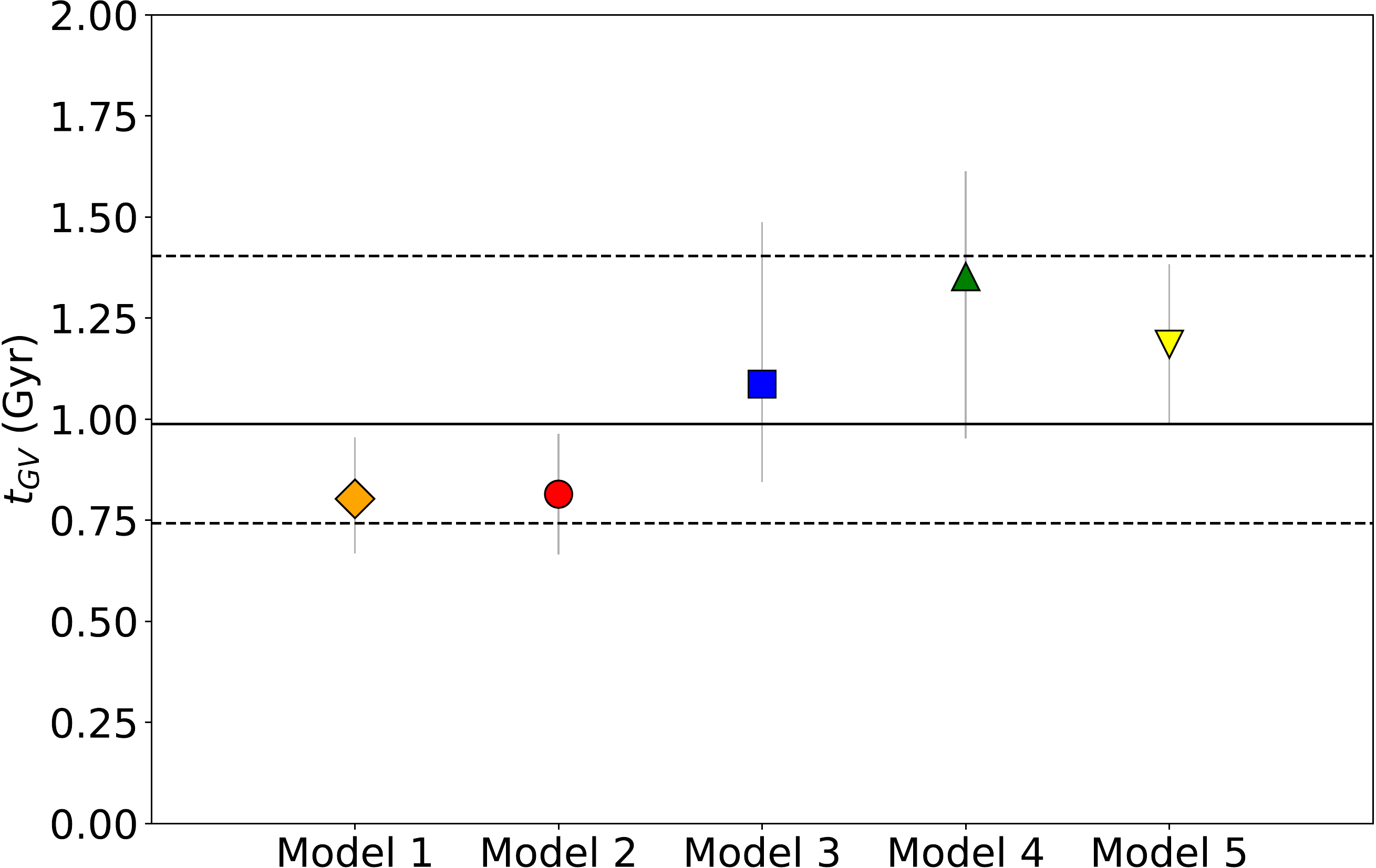}
    \includegraphics[width=8.3cm]{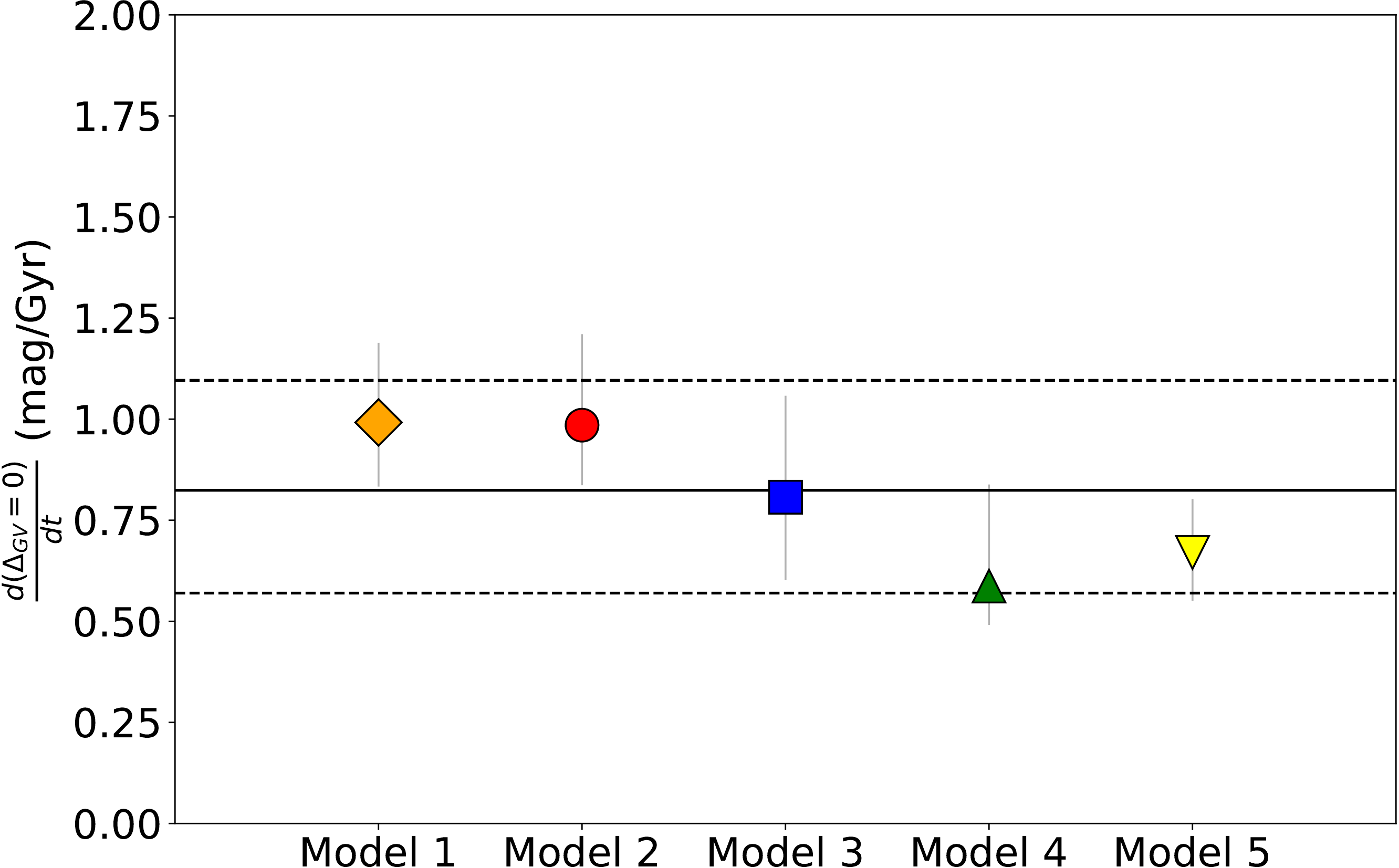}
    \caption{Top panel: measured green-valley crossing time-scales ($t_{GV}$), in Gyr, for each model as described in Sec.~\ref{sec:colourage}. The solid and dashed lines represent the median and $68\%$ interval of all five models combined. We measure a combined green-valley crossing time-scale of $0.99^{+0.42}_{-0.25}$~Gyr.
    Bottom panel: same as the top panel but for our measured crossing rates at the bottom of the green valley ($d(\Delta_{GV} = 0) / dt$), in mag per Gyr. Again, the solid and dashed lines represent the median and $68\%$ interval of all five models combined, and we measure a combined crossing rate at the bottom of the green valley of $0.82^{+0.27}_{-0.25}$~mag/Gyr.
    Table~\ref{tab:gv_time-scales} summarizes the measured time-scales for each model.}
    \label{fig:GV_crossingtime}
\end{figure}

In this Section we use the derived colour-age relations of the fast $\tau$-population to constrain the galaxy transition time-scale through the green valley. The top panel of Figure~\ref{fig:GV_crossingtime} shows the derived green-valley crossing time-scales, $t_{GV}$, of the five models of Sec.~\ref{sec:colourage}. This time-scale is here defined as the time difference between $\Delta_{GV} = 0.3$~mag (i.e., our red-sequence/green-valley boundary) and $\Delta_{GV} = -0.5$~mag (i.e., our green-valley/blue-cloud boundary) in the colour-age relations. While the fits to the S1 galaxy sample (i.e., Models 1 and 2) show slightly faster crossing time-scales compared to the fits to the S2 sample (i.e., Models 3, 4, and 5), all time-scales are roughly consistent within the uncertainties, and we measure
a global, combined $t_{GV}$ of $0.99^{+0.42}_{-0.25}$~Gyr. In the Figure, the data points and uncertainty ranges represent the medians and 68\% intervals of each $t_{GV}$ measurement, derived by measuring $t_{GV}$ on 500 samples randomly-drawn from the posterior distribution of each colour-age relation, separately.
We show the combined median and 68\% interval from the five models as the solid and dashed lines, respectively, derived by randomly drawing 500 samples from the $t_{GV}$ distribution of each individual model.
Our result suggests that on average, all the galaxies populating the green valley at the redshifts of our sample, will have transitioned to the red sequence within $\sim 1$~Gyr. 

One of the complications of discussing green-valley crossing time-scales is that such measurements depend on the width of the green valley, which is likely different in different studies. To present a more universal quantity, we also consider the instantaneous crossing rate at the bottom of the green valley. In the bottom panel of Figure~\ref{fig:GV_crossingtime}, we show the galaxy crossing rate at the bottom of the green valley, $d(\Delta_{GV} = 0) / dt$, measured from our data and reported in mag/Gyr, where magnitudes correspond to $\Delta_{GV}$ colours (i.e., $NUV-r$~magnitudes). This represents the rate at which galaxies are progressing on the colour-colour diagram as they cross the lowest density region between the blue cloud and the red sequence. As in the top panel of Figure~\ref{fig:GV_crossingtime}, individual points represent each of our five fitting models and the solid and dashed lines represent the median and $68\%$ interval of all models combined, all derived similarly as for our $t_{GV}$ measurements. While individual points show some differences, they are again consistent within the uncertainties, and we measure a combined crossing rate at the bottom of the green valley of $0.82^{+0.27}_{-0.25}$~mag/Gyr.
While our measured green-valley transition time-scale of $0.99^{+0.42}_{-0.25}$~Gyr depends on the definition of our green-valley boundaries, we note that the crossing rate at the bottom of the green valley measured here is independent from such choices and may represent a more robust measurement of the crossing rate of galaxies as they transition from the blue, star-forming region to the red, quiescent region of the $NUVrK$ colour-colour diagram.
Table \ref{tab:gv_time-scales} summarizes the measured green-valley crossing time-scales ($t_{GV}$) and crossing rates at the bottom of the green valley ($d(\Delta_{GV} = 0) / dt$) for the five models explored here as well as for all {the} models combined.

\begin{table*}
\centering
\caption{Summary table of the measured green-valley time-scales, for the different fits to the fast-$\tau$ galaxy colour-age relation (Fig.~\ref{fig:GV_crossingtime}). The Models differ in how the colour-age relation is parametrized {and which samples are fitted. Models 1 \&\ 2 are fitted with the S1 sample using three slopes with either free constraints between each slope (Model 1), or with break-points set to the green-valley boundaries (Model 2). Model 3 \&\ 4 follow the parametrizations of Model 1 \&\ 2, respectively, but fitted on the S2 sample. Model 5 is a single slope model fitted on the S2 sample.}. Sample S1 includes all fast-$\tau$ objects while Sample S2 only includes fast-$\tau$ objects with final total masses of $M_{T,f} > 10^{10} M_{\odot}$. See details in Section~\ref{sec:colourage}.}
\label{tab:gv_time-scales}
\begin{tabular}{lcccccc}
\hline
{Model ID} & Number of slopes & Break-points & Sample & Mass range & $t_{GV}$ & $d(\Delta_{GV} = 0) / dt$ \\
& in each Model & constraints & fitted & ($\log_{10}(M_{T,f}/M_{\odot})$) & (Gyr) & (mag/Gry)\\
\hline
Model 1     & 3		& free						& S1    & $> 8$		& $0.80^{+0.15}_{-0.14}$     & $0.99^{+0.20}_{-0.16}$\\[5pt]
Model 2     & 3		& set to green-valley boundaries	& S1    & $> 8$		& $0.81^{+0.15}_{-0.15}$     & $0.99^{+0.23}_{-0.15}$\\[5pt]
Model 3     & 3		& free						& S2    & $> 10$	& $1.09^{+0.40}_{-0.24}$     & $0.81^{+0.25}_{-0.21}$\\[5pt]
Model 4     & 3		& set to green-valley boundaries	& S2    & $> 10$	& $1.35^{+0.26}_{-0.40}$     & $0.59^{+0.25}_{-0.10}$\\[5pt]
Model 5     & 1		& -							& S2	    & $> 10$	& $1.19^{+0.20}_{-0.20}$     & $0.67^{+0.13}_{-0.12}$\\[5pt]
Combined    & - 	& -							& - 	    & -		& $0.99^{+0.42}_{-0.25}$     & $0.82^{+0.27}_{-0.25}$\\
\hline
\end{tabular}
\end{table*}

\subsubsection{Comparison to the literature}\label{sec:crossingTimeLiterature}

The current literature lacks comparable (i.e., $NUVrK$-based) estimates of the transition time-scale of green-valley galaxies at the redshifts probed in this work. {We therefore compare our results to that of other approaches and to that of lower-redshift $NUVrK$ studies}.
At similar redshifts {to ours}, \citet{Pandya2017} estimate transition time-scales based on the identification of star-forming, transitioning, and quiescent galaxies in the sSFR-$M_{\star}$ plane. Using a sample of $M_{\star}>10^9 M_{\odot}$ galaxies within $0.5<z<3$ from CANDELS complemented by the GAMA survey \citep{Driver2011, Liske2015} at low redshifts (i.e., $0.005<z<0.12$), they define such transitioning galaxies as the objects within $1.5\sigma$ to $3.5\sigma$ below the fits to the star-forming main sequence (SFMS) in seven redshift bins up to $z=3$. Estimating the $M_{\star}>10^{10} M_{\odot}$ galaxy number densities of each population (i.e., star-forming, transitioning, and quiescent), they then estimate the galaxy transition time-scale through the transition phase based on the relative change in number densities between their transitioning and quiescent populations with time (see \citealp{Pandya2017} for details). Assuming that all transitioning galaxies only move unidirectionally towards the quiescent population with time, these authors find transition time-scales upper limits of about $3$~Gyr at $z\sim1$ to about $1$~Gyr at $z=2$. While their methodology differs from ours, these results are not inconsistent with our green-valley crossing time-scales. In other words, this does not contradict the idea that green-valley galaxies reach quiescence within $\sim 1$~Gyr of evolution at redshifts of $z=1.0-1.8$, as measured in our work.

Other works in the literature mostly probe lower redshifts, and typically find slower green-valley transition time-scales, as can be expected from simple observations of, e.g., the slow down of the {growth} of the quiescent SMFs at redshifts $z\lesssim1$ \citep[e.g.,][]{Ilbert2010}. For instance, based on a sample of galaxies from the GAMA survey in the redshift range $z=0.1-0.2$, \citet{Phillipps2019} find that green-valley galaxies defined in the ($u-r$) colour\footnote{As already emphasized, note that defining green-valley galaxies in the $NUVrK$ diagram reduces the contamination from the star-forming and quiescent populations compared to other methods employing, e.g., $U$-band to optical colours (see \citealp{Siudek2018, Moutard2020b}). Careful considerations should therefore be taken when comparing time-scales derived using different methodologies, and we here do not attempt to compare different results beyond general trends.} vs.~mass plane and ranging in stellar masses from $10^{9.5} M_{\odot}$ to $10^{11.2} M_{\odot}$ require from 2~Gyr (for their less massive galaxies) to 4~Gyr (for their most massive galaxies) of evolution to reach past the green-valley/red-sequence boundary that they define in the colour-mass plane.

This result is somewhat consistent with {that of} \citet{Rowlands2018} who independently estimate lower limits on the transition time-scales of $M_{\star}>10^{10.6} M_{\odot}$ post-starburst (PSB) and green-valley galaxies from the GAMA survey at $z=0.05-1.0$. In their work, they define PSB and green-valley galaxies from a Principal Component Analysis introduced in \citet{Wild2007,Wild2009} based on spectral indices related to the strength of the $4000$\AA\ break and the excess of the Balmer absorption lines, and derive their time-scales based on the comparison of the SMFs of the two populations to that of the red sequence. Assuming that the growth of the quiescent population is entirely due to their green-valley population (i.e., assuming no contribution from the PSB galaxies), they find a constant transition time-scale of $2.6_{-0.7}^{+1.4}$~Gyr over $z=0.05-1.0$, which they indicate may represent a lower limit estimate as both green-valley and PSB populations may, in reality, be transitioning to the red sequence.
Inversely, the same authors find a transition time-scale through the PSB phase at $z\sim0.7$ of $0.5_{-0.1}^{+0.3}$~Gyr, assuming that the green-valley population does not contribute to the growth of the quiescent population. In other words, according to these authors, PSB galaxies could entirely account for the growth of the quiescent population at $z\sim0.7$ given a visibility time-scale of $\sim 0.5$~Gyr for these galaxies. We further discuss the visibility time-scales of PSB and green-valley galaxies and their implications with respect to the growth of the quiescent population in Section~\ref{sec:SMFgrowth}.

Additionally, \citet{Moutard2016b} estimate green-valley crossing time-scales {at $0.2<z<0.5$ } by comparing the colour evolution of BC03 models to the distribution of $M_{\star}\sim10^{10.6} M_{\odot}$ galaxies in the $NUVrK$ diagram. In their work, they use a sample of galaxies from the VIPERS \citep{Guzzo2014} Multi-Lambda Survey \citep[VIPERS-MLS;][]{Moutard2016a} and, similarly to our work, define the green valley {in relation to}
the lowest density region between the blue cloud and the red sequence in the $NUVrK$ diagram. Exploring a suite of SFHs of constant star-formation followed by an exponential decline and tracking the $NUVrK$ colours of these models with respect to the observed $NUVrK$ distribution, they show that the models most consistent with their observations allow a range of crossing time-scales within about $1$ to $3.5$~Gyr at $0.2<z<0.5$. They conclude that the building of the $M_{\star}\sim10^{10.6} M_{\odot}$ quiescent population must be driven by slow quenching mechanisms at these redshifts. This is consistent with the other {studies} at similar redshifts discussed {above}, and, assuming a gradual lengthening of quenching time-scales over time, {it is in overall agreement}
with our measured crossing time-scale of $0.99^{+0.42}_{-0.25}$~Gyr at $z=1.0-1.8$. Note that these results are also consistent with the individual time-scales we measure for each combination of model and sample (Models 1 to 5, and samples S1 \& S2, see Table~\ref{tab:gv_time-scales}). It is worth mentioning that this includes Models 3, 4, \&~5 for which the range of galaxy stellar masses considered (i.e., the S2 sample, $M_{T,f}>10^{10} M_{\odot}$) might be more similar to the studies mentioned here, compared to Models 1 \& 2 which include our entire fast-$\tau$ population (i.e., the S1 sample). Note, however, that the small size of our S2 sample, and especially its apparent lack of young blue galaxies ($7$ sources with ages $\lesssim2$~Gyr compared to $18$ for the S1 sample), are a potential bias of Models 3, 4, \&~5, and we therefore do not attempt a more detailed comparison of our individual time-scales to that of the literature beyond the global trends mentioned here.

\begin{figure*}
  \centering
    \includegraphics[width=14cm]{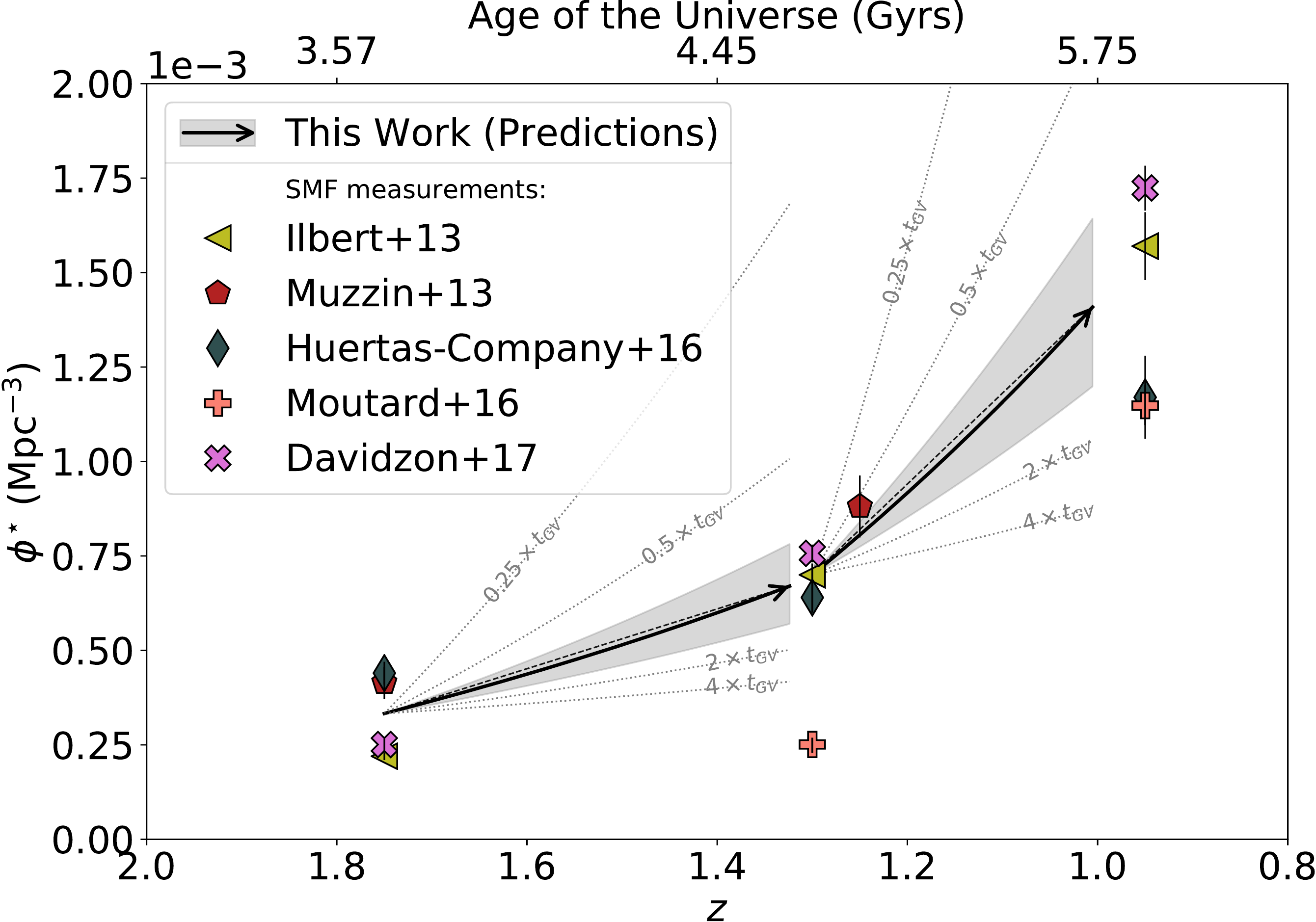}
    \caption{
    Growth of the quiescent $\phi^*$ as a function of redshift.
    The data points show the measured $\phi^*$ of the quiescent SMFs from different surveys in the literature. The black arrows are an independent measurement and represent our predicted $\phi^*$ growth based on our derived green-valley crossing time-scale, $t_{GV}$, and our mass-complete green-valley and red-sequence galaxy number densities (see details in Sec.~\ref{sec:SMFgrowth}). The arrows and shaded areas represent the medians and $68\%$ intervals of our $\phi^*$ growth predictions over $1$~Gyr of evolution, where $1$~Gyr corresponds to the green-valley crossing time-scale from our data (i.e., $t_{GV} = 0.99^{+0.42}_{-0.25}$~Gyr). The arrows assume a linear growth over time, while, for comparison, the thin dashed lines represent our predictions assuming a linear growth in redshift space. There is a remarkable agreement between our predicted $\phi^*$ growth and the measured $\phi^*$ of the quiescent SMFs from the literature. The grey dotted lines represent scenarios where we consider alternative green-valley crossing time-scales of $0.25 \times t_{GV}$, $0.5 \times t_{GV}$, $2 \times t_{GV}$, and $4 \times t_{GV}$, which correspond to crossing time-scales of about $0.25$, $0.5$, $2$, and $4$~Gyr, respectively. In contrast to the predictions based on our actual measured time-scale (black arrows), these alternative scenarios do not reproduce the growth of the quiescent SMF observed in the literature.
    }
    \label{fig:phis_vs_z}
\end{figure*}

\subsection{Growth of the quiescent galaxy stellar mass function}\label{sec:SMFgrowth}

\subsubsection{Growth of the red sequence}\label{subsec:redseqgrowth}

Stellar mass functions (SMFs) have been widely used in the literature to investigate stellar mass build-up through cosmic time and understand differences in the growth of stellar mass for different galaxy populations within a wide range of cosmic epochs \citep[e.g.,][]{Ilbert2010, Ilbert2013,  Muzzin2013, Mortlock2015, Huertas-Company2016, Moutard2016b, Davidzon2017, Arcila-Osejo2019,  Kawinwanichakij2020, McLeod2021}.
At low to intermediate redshifts, it is now fairly well accepted that the growth of the quiescent galaxy SMF is primarily driven by two different processes: mass-quenching at high masses (i.e., $\gtrsim 10^{10}~M_{\odot}$) and environmental quenching at lower masses \citep[e.g.,][]{Peng2010}, which result in quiescent SMFs that are comprised of two components, each described by a \cite{Schechter1976} function. 
In this scenario, it can be assumed that the shape of each component is independent of redshift and follows the Schechter functional form, with fixed turnover mass ($M_{\star}^*$) and low-mass slope coefficients ($\alpha_1$, $\alpha_2$), while only the normalizations of the quiescent galaxy number densities ($\phi^*_1$, $\phi^*_2$) evolve, separately, for each component \citep{Peng2010}. Evidence for this growth scenario of the quiescent SMF is seen observationally both in the local Universe \citep{Peng2010, Baldry2012} and up to intermediate redshifts \citep[$z\sim1.5$,][]{Arcila-Osejo2013, Arcila-Osejo2019, McLeod2021}, and these SMF results support the idea that mass-quenching has been the dominant mechanism for the growth of the massive quiescent galaxy population {over} the past $\sim 9$~Gyr of cosmic time. 

However, this scenario {\emph{assumes}} that galaxies move from the blue cloud to populate the quiescent SMF, whereby, by {construction}, they must first transition through the green valley. {Many of the crossing time measurements in the literature rely on this assumption to derive their crossing time-scales (Sec.~\ref{sec:crossingTimeLiterature}), while our own crossing time measurements do not (Sec.~\ref{sec:crossingTimeMeasurement})}. {We can therefore test this growth scenario by checking} whether there exists sufficiently many such transitioning galaxies to account for the growth of the quiescent galaxy population in the young Universe. As seen in Sec.~\ref{sec:crossingTimeMeasurement}, galaxies currently in the green valley at our redshifts will have all transitioned to the red sequence on the time-scale $t=t_{GV} \sim 1$~Gyr. Using this time-scale and our green-valley and red-sequence number densities we can therefore predict the growth rate of the red-sequence population within $t=1$~Gyr of evolution at our redshifts, and compare it to the growth rate of the quiescent SMFs from the literature at these epochs.

We derive our galaxy number densities for our green-valley and red-sequence populations using our final photometric sample down to stellar masses of $10^{10}~M_{\odot}$. {As mentioned in Sec.~\ref{sec:grism_sel}}, this threshold ensures a conservative mass completeness over our redshift range for all galaxy populations \citep[see, e.g., Fig.~1 in][]{Grazian2015}. To test for systematic effects, we also use a more relaxed ($10^{9.5}~M_{\odot}$) and a more conservative ($10^{10.5}~M_{\odot}$) threshold, and find that {doing so} doesn't change our results. We therefore only report results using our $10^{10}~M_{\odot}$ mass threshold. We find equivalent galaxy number densities between our green-valley and red-sequence populations ($150$ and $174$ sources over our photometric volume within $z_{phot} = 1.0-1.8$, respectively). Albeit statistically poorer, we also find equivalent number densities between both populations when using our spectroscopic sample down to the same stellar mass threshold ($10$ and $12$ sources over our spectroscopic volume within $z_{grism} = 1.0-1.8$, respectively). These green-valley and red-sequence number densities suggest that our red-sequence population would overall grow by a factor of $\sim 2$ once all green-valley galaxies will have transitioned to the red sequence. Taken at face value together with our green-valley crossing time-scale measured in Sec.~\ref{sec:gvcrossingtime}, this implies a factor of $2.01^{+0.33}_{-0.30}$ number density growth of the red sequence over $1$~Gyr of evolution at these redshifts.

This, however, assumes that all green-valley galaxies in our mass-complete sample will move on to the (mass-complete) red sequence. As mentioned earlier, this does not account for the possibility of rejuvenation (either in-situ or external in origin), nor does it account for the possibility of green-valley/green-valley or green-valley/red-sequence (dry) major mergers en route to quiescence. Albeit likely not significant \citep[e.g.,][]{Mendez2011, Weigel2017, Chauke2019}, if taken into account, these effects would potentially (slightly) decrease our estimated growth rate of the red sequence. Also, our analysis ignores the possibility of red-sequence/red-sequence major mergers within our $\sim1$~Gyr transition time-scale, which would also potentially decrease our estimated growth rate of the red sequence if taken into account. 
On the other hand, note that we are applying the same stellar mass threshold to estimate our (mass-complete) green-valley and red-sequence number densities. Green-valley galaxies might acquire some stellar mass (albeit not significantly, see Section~\ref{sec:agetau}) as their SFRs decline to quiescence. 
Additionally, a number of red-sequence galaxies above our mass-complete threshold of $10^{10}~M_{\odot}$ might have potentially quenched as a result of (dry) major mergers between green-valley/green-valley or green-valley/red-sequence galaxies of (pre-merger) stellar masses below $10^{10}~M_{\odot}$. Therefore, our green-valley number density estimate might not be accounting for a (small) number of green-valley galaxies (of masses slightly below $10^{10}~M_{\odot}$) that would be the progenitors (either through remaining star-formation or major dry mergers) of some red-sequence galaxies with stellar masses near our stellar mass completeness threshold. Similarly, major merger events between $M_{\star} < 10^{10}~M_{\odot}$ red-sequence/red-sequence galaxies can in principle contribute to the build-up of the $M_{\star} > 10^{10}~M_{\odot}$ red sequence within our green-valley transition time-scale of $\sim1$~Gyr.
Taken into account, these effects would potentially (slightly) increase our estimated growth rate of the red sequence. While the net effect of all these potential biases remain to be determined, they likely do not significantly affect our results as they are usually small effects \citep{Weigel2017, Chauke2019}.  {Moreover, these effects} would individually contribute in opposite directions to our red-sequence growth rate estimate (i.e., both towards its increase and towards its decrease). We defer the further exploration of these effects to a future analysis.

\subsubsection{Prediction of the growth of the quiescent SMF}

If the red sequence is populated by galaxies crossing the green valley, the growth rate of the quiescent galaxy SMFs must match the growth rate we predict from our crossing time measurements (i.e., $2.01^{+0.33}_{-0.30}$~Gyr$^{-1}$, see Sec.~\ref{subsec:redseqgrowth}). To compare to the SMFs in the literature, we first compile the normalization parameters $\phi^*$ of the quiescent SMFs measured within $z=0.8-2.0$ from different surveys \citep[namely,][]{Ilbert2013,  Muzzin2013, Huertas-Company2016, Moutard2016b, Davidzon2017}. These surveys all derive their quiescent SMFs in our redshift range of interest with single Schechter functions down to roughly similar mass completenesses of $\sim 10^{9.8}~M_{\odot}$ and turnover masses of $M_\star^*\sim 10^{10.7}~M_{\odot}$, with only small variations depending on redshifts and surveys; these mass ranges correspond to the mass-quenching regime of the SMFs. While these surveys cover different total sky areas (i.e., from $\sim 0.24$~deg$^{2}$ to $\sim 22$~deg$^2$) and use different colour-selection techniques to separate the star-forming and quiescent populations (i.e., $UVJ$, $NUVrJ$, or $NUVrK$ definitions), there is an overall good agreement and consistent trend of the evolution of $\phi^*$ as a function of redshift. 
{In Figure~\ref{fig:phis_vs_z} we show the $\phi^*$ measurements from these surveys at different redshifts (data points). There is a clear trend whereby the number density of quiescent galaxies, $\phi^*$, grows with decreasing redshift.} Only \citet{Moutard2016b} report lower values of $\phi^*$ in their highest redshift bin of $z=1.1-1.5$ compared to the other surveys. While derived over a much larger sky-area than the other surveys ($\sim 22$~deg$^2$ compared to $\lesssim 1.5$~deg$^2$ for the others), this discrepancy might partially be a result of their more stringent selection of the quiescent population as they identify and separate green-valley galaxies from the star-forming and quiescent populations, while the other surveys compiled here only identify the latter two populations. Despite this difference, their results follow a similar trend than the other surveys for the evolution of $\phi^*$ with redshift (see \citealp{Moutard2016b} for a more detailed discussion of their results and comparison to other works).

In Figure~\ref{fig:phis_vs_z}  we also show our independent predictions of the growth of $\phi^*$ over time at these redshifts (black arrows, with shaded grey areas indicating uncertainty ranges). We anchor our predictions by normalizing our red-sequence growth rate of $2.01^{+0.33}_{-0.30}$~Gyr$^{-1}$ to the median of the direct $\phi^*$ measurements at $z=1.75$ and, separately, at $z\simeq1.3$. For both 1~Gyr intervals (i.e., $z=1.75$ to $z\simeq1.3$, and $z=1.3$ to $z\simeq1.0$), the arrows and shaded areas represent the medians and $68\%$ intervals of our $\phi^*$ growth predictions over $1$~Gyr of evolution, respectively. In the Figure, we show our predictions assuming both a linear growth over time (black arrows), as well as in redshift space (thin dashed lines), which are almost identical. We also consider alternative growth scenarios using green-valley crossing time-scales of $0.25 \times t_{GV}$, $0.5 \times t_{GV}$, $2 \times t_{GV}$, and $4 \times t_{GV}$, which we show as grey dotted lines in the Figure. These correspond to crossing time-scales of about $0.25$, $0.5$, $2$, and $4$~Gyr, respectively, as opposed to $\sim 1$~Gyr for our predicted $\phi^*$ growth. As seen in the Figure, these alternative scenarios do not reproduce the growth of the quiescent SMFs from the literature.

The agreement seen in Fig.~\ref{fig:phis_vs_z} between the redshift-dependent quiescent number densities measured by SMF studies (data points), and our predictions based on our time-scale measurement (black arrows) is remarkable. It seems to account completely for the growth of the quiescent SMF at these redshifts and supports the idea that green-valley galaxies move on to the red sequence. Consequently, it supports the scenario whereby the massive quiescent galaxy population at these redshifts grows via the mass-quenching of star-forming galaxies through the green valley, within crossing time-scales as measured in Sec.~\ref{sec:gvcrossingtime} (i.e., $t_{GV} = 0.99^{+0.42}_{-0.25}$~Gyr).

\subsubsection{{Contribution from PSB galaxies}}

As mentioned in Section~\ref{sec:crossingTimeLiterature}, other works in the literature have suggested that the growth of the red sequence could be entirely accounted for by galaxies going through a PSB phase characterized by rapid quenching, assuming that such transition phase be of the order of, e.g., $\sim 0.5$~Gyr at $z\sim0.7$ \citep{Rowlands2018}, or shorter at even higher redshifts. Indeed, \citet{Wild2016} compare the post-starburst SMFs to that of the quiescent population at different redshift intervals within $z=0.5-2.0$ for a sample of $M_{\star}>10^{10} M_{\odot}$ galaxies selected from the UKIDSS Ultra Deep Survey \citep{Lawrence2007} and show that very short PSB visibility time-scales of the order of $250$~Myrs are required at $z=0.5-1.5$ for the red-sequence growth to be entirely accounted for by the PSB galaxies.

However, based on the SED-fitting of deep Keck MOSFIRE and LRIS spectra of a sample of $M_{\star} = 10^{10.5}-10^{11.5} M_{\odot}$ galaxies at $z=1.0-2.5$, \citet{Belli2019} derive PSB visibility times of $\sim0.5$~Gyr based on the derived ageing of PSB galaxies along the blue side of the quiescent population in the $UVJ$ diagram. Using the PSB and red-sequence number densities from the UltraVISTA survey \citep{McCracken2012}, \citet{Belli2019} subsequently indicate that this fast quenching through the PSB phase could potentially only account for up to half of the growth of the red sequence at $z\sim2.2$ and up to only one fifth of the growth of the red sequence at $z\sim1.4$ \citep[and therefore probably even lower at lower redshifts, including the redshift range considered in][]{Wild2016}.
In a recent work, \citet{Wild2020} also measure PSB visibility times based on the SED-fitting of VLT/VIMOS and FORS2 spectra of 39 galaxies within $z=0.5-1.3$ and $M_{\star} > 10^{10} M_{\odot}$, and find values of $0.5$ to $1$~Gyr. Based on their previous work \citep[i.e.,][]{Wild2016}, the authors suggest that only up to 25 to 50\% of the growth of the red sequence may actually be accounted for by the quenching of galaxies through a PSB phase at redshifts slightly lower than one. {This indicates that} PSB galaxies may not contribute as much to the growth of the red sequence as previously thought, at least up to intermediate redshifts. 

However, note that different works in the literature not only employ different methodologies to estimate transition time-scales and to determine contributions to the growth of the quiescent population, but also often use different definitions to identify transitioning galaxies (whether PSB or green-valley). PSB definitions typically rely on either: {\it (i)} spectroscopic features via the strengths of the $H\delta$ absorption line, the 4000\AA\ break, and the \Oii\ emission line \citep[e.g.,][]{Goto2007, Wild2007, Muzzin2014} or other spectral measurement (e.g., the $A/K$ stellar-template ratio and $H\alpha$ equivalent width, \citealp{Quintero2004}), or {\it (ii)} photometric features via super-colours \citep[e.g.,][]{Wild2020} or galaxy colours in the rest-frame $UVJ$ diagram \citep[e.g.,][]{Whitaker2012, Belli2019}. 
Galaxies defined in these ways may contribute in various amounts to the population of ($M_{\star} > 10^{10} M_{\odot}$) green-valley galaxies as defined in our work. However, based on a sample of $z=1.0-1.4$ cluster galaxies from the GOGREEN survey \citep{Balogh2017}, \citet{McNab2021} show that the population of spectroscopic PSB and photometric ($UVJ$) PSB galaxies (referred to as Blue Quiescent galaxies in their work) do not significantly overlap with their green-valley population as defined using rest-frame $NUV-V$ vs. $V-J$ colours. 
This suggests that our $NUVrK$ green-valley region is likely not comprised of a significant number of PSB galaxies, and that the contribution from galaxies going through a PSB phase probably remains low in (a) our derived green-valley crossing time-scale, (b) red-sequence growth rate, and (c) predictions of the quiescent $\phi^{*}$ growth. However, we note that precisely understanding the level of such contribution in our work would require a detailed investigation of the spectral features of each galaxy identified in our green-valley region.

\section{Conclusions}\label{sec:conclusion}

In this paper, we constrain the colour evolution and quenching time-scales of $z=1.0-1.8$ galaxies across the green valley. We first derive rest-frame $NUVrK$ colours (as well as stellar masses and other properties) by SED-fitting CANDELS GOODS-S and UDS broadband data. We then photometrically identify blue-cloud, green-valley, and red-sequence galaxies based on the galaxy bi-modal distribution in the $NUVrK$ colour-colour diagram, which better resolves the green valley and better distinguishes between star-forming and quiescent galaxies compared to colour diagnostics covering shorter wavelength windows such as the $UVJ$. We additionally introduce a new colour parameter, $\Delta_{GV}$, which is the $NUV-r$ colour-distance to the bottom of the green valley (in AB~mag), and encodes how likely a galaxy is a transitioning (i.e., green-valley) galaxy, independently from any green-valley boundary definition. 
Independently from our broadband SED-fitting, we derive SFH parameters (i.e., ages, $\tau$) based on the fitting of deep {\it HST} G102 \&\ G141 grism spectroscopic data available in the \grizli\ database over the GOODS-S and UDS footprints. 
Combining our photometric $NUVrK$ classification with our independent spectroscopic SFH measurements, we investigate trends in the galaxy populations and reach the following conclusions: 

\begin{itemize}

\item[1.] We first investigate the relation between galaxy photometric colours and spectroscopically-derived SFH parameters, and reveal that galaxies follow different evolutionary tracks. We find a multi-modal $\tau$ distribution, and identify three $\tau$-populations: fast (i.e., $\tau<0.5$~Gyr), intermediate (i.e., $0.5<\tau<1.5$~Gyr), and slow (i.e., $\tau>1.5$~Gyr). Most galaxies are on the fast and slow tracks, with only a small number on the intermediate track.  We see that galaxies on a given spectroscopic $\tau$-track form a photometric colour-sequence (also correlated to spectroscopic age) from the blue cloud to the red sequence via intermediate, green-valley, colours, although only galaxies on fast $\tau$-tracks are seen on the red sequence at these redshifts. Galaxies on slower tracks did not yet leave their star-forming phase, which is not surprising given their long $\tau$ time-scales and the short Hubble times at these redshifts. 

\item[2.] We show that fast $\tau$-track galaxies only leave the blue cloud after they have spent a significant amount of time on the declining phase of their SFHs, of up to $\sim 80\%$ of their lifetime (i.e., age) {up to that point}, which corresponds to $\sim 1.5$~Gyr for these galaxies. This indicates that galaxy stellar mass build-up is minimal after a galaxy leaves the blue cloud, and may contribute to no more than $\sim 5\%$ of these galaxies' total, final masses. This also suggests that quenching time-scale definitions based on the fraction of mass a galaxy has formed should be carefully chosen and may be as high as $95\%$ for the {\emph{lower}} bound thresholds only to correspond to time-scales derived from colour definitions such as in this work.

\item[3.] We visually identify strong spectral features in the stacked grism spectra characteristic of the ageing of the stellar populations and the steady decline of the star formation along the 
\BGR\ track. This confirms that green-valley galaxies are intermediate in evolutionary state between the star-forming and quiescent populations.

\item[4.] We derive the colour-age relation of galaxies along the different $\tau$ tracks. We show that galaxies on intermediate and slow tracks undergo small and steady colour evolution, of less than $\sim 1$~mag within $\sim 4$~Gyr, and remain in the blue cloud over this time. In contrast, for galaxies on the fast tracks (i.e., $\tau<0.5$~Gyr), we show that their colour-age relation is steady in the blue cloud, accelerates in the green valley, and subsequently flattens in the red sequence within the same (i.e., $\sim 4$~Gyr) amount of time. For this group of galaxies, we measure a green-valley transition time-scale of $0.99^{+0.42}_{-0.25}$~Gyr within our green-valley boundaries, and we measure a $\Delta_{GV}$ (i.e., $NUV-r$) crossing rate at the bottom of the green valley of $0.82^{+0.27}_{-0.25}$~mag/Gyr (independent {of green-valley} boundary definitions).

\item[5.] Based on our green-valley transition time-scale for fast $\tau$ galaxies and our mass-complete (i.e., $M_{\star} > 10^{10} M_{\odot}$) green-valley and red-sequence number densities, we estimate the growth of the red sequence at our redshifts and find a factor of $2.01^{+0.33}_{-0.30}$ growth in number density per~Gyr. Using this growth estimate, we predict the growth of the quiescent galaxy characteristic number density, $\phi^{*}$,  within $z=1.8-1.0$ and find remarkable agreement between our prediction and the evolution of $\phi^{*}$ from direct SMF measurements in the literature. This remarkable agreement supports the scenario whereby the $M_{\star} > 10^{10} M_{\odot}$ quiescent galaxy population grows via the mass-quenching of star-forming galaxies through the green valley.

\end{itemize}

Altogether, our analysis is consistent with the scenario in which the quiescent galaxy population at $z=1.0-1.8$ grows primarily through the quenching of star-forming galaxies. This is supported by our results that show: {\it (i)} a remarkable agreement between the growth of the quiescent galaxy SMF measurements reported in the literature and the predicted number density growth of the red-sequence population that we derive from our spectroscopic measurements of the green-valley crossing time-scales, and {\it (ii)} spectroscopic features consistent with these galaxies following a time progression from the blue cloud, to the green valley, and on to the red sequence.

Our results offer a new approach to studying galaxy quenching and the build-up of the red-sequence population over time. In the future, we plan to extend our analysis by employing flexible star formation histories \citep[e.g.,][]{Iyer2019, Leja2019} to better constrain the evolutionary pathways of quenching galaxies, and by examining the physical sizes of galaxies in the green valley. This work with {\it HST} grisms represents a pathfinder study for future slitless-spectroscopic grism surveys using, e.g., the {\it JWST}'s Near Infrared Imager and Slitless Spectrograph (NIRISS; \citealp{Willott2022}; R.~Doyon et al., in prep), {\it Euclid}'s Near Infrared Spectro-Photometer \citep[NISP;][]{Costille2018}, the {\it Roman Space Telescope}'s grism spectrometer \citep{Gong2020} or the grism spectrograph onboard the planned {\it Cosmological Advanced Survey Telescope for Optical and UV Research} ({\it CASTOR}; \citealp{Cote2019}).

\section*{Acknowledgements}

{We thank the referee for their useful comments and suggestions that improved the clarity of the paper.}
We would like to thank the SMU extragalactic group for productive discussions about this work.
We acknowledge funding support from the Natural Sciences and Engineering Research Council (NSERC) of Canada through a Discovery Grant and Discovery Accelerator Supplement, and from the Canadian Space Agency through grant 18JWST-GTO1. 
CP is supported by the Canadian Space Agency under a contract with NRC Herzberg Astronomy and Astrophysics. 
This research has made use of data products obtained from the Grism Redshift \&~Line Analysis Software (\grizli) database, generously operated and maintained by Gabriel Brammer. This research has made use of data obtained from the Mikulski Archive for Space Telescopes (MAST). STScI is operated by the Association of Universities for Research in Astronomy, Inc., under NASA contract NAS5-26555. Finally, GN would like to thank Ludovic Montier for his expertise in linearity.

{\it Software}: {\tt ASTROPY} \citep{Astropy2013, Astropy2018}, {\tt MATPLOTLIB} \citep{Hunter2007}, {\tt NUMPY} \citep{Harris2020}, {\tt SCIKIT-LEARN} \citep{Pedregosa2012}, {\tt SCIPY} \citep{Virtanen2020}.

\section*{DATA AVAILABILITY}
The public datasets used in this research are available online as described in Section~\ref{sec:parentdata}. The derived data products underlying this research will be shared on reasonable request to the corresponding author.





\begin{thebibliography}{99}
\bibitem[Abadi et al.(1999)]{Abadi1999} Abadi, M.~G., Moore, B., \& Bower, R.~G.\ 1999, \mnras, 308, 947. 
\bibitem[Abramson et al.(2016)]{Abramson2016} Abramson, L.~E., Gladders, M.~D., Dressler, A., et al.\ 2016, \apj, 832, 7. 
\bibitem[Abramson et al.(2020)]{Abramson2020} Abramson, L.~E., Brammer, G.~B., Schmidt, K.~B., et al.\ 2020, \mnras, 493, 952. 
\bibitem[Acquaviva et al.(2011)]{Acquaviva2011} Acquaviva, V., Gawiser, E., \& Guaita, L.\ 2011, \apj, 737, 47. 
\bibitem[Ambikasaran et al.(2015)]{Ambikasaran2015} Ambikasaran, S., Foreman-Mackey, D., Greengard, L., et al.\ 2015, IEEE Transactions on Pattern Analysis and Machine Intelligence, 38, 252. 
\bibitem[Arcila-Osejo \& Sawicki(2013)]{Arcila-Osejo2013} Arcila-Osejo, L. \& Sawicki, M.\ 2013, \mnras, 435, 845. 
\bibitem[Arcila-Osejo et al.(2019)]{Arcila-Osejo2019} Arcila-Osejo, L., Sawicki, M., Arnouts, S., et al.\ 2019, \mnras, 486, 4880. 
\bibitem[Arnouts et al.(2007)]{Arnouts2007} Arnouts, S., Walcher, C.~J., Le F{\`e}vre, O., et al.\ 2007, \aap, 476, 137. 
\bibitem[Arnouts et al.(2013)]{Arnouts2013} Arnouts, S., Le Floc'h, E., Chevallard, J., et al.\ 2013, \aap, 558, A67. 
\bibitem[Asplund et al.(2009)]{Asplund2009} Asplund, M., Grevesse, N., Sauval, A.~J., et al.\ 2009, \araa, 47, 481. 
\bibitem[Astropy Collaboration et al.(2013)]{Astropy2013} Astropy Collaboration, Robitaille, T.~P., Tollerud, E.~J., et al.\ 2013, \aap, 558, A33. 
\bibitem[Astropy Collaboration et al.(2018)]{Astropy2018} Astropy Collaboration, Price-Whelan, A.~M., Sip{\H{o}}cz, B.~M., et al.\ 2018, \aj, 156, 123. 
\bibitem[Atek et al.(2010)]{Atek2010} Atek, H., Malkan, M., McCarthy, P., et al.\ 2010, \apj, 723, 104. 
\bibitem[Baldry et al.(2012)]{Baldry2012} Baldry, I.~K., Driver, S.~P., Loveday, J., et al.\ 2012, \mnras, 421, 621. 
\bibitem[Balogh et al.(2017)]{Balogh2017} Balogh, M.~L., Gilbank, D.~G., Muzzin, A., et al.\ 2017, \mnras, 470, 4168. 
\bibitem[Beifiori et al.(2017)]{Beifiori2017} Beifiori, A., Mendel, J.~T., Chan, J.~C.~C., et al.\ 2017, \apj, 846, 120. 
\bibitem[Belli et al.(2019)]{Belli2019} Belli, S., Newman, A.~B., \& Ellis, R.~S.\ 2019, \apj, 874, 17. 
\bibitem[Bertin \& Arnouts(1996)]{Bertin1996} Bertin, E. \& Arnouts, S.\ 1996, \aaps, 117, 393. 
\bibitem[Bower et al.(2017)]{Bower2017} Bower, R.~G., Schaye, J., Frenk, C.~S., et al.\ 2017, \mnras, 465, 32. 
\bibitem[Brammer et al.(2008)]{Brammer2008} Brammer, G.~B., van Dokkum, P.~G., \& Coppi, P.\ 2008, \apj, 686, 1503. 
\bibitem[Brammer et al.(2009)]{Brammer2009} Brammer, G.~B., Whitaker, K.~E., van Dokkum, P.~G., et al.\ 2009, \apjl, 706, L173. 
\bibitem[Brammer et al.(2012)]{Brammer2012} Brammer, G.~B., van Dokkum, P.~G., Franx, M., et al.\ 2012, \apjs, 200, 13. 
\bibitem[Brammer(2019)]{Brammer2019} Brammer, G.\ 2019, Astrophysics Source Code Library. ascl:1905.001
\bibitem[Bressan et al.(2012)]{Bressan2012} Bressan, A., Marigo, P., Girardi, L., et al.\ 2012, \mnras, 427, 127. 
\bibitem[Bruzual A.(1983)]{Bruzual1983} Bruzual A., G.\ 1983, \apj, 273, 105. 
\bibitem[Bruzual A. \& Charlot(1993)]{Bruzual1993} Bruzual A., G. \& Charlot, S.\ 1993, \apj, 405, 538. 
\bibitem[Bruzual \& Charlot(2003)]{Bruzual2003} Bruzual, G. \& Charlot, S.\ 2003, \mnras, 344, 1000. 
\bibitem[Buchner et al.(2014)]{Buchner2014} Buchner, J., Georgakakis, A., Nandra, K., et al.\ 2014, \aap, 564, A125. 
\bibitem[Buchner(2021)]{Buchner2021} Buchner, J.\ 2021, The Journal of Open Source Software, 6, 3001. 
\bibitem[Calzetti et al.(2000)]{Calzetti2000} Calzetti, D., Armus, L., Bohlin, R.~C., et al.\ 2000, \apj, 533, 682. 
\bibitem[Carnall et al.(2018)]{Carnall2018} Carnall, A.~C., McLure, R.~J., Dunlop, J.~S., et al.\ 2018, \mnras, 480, 4379. 
\bibitem[Carnall et al.(2019a)]{Carnall2019a} Carnall, A.~C., Leja, J., Johnson, B.~D., et al.\ 2019a, \apj, 873, 44. 
\bibitem[Carnall et al.(2019b)]{Carnall2019b} Carnall, A.~C., McLure, R.~J., Dunlop, J.~S., et al.\ 2019b, \mnras, 490, 417. 
\bibitem[Carnall et al.(2020)]{Carnall2020} Carnall, A.~C., Walker, S., McLure, R.~J., et al.\ 2020, \mnras, 496, 695. 
\bibitem[Castellano et al.(2007)]{Castellano2007} Castellano, M., Salimbeni, S., Trevese, D., et al.\ 2007, \apj, 671, 1497. 
\bibitem[Cimatti et al.(2006)]{Cimatti2006} Cimatti, A., Daddi, E., \& Renzini, A.\ 2006, \aap, 453, L29. 
\bibitem[Chauke et al.(2019)]{Chauke2019} Chauke, P., van der Wel, A., Pacifici, C., et al.\ 2019, \apj, 877, 48. 
\bibitem[Chevallard \& Charlot(2016)]{Chevallard2016} Chevallard, J. \& Charlot, S.\ 2016, \mnras, 462, 1415. 
\bibitem[Choi et al.(2016)]{Choi2016} Choi, J., Dotter, A., Conroy, C., et al.\ 2016, \apj, 823, 102. 
\bibitem[Conroy et al.(2009)]{Conroy2009} Conroy, C., Gunn, J.~E., \& White, M.\ 2009, \apj, 699, 486. 
\bibitem[Conroy \& Gunn(2010)]{Conroy2010} Conroy, C. \& Gunn, J.~E.\ 2010, \apj, 712, 833. 
\bibitem[Costille et al.(2018)]{Costille2018} Costille, A., Caillat, A., Rossin, C., et al.\ 2018, \procspie, 10698, 106982B. 
\bibitem[Cote et al.(2019)]{Cote2019} Cote, P., Abraham, B., Balogh, M., et al.\ 2019, clrp, 2020, 18. 
\bibitem[Cowie et al.(1996)]{Cowie1996} Cowie, L.~L., Songaila, A., Hu, E.~M., et al.\ 1996, \aj, 112, 839. 
\bibitem[Dahlen et al.(2013)]{Dahlen2013} Dahlen, T., Mobasher, B., Faber, S.~M., et al.\ 2013, \apj, 775, 93. 
\bibitem[Dahlen et al.(2005)]{Dahlen2005} Dahlen, T., Mobasher, B., Somerville, R.~S., et al.\ 2005, \apj, 631, 126. 
\bibitem[Darvish et al.(2016)]{Darvish2016} Darvish, B., Mobasher, B., Sobral, D., et al.\ 2016, \apj, 825, 113. 
\bibitem[Davidzon et al.(2017)]{Davidzon2017} Davidzon, I., Ilbert, O., Laigle, C., et al.\ 2017, \aap, 605, A70. 
\bibitem[Dempster et al.(1977)]{Dempster1977} Dempster, A., Laird, N., \& Rubin, D.\ 1977, J. R. Stat. Soc. Series B, 39, 1.
\bibitem[Dotter(2016)]{Dotter2016} Dotter, A.\ 2016, \apjs, 222, 8. 
\bibitem[Draine \& Li(2007)]{DraineLi2007} Draine, B.~T. \& Li, A.\ 2007, \apj, 657, 810. 
\bibitem[Driver et al.(2011)]{Driver2011} Driver, S.~P., Hill, D.~T., Kelvin, L.~S., et al.\ 2011, \mnras, 413, 971. 
\bibitem[Ekstr{\"o}m et al.(2012)]{Ekstrom2012} Ekstr{\"o}m, S., Georgy, C., Eggenberger, P., et al.\ 2012, \aap, 537, A146. 
\bibitem[Eldridge \& Stanway(2009)]{Eldridge2009} Eldridge, J.~J. \& Stanway, E.~R.\ 2009, \mnras, 400, 1019. 
\bibitem[Erb et al.(2006)]{Erb2006} Erb, D.~K., Shapley, A.~E., Pettini, M., et al.\ 2006, \apj, 644, 813. 
\bibitem[Estrada-Carpenter et al.(2020)]{Estrada2020} Estrada-Carpenter, V., Papovich, C., Momcheva, I., et al.\ 2020, \apj, 898, 171. 
\bibitem[Fang et al.(2018)]{Fang2018} Fang, J.~J., Faber, S.~M., Koo, D.~C., et al.\ 2018, \apj, 858, 100. 
\bibitem[Ferland et al.(2017)]{Ferland2017} Ferland, G.~J., Chatzikos, M., Guzm{\'a}n, F., et al.\ 2017, \rmxaa, 53, 385.
\bibitem[Feroz et al.(2009)]{Feroz2009} Feroz, F., Hobson, M.~P., \& Bridges, M.\ 2009, \mnras, 398, 1601. 
\bibitem[Feroz et al.(2019)]{Feroz2019} Feroz, F., Hobson, M.~P., Cameron, E., et al.\ 2019, The Open Journal of Astrophysics, 2, 10. 
\bibitem[Fontanot et al.(2009)]{Fontanot2009} Fontanot, F., De Lucia, G., Monaco, P., et al.\ 2009, \mnras, 397, 1776. 
\bibitem[Foreman-Mackey et al.(2013)]{Foreman2013} Foreman-Mackey, D., Hogg, D.~W., Lang, D., et al.\ 2013, \pasp, 125, 306. 
\bibitem[Gaia Collaboration et al.(2018)]{Gaia2018} Gaia Collaboration, Brown, A.~G.~A., Vallenari, A., et al.\ 2018, \aap, 616, A1. 
\bibitem[Galametz et al.(2013)]{Galametz2013} Galametz, A., Grazian, A., Fontana, A., et al.\ 2013, \apjs, 206, 10. 
\bibitem[Gawiser et al.(2007)]{Gawiser2007} Gawiser, E., Francke, H., Lai, K., et al.\ 2007, \apj, 671, 278. 
\bibitem[Gong et al.(2020)]{Gong2020} Gong, Q., Bergkoetter, M., Berrier, J., et al.\ 2020, JATIS, 6, 045008. 
\bibitem[Goto(2007)]{Goto2007} Goto, T.\ 2007, \mnras, 381, 187. 
\bibitem[Grazian et al.(2015)]{Grazian2015} Grazian, A., Fontana, A., Santini, P., et al.\ 2015, \aap, 575, A96. 
\bibitem[Grogin et al.(2011)]{Grogin2011} Grogin, N.~A., Kocevski, D.~D., Faber, S.~M., et al.\ 2011, \apjs, 197, 35. 
\bibitem[Gunn \& Gott(1972)]{Gunn1972} Gunn, J.~E. \& Gott, J.~R.\ 1972, \apj, 176, 1. 
\bibitem[Guo et al.(2013)]{Guo2013} Guo, Y., Ferguson, H.~C., Giavalisco, M., et al.\ 2013, \apjs, 207, 24. 
\bibitem[Guzzo et al.(2014)]{Guzzo2014} Guzzo, L., Scodeggio, M., Garilli, B., et al.\ 2014, \aap, 566, A108. 
\bibitem[Hamilton(1985)]{Hamilton1985} Hamilton, D.\ 1985, \apj, 297, 371. 
\bibitem[Harris et al.(2020)]{Harris2020} Harris, C.~R., Millman, K.~J., van der Walt, S.~J., et al.\ 2020, \nat, 585, 357. 
\bibitem[Hatch et al.(2016)]{Hatch2016} Hatch, N.~A., Muldrew, S.~I., Cooke, E.~A., et al.\ 2016, \mnras, 459, 387. 
\bibitem[Hogg et al.(2002)]{Hogg2002} Hogg, D.~W., Baldry, I.~K., Blanton, M.~R., et al.\ 2002, arXiv, astro-ph/0210394
\bibitem[Hopkins et al.(2008)]{Hopkins2008} Hopkins, P.~F., Hernquist, L., Cox, T.~J., et al.\ 2008, \apjs, 175, 356. 
\bibitem[Huertas-Company et al.(2016)]{Huertas-Company2016} Huertas-Company, M., Bernardi, M., P{\'e}rez-Gonz{\'a}lez, P.~G., et al.\ 2016, \mnras, 462, 4495. 
\bibitem[Hunter(2007)]{Hunter2007} Hunter, J.~D.\ 2007, Computing in Science and Engineering, 9, 90. 
\bibitem[Ilbert et al.(2005)]{Ilbert2005} Ilbert, O., Tresse, L., Zucca, E., et al.\ 2005, \aap, 439, 863. 
\bibitem[Ilbert et al.(2010)]{Ilbert2010} Ilbert, O., Salvato, M., Le Floc'h, E., et al.\ 2010, \apj, 709, 644. 
\bibitem[Ilbert et al.(2013)]{Ilbert2013} Ilbert, O., McCracken, H.~J., Le F{\`e}vre, O., et al.\ 2013, \aap, 556, A55. 
\bibitem[Ilbert et al.(2015)]{Ilbert2015} Ilbert, O., Arnouts, S., Le Floc'h, E., et al.\ 2015, \aap, 579, A2. 
\bibitem[Iyer \& Gawiser(2017)]{Iyer2017} Iyer, K. \& Gawiser, E.\ 2017, \apj, 838, 127. 
\bibitem[Iyer et al.(2019)]{Iyer2019} Iyer, K.~G., Gawiser, E., Faber, S.~M., et al.\ 2019, \apj, 879, 116. 
\bibitem[Jaschek \& Jaschek(1995)]{Jaschek1995} Jaschek, C. \& Jaschek, M.\ 1995, The Behavior of Chemical Elements in Stars, by Carlos Jaschek and Mercedes Jaschek, pp. 338. ISBN 052141136X. Cambridge, UK: Cambridge University Press, June 1995, 338.
\bibitem[Jian et al.(2020)]{Jian2020} Jian, H.-Y., Lin, L., Koyama, Y., et al.\ 2020, \apj, 894, 125. 
\bibitem[Johnson et al.(2021)]{Johnson2021} Johnson, B.~D., Leja, J., Conroy, C., et al.\ 2021, \apjs, 254, 22. 
\bibitem[Kawinwanichakij et al.(2020)]{Kawinwanichakij2020} Kawinwanichakij, L., Papovich, C., Ciardullo, R., et al.\ 2020, \apj, 892, 7. 
\bibitem[Kennicutt(1998)]{Kennicutt1998} Kennicutt, R.~C.\ 1998, \araa, 36, 189. 
\bibitem[Kewley et al.(2004)]{Kewley2004} Kewley, L.~J., Geller, M.~J., \& Jansen, R.~A.\ 2004, \aj, 127, 2002. 
\bibitem[Koekemoer et al.(2011)]{Koekemoer2011} Koekemoer, A.~M., Faber, S.~M., Ferguson, H.~C., et al.\ 2011, \apjs, 197, 36. 
\bibitem[Krishnan et al.(2017)]{Krishnan2017} Krishnan, C., Hatch, N.~A., Almaini, O., et al.\ 2017, \mnras, 470, 2170. 
\bibitem[Kroupa(2001)]{Kroupa2001} Kroupa, P.\ 2001, \mnras, 322, 231. 
\bibitem[Kurk et al.(2009)]{Kurk2009} Kurk, J., Cimatti, A., Zamorani, G., et al.\ 2009, \aap, 504, 331. 
\bibitem[Labb{\'e} et al.(2005)]{Labbe2005} Labb{\'e}, I., Huang, J., Franx, M., et al.\ 2005, \apjl, 624, L81. 
\bibitem[Laidler et al.(2007)]{Laidler2007} Laidler, V.~G., Papovich, C., Grogin, N.~A., et al.\ 2007, \pasp, 119, 1325. 
\bibitem[Laigle et al.(2016)]{Laigle2016} Laigle, C., McCracken, H.~J., Ilbert, O., et al.\ 2016, \apjs, 224, 24. 
\bibitem[Lawrence et al.(2007)]{Lawrence2007} Lawrence, A., Warren, S.~J., Almaini, O., et al.\ 2007, \mnras, 379, 1599. 
\bibitem[Lee et al.(2018)]{Lee2018} Lee, B., Giavalisco, M., Whitaker, K., et al.\ 2018, \apj, 853, 131. 
\bibitem[Leja et al.(2017)]{Leja2017} Leja, J., Johnson, B.~D., Conroy, C., et al.\ 2017, \apj, 837, 170. 
\bibitem[Leja et al.(2019)]{Leja2019} Leja, J., Carnall, A.~C., Johnson, B.~D., et al.\ 2019, \apj, 876, 3. 
\bibitem[Liddle(2007)]{Liddle2007} Liddle, A.~R.\ 2007, \mnras, 377, L74. 
\bibitem[Liske et al.(2015)]{Liske2015} Liske, J., Baldry, I.~K., Driver, S.~P., et al.\ 2015, \mnras, 452, 2087. 
\bibitem[Lotz et al.(2017)]{Lotz2017} Lotz, J.~M., Koekemoer, A., Coe, D., et al.\ 2017, \apj, 837, 97. 
\bibitem[Lower et al.(2020)]{Lower2020} Lower, S., Narayanan, D., Leja, J., et al.\ 2020, \apj, 904, 33. 
\bibitem[Marigo et al.(2008)]{Marigo2008} Marigo, P., Girardi, L., Bressan, A., et al.\ 2008, \aap, 482, 883. 
\bibitem[Martig et al.(2009)]{Martig2009} Martig, M., Bournaud, F., Teyssier, R., et al.\ 2009, \apj, 707, 250. 
\bibitem[McCracken et al.(2012)]{McCracken2012} McCracken, H.~J., Milvang-Jensen, B., Dunlop, J., et al.\ 2012, \aap, 544, A156. 
\bibitem[McLeod et al.(2021)]{McLeod2021} McLeod, D.~J., McLure, R.~J., Dunlop, J.~S., et al.\ 2021, \mnras, 503, 4413. 
\bibitem[McNab et al.(2021)]{McNab2021} McNab, K., Balogh, M.~L., van der Burg, R.~F.~J., et al.\ 2021, \mnras, 508, 157. 
\bibitem[Mendez et al.(2011)]{Mendez2011} Mendez, A.~J., Coil, A.~L., Lotz, J., et al.\ 2011, \apj, 736, 110. 
\bibitem[Momcheva et al.(2016)]{Momcheva2016} Momcheva, I.~G., Brammer, G.~B., van Dokkum, P.~G., et al.\ 2016, \apjs, 225, 27. 
\bibitem[Moore et al.(1996)]{Moore1996} Moore, B., Katz, N., Lake, G., et al.\ 1996, \nat, 379, 613. 
\bibitem[Mortlock et al.(2015)]{Mortlock2015} Mortlock, A., Conselice, C.~J., Hartley, W.~G., et al.\ 2015, \mnras, 447, 2. 
\bibitem[Moustakas et al.(2006)]{Moustakas2006} Moustakas, J., Kennicutt, R.~C., \& Tremonti, C.~A.\ 2006, \apj, 642, 775. 
\bibitem[Moutard et al.(2016a)]{Moutard2016a} Moutard, T., Arnouts, S., Ilbert, O., et al.\ 2016a, \aap, 590, A102. 
\bibitem[Moutard et al.(2016b)]{Moutard2016b} Moutard, T., Arnouts, S., Ilbert, O., et al.\ 2016b, \aap, 590, A103. 
\bibitem[Moutard et al.(2020a)]{Moutard2020a} Moutard, T., Sawicki, M., Arnouts, S., et al.\ 2020a, \mnras, 494, 1894. 
\bibitem[Moutard et al.(2020b)]{Moutard2020b} Moutard, T., Malavasi, N., Sawicki, M., et al.\ 2020b, \mnras, 495, 4237. 
\bibitem[Muzzin et al.(2013)]{Muzzin2013} Muzzin, A., Marchesini, D., Stefanon, M., et al.\ 2013, \apj, 777, 18. 
\bibitem[Muzzin et al.(2014)]{Muzzin2014} Muzzin, A., van der Burg, R.~F.~J., McGee, S.~L., et al.\ 2014, \apj, 796, 65. 
\bibitem[Nantais et al.(2017)]{Nantais2017} Nantais, J.~B., Muzzin, A., van der Burg, R.~F.~J., et al.\ 2017, \mnras, 465, L104. 
\bibitem[Neistein et al.(2006)]{Neistein2006} Neistein, E., van den Bosch, F.~C., \& Dekel, A.\ 2006, \mnras, 372, 933. 
\bibitem[Nelson et al.(2018)]{Nelson2018} Nelson, D., Pillepich, A., Springel, V., et al.\ 2018, \mnras, 475, 624. 
\bibitem[Noirot et al.(2016)]{Noirot2016} Noirot, G., Vernet, J., De Breuck, C., et al.\ 2016, \apj, 830, 90. 
\bibitem[Noirot et al.(2018)]{Noirot2018} Noirot, G., Stern, D., Mei, S., et al.\ 2018, \apj, 859, 38. 
\bibitem[Pacifici et al.(2015)]{Pacifici2015} Pacifici, C., da Cunha, E., Charlot, S., et al.\ 2015, \mnras, 447, 786. 
\bibitem[Pandya et al.(2017)]{Pandya2017} Pandya, V., Brennan, R., Somerville, R.~S., et al.\ 2017, \mnras, 472, 2054. 
\bibitem[Papovich et al.(2001)]{Papovich2001} Papovich, C., Dickinson, M., \& Ferguson, H.~C.\ 2001, \apj, 559, 620. 
\bibitem[Papovich et al.(2010)]{Papovich2010} Papovich, C., Momcheva, I., Willmer, C.~N.~A., et al.\ 2010, \apj, 716, 1503. 
\bibitem[Patel et al.(2011)]{Patel2011} Patel, S.~G., Kelson, D.~D., Holden, B.~P., et al.\ 2011, \apj, 735, 53. 
\bibitem[Patel et al.(2012)]{Patel2012} Patel, S.~G., Holden, B.~P., Kelson, D.~D., et al.\ 2012, \apjl, 748, L27. 
\bibitem[Paxton et al.(2011)]{Paxton2011} Paxton, B., Bildsten, L., Dotter, A., et al.\ 2011, \apjs, 192, 3. 
\bibitem[Paxton et al.(2013)]{Paxton2013} Paxton, B., Cantiello, M., Arras, P., et al.\ 2013, \apjs, 208, 4. 
\bibitem[Paxton et al.(2015)]{Paxton2015} Paxton, B., Marchant, P., Schwab, J., et al.\ 2015, \apjs, 220, 15. 
\bibitem[Pedregosa et al.(2012)]{Pedregosa2012} Pedregosa, F., Varoquaux, G., Gramfort, A., et al.\ 2012, arXiv:1201.0490
\bibitem[Peng et al.(2010)]{Peng2010} Peng, Y., Lilly, S.~J., Kova{\v{c}}, K., et al.\ 2010, \apj, 721, 193. 
\bibitem[Peng et al.(2012)]{Peng2012} Peng, Y.-. jie ., Lilly, S.~J., Renzini, A., et al.\ 2012, \apj, 757, 4. 
\bibitem[Peng et al.(2015)]{Peng2015} Peng, Y., Maiolino, R., \& Cochrane, R.\ 2015, \nat, 521, 192. 
\bibitem[Phillipps et al.(2019)]{Phillipps2019} Phillipps, S., Bremer, M.~N., Hopkins, A.~M., et al.\ 2019, \mnras, 485, 5559. 
\bibitem[Pietrinferni et al.(2004)]{Pietrinferni2004} Pietrinferni, A., Cassisi, S., Salaris, M., et al.\ 2004, \apj, 612, 168. 
\bibitem[Poggianti et al.(2017)]{Poggianti2017} Poggianti, B.~M., Moretti, A., Gullieuszik, M., et al.\ 2017, \apj, 844, 48. 
\bibitem[Pontzen et al.(2017)]{Pontzen2017} Pontzen, A., Tremmel, M., Roth, N., et al.\ 2017, \mnras, 465, 547. 
\bibitem[Powell et al.(2017)]{Powell2017} Powell, M.~C., Urry, C.~M., Cardamone, C.~N., et al.\ 2017, \apj, 835, 22. 
\bibitem[Quintero et al.(2004)]{Quintero2004} Quintero, A.~D., Hogg, D.~W., Blanton, M.~R., et al.\ 2004, \apj, 602, 190. 
\bibitem[Rodr{\'\i}guez Montero et al.(2019)]{Rodriguez2019} Rodr{\'\i}guez Montero, F., Dav{\'e}, R., Wild, V., et al.\ 2019, \mnras, 490, 2139. 
\bibitem[Rowlands et al.(2018)]{Rowlands2018} Rowlands, K., Wild, V., Bourne, N., et al.\ 2018, \mnras, 473, 1168. 
\bibitem[Rudnick et al.(2003)]{Rudnick2003} Rudnick, G., Rix, H.-W., Franx, M., et al.\ 2003, \apj, 599, 847. 
\bibitem[Sales et al.(2015)]{Sales2015} Sales, L.~V., Vogelsberger, M., Genel, S., et al.\ 2015, \mnras, 447, L6. 
\bibitem[Salim et al.(2005)]{Salim2005} Salim, S., Charlot, S., Rich, R.~M., et al.\ 2005, \apjl, 619, L39. 
\bibitem[Salim(2014)]{Salim2014} Salim, S.\ 2014, Serbian Astronomical Journal, 189, 1. 
\bibitem[Salimbeni et al.(2009)]{Salimbeni2009} Salimbeni, S., Castellano, M., Pentericci, L., et al.\ 2009, \aap, 501, 865. 
\bibitem[S{\'a}nchez-Bl{\'a}zquez et al.(2006)]{Sanchez2006} S{\'a}nchez-Bl{\'a}zquez, P., Peletier, R.~F., Jim{\'e}nez-Vicente, J., et al.\ 2006, \mnras, 371, 703. 
\bibitem[Santini et al.(2015)]{Santini2015} Santini, P., Ferguson, H.~C., Fontana, A., et al.\ 2015, \apj, 801, 97. 
\bibitem[Santos et al.(2014)]{Santos2014} Santos, J.~S., Altieri, B., Tanaka, M., et al.\ 2014, \mnras, 438, 2565. 
\bibitem[Sawicki \& Yee(1998)]{Sawicki1998} Sawicki, M. \& Yee, H.~K.~C.\ 1998, \aj, 115, 1329. 
\bibitem[Sawicki(2001)]{Sawicki2001} Sawicki, M.\ 2001, \aj, 121, 2405. 
\bibitem[Sawicki(2012a)]{Sawicki2012a} Sawicki, M.\ 2012a, \mnras, 421, 2187. 
\bibitem[Sawicki(2012b)]{Sawicki2012b} Sawicki, M.\ 2012b, \pasp, 124, 1208. 
\bibitem[Schawinski et al.(2014)]{Schawinski2014} Schawinski, K., Urry, C.~M., Simmons, B.~D., et al.\ 2014, \mnras, 440, 889. 
\bibitem[Schechter(1976)]{Schechter1976} Schechter, P.\ 1976, \apj, 203, 297. 
\bibitem[Schwarz(1978)]{Schwarz1978} Schwarz, G.\ 1978, Annals of Statistics, 6, 461
\bibitem[Shapley et al.(2005)]{Shapley2005} Shapley, A.~E., Steidel, C.~C., Erb, D.~K., et al.\ 2005, \apj, 626, 698. 
\bibitem[Siudek et al.(2018)]{Siudek2018} Siudek, M., Ma{\l}ek, K., Pollo, A., et al.\ 2018, \aap, 617, A70. 
\bibitem[\protect\citeauthoryear{Skilling}{2004}]{Skilling2004} Skilling J., 2004, AIPC, 735, 395. 
\bibitem[Speagle(2020)]{Speagle2020} Speagle, J.~S.\ 2020, \mnras, 493, 3132. 
\bibitem[Stanway \& Eldridge(2018)]{Stanway2018} Stanway, E.~R. \& Eldridge, J.~J.\ 2018, \mnras, 479, 75. 
\bibitem[Taylor et al.(2009)]{Taylor2009} Taylor, E.~N., Franx, M., van Dokkum, P.~G., et al.\ 2009, \apjs, 183, 295. 
\bibitem[Treu et al.(2005)]{Treu2005} Treu, T., Ellis, R.~S., Liao, T.~X., et al.\ 2005, \apjl, 622, L5. 
\bibitem[Trussler et al.(2020)]{Trussler2020} Trussler, J., Maiolino, R., Maraston, C., et al.\ 2020, \mnras, 491, 5406. 
\bibitem[Ueda et al.(2008)]{Ueda2008} Ueda, Y., Watson, M.~G., Stewart, I.~M., et al.\ 2008, \apjs, 179, 124. 
\bibitem[van den Bosch et al.(2008)]{Bosch2008} van den Bosch, F.~C., Aquino, D., Yang, X., et al.\ 2008, \mnras, 387, 79. 
\bibitem[Vergani et al.(2018)]{Vergani2018} Vergani, D., Garilli, B., Polletta, M., et al.\ 2018, \aap, 620, A193. 
\bibitem[Villaume et al.(2015)]{Villaume2015} Villaume, A., Conroy, C., \& Johnson, B.~D.\ 2015, \apj, 806, 82. 
\bibitem[Virtanen et al.(2020)]{Virtanen2020} Virtanen, P., Gommers, R., Oliphant, T.~E., et al.\ 2020, Nature Methods, 17, 261. 
\bibitem[Weigel et al.(2017)]{Weigel2017} Weigel, A.~K., Schawinski, K., Caplar, N., et al.\ 2017, \apj, 845, 145. 
\bibitem[Westera et al.(2002)]{Westera2002} Westera, P., Lejeune, T., Buser, R., et al.\ 2002, \aap, 381, 524. 
\bibitem[Whitaker et al.(2011)]{Whitaker2011} Whitaker, K.~E., Labb{\'e}, I., van Dokkum, P.~G., et al.\ 2011, \apj, 735, 86. 
\bibitem[Whitaker et al.(2012)]{Whitaker2012} Whitaker, K.~E., Kriek, M., van Dokkum, P.~G., et al.\ 2012, \apj, 745, 179. 
\bibitem[Wild et al.(2007)]{Wild2007} Wild, V., Kauffmann, G., Heckman, T., et al.\ 2007, \mnras, 381, 543. 
\bibitem[Wild et al.(2009)]{Wild2009} Wild, V., Walcher, C.~J., Johansson, P.~H., et al.\ 2009, \mnras, 395, 144. 
\bibitem[Wild et al.(2016)]{Wild2016} Wild, V., Almaini, O., Dunlop, J., et al.\ 2016, \mnras, 463, 832. 
\bibitem[Wild et al.(2020)]{Wild2020} Wild, V., Taj Aldeen, L., Carnall, A., et al.\ 2020, \mnras, 494, 529. 
\bibitem[Williams et al.(2009)]{Williams2009} Williams, R.~J., Quadri, R.~F., Franx, M., et al.\ 2009, \apj, 691, 1879. 
\bibitem[Willott et al.(2022)]{Willott2022} Willott, C.~J., Doyon, R., Albert, L., et al.\ 2022, arXiv:2202.01714
\bibitem[Wright et al.(2019)]{Wright2019} Wright, R.~J., Lagos, C. del P., Davies, L.~J.~M., et al.\ 2019, \mnras, 487, 3740. 
\bibitem[Wuyts et al.(2007)]{Wuyts2007} Wuyts, S., Labb{\'e}, I., Franx, M., et al.\ 2007, \apj, 655, 51. 
\bibitem[Xue et al.(2011)]{Xue2011} Xue, Y.~Q., Luo, B., Brandt, W.~N., et al.\ 2011, \apjs, 195, 10. 
\bibitem[Yabe et al.(2009)]{Yabe2009} Yabe, K., Ohta, K., Iwata, I., et al.\ 2009, \apj, 693, 507. 
\bibitem[Yuan et al.(2010)]{Yuan2010} Yuan, T.-T., Kewley, L.~J., \& Sanders, D.~B.\ 2010, \apj, 709, 884. 
\bibitem[Zolotov et al.(2015)]{Zolotov2015} Zolotov, A., Dekel, A., Mandelker, N., et al.\ 2015, \mnras, 450, 2327. 
\end{thebibliography}




\appendix

\section{Uncorrected rest-frame colours}\label{app:colours}

\begin{figure}
  \centering
    \includegraphics[width=8.3cm]{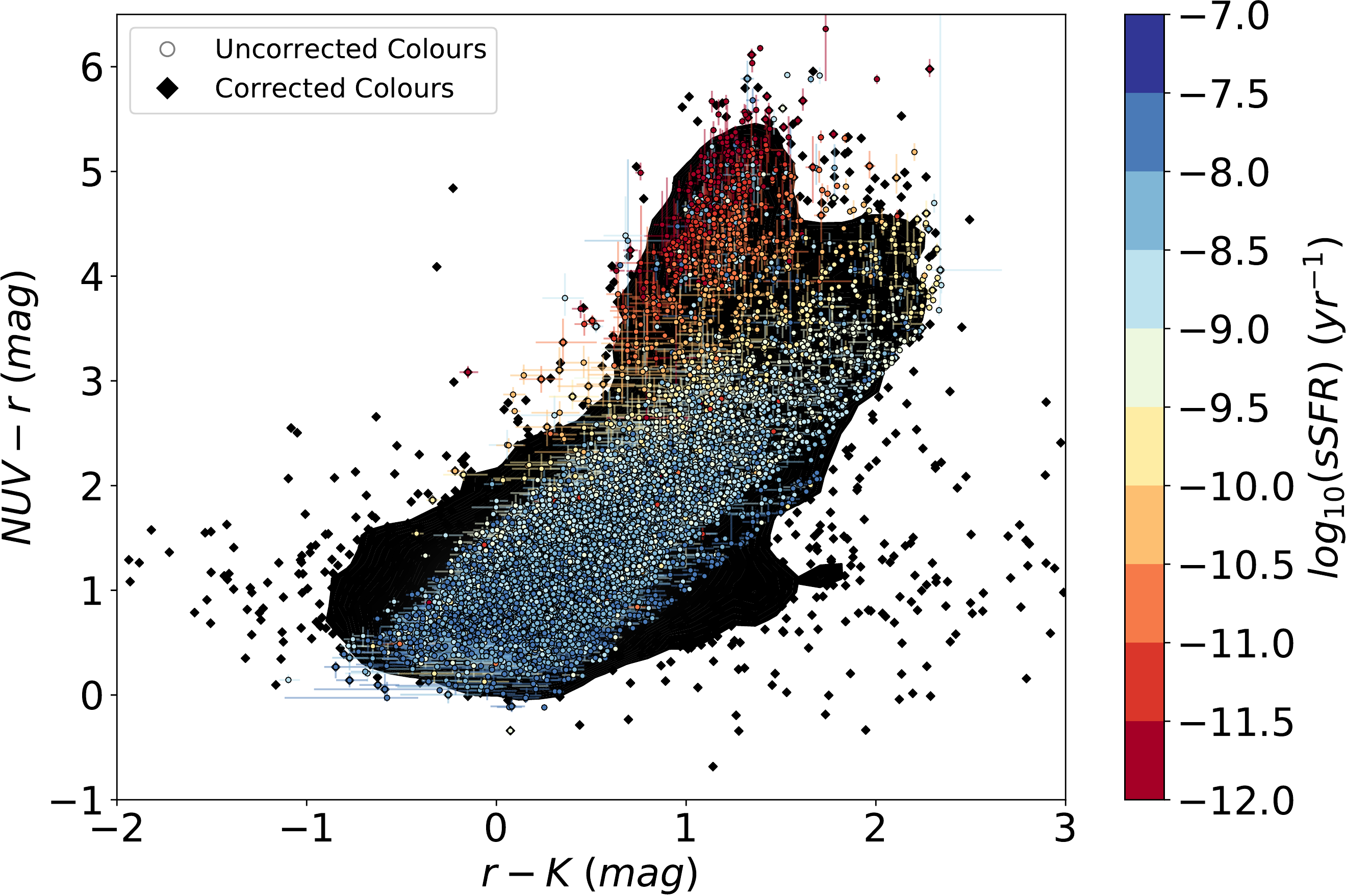}
    \caption{
   {$NUVrK$ distribution of corrected and uncorrected rest-frame colours. Black contours and diamonds indicate the $NUVrK$ distribution of our final photometric sample (i.e., they are the galaxies with corrected colours, shown in the top panel of Fig.~\ref{fig:nuvrk}). The circles show the $NUVrK$ distribution for uncorrected rest-frame colours, colour-coded by sSFR. Uncorrected colours, as opposed to corrected colours, show underestimated and artificially bounded scatter.}
    }
    \label{fig:nuvrk_comparison}
\end{figure}

{
Figure~\ref{fig:nuvrk_comparison} shows the $NUVrK$ distribution of corrected and uncorrected rest-frame colours. We indicate with black contours, and black diamonds outside of the contours, the $NUVrK$ distribution of our final photometric sample; i.e., the galaxies with rest-frame colours as derived in Sec.~\ref{sec:restframecolours} and shown in the top panel of Fig.~\ref{fig:nuvrk}. Overlaid on the contours, the Figure also shows the $NUVrK$ distribution for uncorrected rest-frame colours (circles, colour-coded by sSFR). These uncorrected colours are directly derived from the posterior SED models by integrating the rest-frame posterior SEDs over the filters of interest ($NUV$, $r$, and $K$ here). These are the colours usually calculated by SED-fitting codes, including the new generation of SED-fitting codes which do not suffer from parameter space gridding (i.e., \bagpipes\ here). As mentioned in Sec.~\ref{sec:restframecolours}, such rest-frame colour estimation is often unrealistic, likely due to improper propagation of parameter uncertainties (e.g. redshift), limitations from the explored parameter ranges, or modelling assumptions. These issues can lead to underestimated scatter and uncertainties, and boxy patterns in colour-colour space. As seen in Figure~\ref{fig:nuvrk_comparison}, the uncorrected colours of our photometric sample (circles) indeed show unrealistically small scatter, artificially bounded to a narrow range of $r-K$ colours of about $1$~mag over a wide range of $NUV-r$ colours.
}

\begin{figure}
  \centering
    \includegraphics[width=8.3cm]{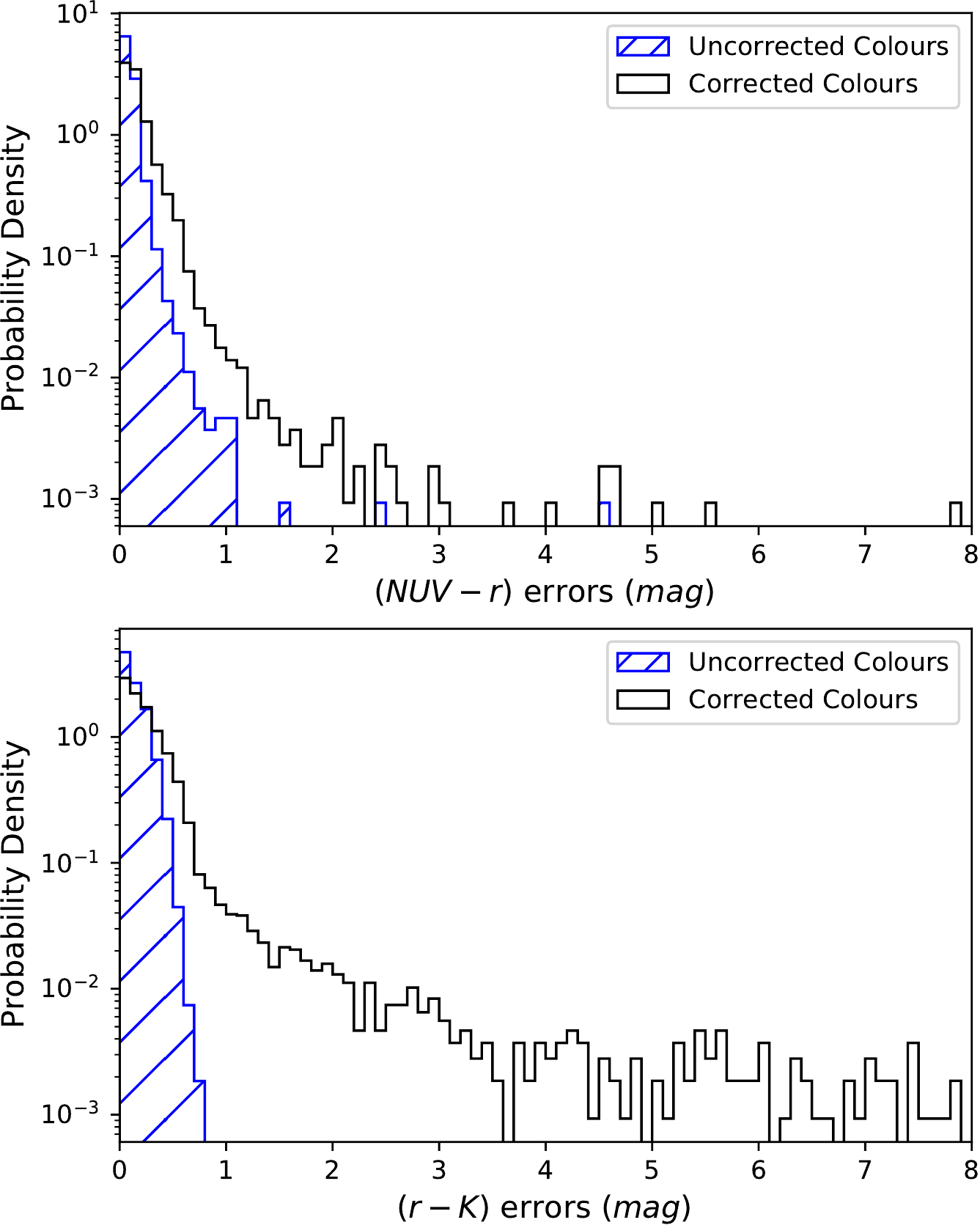}
    \caption{
   {Distribution of rest-frame colour uncertainties. The top and bottom panels show the distribution of $NUV-r$ and $r-K$ uncertainties, respectively. In both panels, the hatched blue histogram indicates the distribution for uncorrected colours, while the black histogram shows the distribution for corrected colours. As seen in both panels, uncorrected rest-frame colours have underestimated uncertainties.}
    }
    \label{fig:nuvrkerrors_comparison}
\end{figure}

{
Figure~\ref{fig:nuvrkerrors_comparison} shows the distribution of $NUV-r$ (top panel) and $r-K$ (bottom panel) uncertainties. Each panel shows the distribution for uncorrected and corrected colours (hatched blue and black histograms, respectively). As expected, the panels show the clear underestimation of colour uncertainties for uncorrected colours, with unrealistically sharp cut-offs below $1$~mag. Note that for $r-K$ colours, this sharp cut-off is likely correlated to the narrow range over which colours are artificially constrained. To alleviate  such undesirable effects and derive realistic rest-frame colours, we therefore use our corrected rest-frame colours throughout the paper. Note that the difference between the median corrected and uncorrected colours is negligible ($<0.1$~mag and $\sim0.1$~mag for $NUV-r$ and $r-K$ colours, respectively), which suggests that the colour correction does not introduce any bias with respect to the uncorrected colours, and only mitigates the underestimated scatter and uncertainties. We refer the reader to Sec.~\ref{sec:restframecolours} for the details of our colour correction. 
}

\bsp	
\label{lastpage}
\end{document}